\documentclass[10pt]{article}

\newtheorem{lem}{Lemma}
\newtheorem{tm}{Theorem}

\newtheorem{de}{Definition}
\newtheorem{cl}{Collorary}
\newtheorem{lw}{Law}

\begin{document}
\pagestyle{plain}
\begin{center}
\vskip10mm
   {\Large {\bf Structure behind Mechanics}}
\vskip7mm
\renewcommand{\thefootnote}{\dag}
{\large Toshihiko Ono}\footnote{ 
e-mail: BYQ02423@nifty.ne.jp \ \ or \ \ 
tono@swift.phys.s.u-tokyo.ac.jp}
\vskip5mm
\par\noindent
{\it 703 Shuwa Daiich Hachioji Residence,\\
4-2-7 Myojin-cho, Hachioji-shi, Tokyo 192-0046, Japan}
\vskip5mm
\end{center}
\vskip10mm
\setcounter{footnote}{0}
\renewcommand{\thefootnote}{\arabic{footnote}}

\begin{abstract}
This paper proposes a basic theory on
physical reality,
 and
a new foundation for quantum mechanics
and classical mechanics.
It does not only solve the problem
of the arbitrariness on the
operator ordering for the quantization procedure,
but also clarifies how the classical-limit occurs.
It further compares the new theory with the
known quantization methods, and
proposes a self-consistent interpretation
for quantum mechanics.
It also provides the internal structure inducing half-integer spin 
of a particle, the 
sense of the regularization
 in the quantum field theory, 
the quantization of a phenomenological system,
the causality in quantum mechanics and 
 the origin of the thermodynamic irreversibility
under the new insight.
\end{abstract}
\vskip10mm

\section{INTRODUCTION}

Seventeenth century saw
 Newtonian mechanics,
published as "{\it Principia:
Mathematical principles of natural philosophy},"
the first attempt
to understand this world under few principles
rested on observation and experiment.
It bases itself  on the concept of the {\it force}
acting on a body and on the laws relating it with the motion.
In eighteenth century,
Lagrange's {\it analytical
mechanics},
originated by Mautertuis' theological work,
 built the theory of motion
on an analytic basis, and replaced forces by potentials;
in the next century, Hamilton 
completed the foundation of analytical mechanics
on the principle of least action in stead of Newton's laws.
Besides,
Maxwell's theory of the electromagnetism
has the Lorentz invariance
inconsistent with the invariance
under Galilean transformation,
that Newtonian mechanics obeys.
Twentieth century dawned with
Einstein's relativity 
changing the ordinary belief on the nature of time,
to reveal the four-dimensional {\it spacetime} 
structure of the world.
Relativity
improved Newtonian mechanics
based on the fact that the speed of light $c$
is an invariant constant,
and revised the self-consistency
of the classical mechanics.
Notwithstanding such a revolution,
 Hamiltonian mechanics was
still effective not only for Newtonian mechanics
but also for the Maxwell-Einstein theory,
and the concept of
{\it energy and momentum}
played the most important role in the physics
instead of force for Newtonian mechanics.

Experiments, however, indicated
that microscopic systems seemed not to obey
such classical mechanics so far. Almost one century has passed
since Planck found his constant $h$;
and almost three fourth since
Heisenberg \cite{Heisenberg},  Schr\"odinger
\cite{Schrodinger} and their contemporaries
constructed
the basic formalism of quantum mechanics
after the early days of Einstein and Bohr.
The quantum mechanics
based itself on the concept of {\it wave functions} 
instead of classical energy and momentum,
or that of operators called as observables.
This mechanics
reconstructed the classical field theories
except the general relativity.
Nobody  denies how quantum mechanics,
especially quantum electrodynamics,
succeeded in twentieth century
and developed in the form of the standard model for the
quantum field theories
through the process to find new particles in the nature.

Quantum mechanics,
however,
seems to have left some fundamental open problems
on its formalism and its interpretation:
the problem
on the ambiguity of the operator ordering
in quantum mechanics \cite{Groenwald,van Hove},
which is crucial to quantize the Einstein gravity for instance,
and that on the reality,
which seems incompatible with the causality \cite{EPR,Bell,Aspect}.
These difficulties come from the problem
how and why quantum mechanics relates itself with classical mechanics:
the relationship between the quantization that
constructs quantum mechanics based on
classical mechanics
and the classical-limit
that induces classical mechanics
from quantum mechanics as an approximation
with Planck's constant $h$ taken to be zero;
the incompatibility between
the ontological feature
of classical mechanics and 
the epistemological feature
of quantum mechanics
in the Copenhagen interpretation \cite{Bohr}.

Now,
this paper
proposes a basic theory on 
physical reality, and
introduces a foundation
for quantum mechanics and classical mechanics,
named as {\it protomechanics}, 
that is motivated in the previous letter \cite{Ono}.\footnote{ The author 
of paper \cite{Ono}, "Tosch Ono," 
is the same person as that of the present paper,
"Toshihiko Ono."}
It also
attempts to revise the
nonconstructive idea
that the basic theory of motion is valid in a
way independent of the describing 
scale, though the quantum mechanics has once
destroyed 
such an idea that Newtonian mechanics held in eighteenth century.
The present theory
supposes that a field or a particle $X$
on the four-dimensional spacetime
has its internal-time $\tilde o_A(X)$ relative
to an domain $A$ of the spacetime,
whose boundary and interior
represent the present and the past, respectively.
It further considers that
object $X$ also has the 
external-time $\tilde o_A^*(X)$ relative
to  $A$  which is the internal-time of all the rest but
$X$ in the universe.
Object $X$ gains the actual existence on  $A$
if 
and only if the internal-time
coincide with the external-time:
\begin{equation}\label{AX=XA}
\tilde o _A(X)= \tilde o_A^*(X) .
\end{equation}
This condition discretizes or quantizes the ordinary time passing from 
the past to the future,
and
enables the deterministic structure of
the basic theory to produce the
nondeterministic characteristics of
quantum mechanics.
The both sides of relation (\ref{AX=XA}) further
obey the variational principle as
\begin{equation}
 \label{dAX=0/dXA=0}
\delta \tilde o_A\left( X \right)  = 0
\ \ \ , \ \ \ \
\delta \tilde o^*_{A } (X) = 0 .
\end{equation}
This relation reveals a geometric
structure behind Hamiltonian mechanics
based on the modified  Einstein-de Broglie relation,
and produces the conservation law
of the emergence-frequency of a particle or a field
based on the introduced quantization law of time.
The obtained mechanics,
protomechanics,
rests on the concept of the 
{\it synchronicity}\footnote{This naming
of synchronicity is originated by Jung \cite{Jung&Pauli}.}
instead of energy-momentum or wave-functions,
that synchronizes two intrinsic local clocks
located at different points in the space of
the objects on a present
surface in the spacetime.
It will finally
solve the problem
on the ambiguity of the operator ordering,
and also give a self-consistent interpretation
of quantum mechanics as an ontological theory.

The next section
explains the basic laws on reality
as discussed above,
and leads to the protomechanics in
Section 3, that produces
the conservation laws of momentum
and that of emergence-frequency.
Section 4 presents the dynamical
construction 
for the introduced protomechanics
by utilizing the group-theoretic method
called Lie-Poisson mechanics (consult {\it APPENDIX}).
It provides the difference between classical
mechanics
and quantum mechanics as that of their {\bf function spaces}:
the function space of the observables 
for quantum
 mechanics includes
that for  classical mechanics;
 the dual space of the emergence-measures
for classical 
mechanics includes
that
for quantum
mechanics, viceversa.
Section 5 and Section
6 explain how  protomechanics
deduces classical mechanics and quantum mechanics,
respectively.
In these sections,
protomechanics proves to include both quantum mechanics
and classical mechanics; in conjunction with
the result in Section 4,
it clarifies how
the quantization and the classical-limit occur.
Section 6 additionally presents
a consequent interpretation for the half-integer spin
of a particle.
Section 7 compares the present theory
with the other known quantization methods
from both the group-theoretic view point
and the statistical one,
and further introduces an interpretation
of the regularization method adopted at quantum field theories;
and it will prove applicable for general phenomenological systems.
On the other hand,
 Section 8 considers how it
gives a self-consistent interpretation of reality
or solves the measurement problem,
and interprets the origin of  thermodynamic irreversibility
in the nature; and
it will prove to keep causality even under an
EPR-experiment.
A brief statement of
the conclusion immediately follows.

Let me summarize the construction of the present paper
in the following diagram.

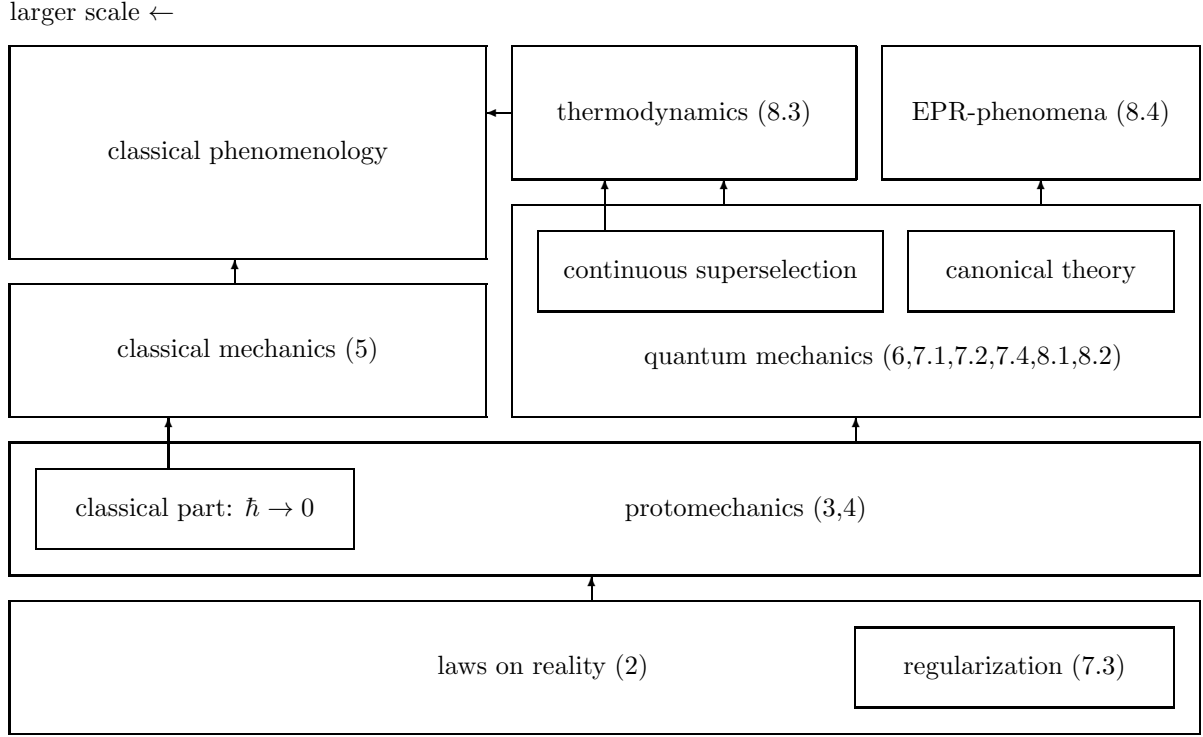
\begin{picture}(480,330)(15,20)

\put(10,220){\framebox(180,80){classical phenomenology}}
\put(10,160){\framebox(180,50){classical mechanics (5)}}

\put(200,250){\framebox(130,50){thermodynamics (8.3)}}
\put(340,250){\framebox(120,50){EPR-phenomena (8.4)}}

\put(200,160){\framebox(260,80){}}
\put(210,200){\framebox(130,30){continuous superselection}}
\put(350,200){\framebox(100,30){canonical theory}}
\put(250,180){quantum mechanics (6,7.1,7.2,7.4,8.1,8.2)}

\put(10,100){\framebox(450,50){\ \ \ \ \ \ \ \ \ \ \ \ \ \ \ \ \ \  
\ \ \ \ \ \ \ \ \ \ \ \ \ \  protomechanics (3,4)}}
\put(20,110){\framebox(120,30){classical part: $\hbar \to 0$}}

\put(10,40){\framebox(450,50){laws on reality (2) \ \ \ \ \ \ \ \ \ \ \ \ \ }}
\put(330,50){\framebox(120,30){regularization (7.3)}}

\put(230,90){\vector(0,1){10}}
\put(70,140){\vector(0,1){20}}
\put(330,150){\vector(0,1){10}}
\put(235,230){\vector(0,1){20}}
\put(280,240){\vector(0,1){10}}
\put(400,240){\vector(0,1){10}}
\put(200,275){\vector(-1,0){10}}
\put(95,210){\vector(0,1){10}}

\put(10,310){larger scale $ \leftarrow $}
\put(380,20){more fundamental $ \downarrow $}
\put(20,20){{\small * Numbers in bracket $( \ )$ refer those of sections.}}
\end{picture}
{}\\
{}\\

In this paper, 
$c$ and
 $h$
denote the speed of light and Planck's constant, respectively.
I will
use Einstein's rule
in the tensor calculus
for Roman indices'
$i, j, k \in {\bf N}^N$
and Greek indices' 
$\nu , \mu  \in {\bf N}^N$,
and not for Greek indices' 
$\alpha , \beta , \gamma \in {\bf N}^N$,
and I
further denote the trace (or supertrace) operation
of a quantum observable $\hat F$ as
$ \langle \hat F  \rangle $
that is only one difference 
from the ordinary notations in quantum mechanics,
where $i=\sqrt{-1}$.
Consult
the brief review on
the differential
geometry in {\it APPENDIX A}
and that on Lie-Poisson mechanics
in {\it APPENDIX B},
which the employed notations follow.
In addition,
notice that the basic theory
uses so-called {\bf c-numbers},
while it also utilize {\bf q-numbers} 
to deduce the quantum mechanics in 
Section 6 for the help of calculations.\footnote{
Such distinction between c-numbers and q-numbers
 does not play an important
role in the present theory.}

\section{LAWS ON REALITY}

Let $M^{(4)}$  represent 
the spacetime, being a four-dimensional oriented
$C^{\infty }$ manifold,
that has the topology
or the family $\tilde {\cal O}={\cal O}_{M^{(4)}}  $ of its open subsets, 
the topological $\sigma $-algebra
${\cal B}\left( {\cal O}_{M^{(4)}} \right) $,
and the volume measure $v^{(4)}$ induced from the metric $g$
on $M^{(4)}$.\footnote{Spacetime $M^{(4)}$ 
may be endowed with some additional structure.}
We shall certainly choose an
arbitrary domain $A\in \tilde {\cal O} $
in the discussion below,
but we are interested in
the case that domain
$A$ represents the past at a moment whose boundary
$\partial A$ is a three-dimensional present hypersurface
in $M^{(4)}$.

The space ${\tilde M} $ represents
that of the objects
whose motion will be described,
and has a projection operator $ \chi _A:{\tilde M} \to {\tilde M}$
for every domain $A\in \tilde {\cal O}$
such that $\chi_A^2=  \chi _A$.
Every object $X\in {\tilde M} $ has
its own domain $D(X)$ such that
\begin{equation}
\chi_{D(X) \setminus A } (X) =X \ \ \ \ \Longleftrightarrow 
\ \ \ \ D(X) \cap A =\emptyset .
\end{equation}
In particle theories,
${\tilde M} $ is identified with the space
 of all the one-dimensional
timelike mani-folds or curves in $M^{(4)}$,
where 
 $\chi_A \left( l\right) =l \cap A $ for every domain $A$
and $D(l) =l$.
In field theories,
the space $\Psi \left( M^{(4)}, V\right) $
of the complex valued or
${\bf Z}_2$-graded fields over $M^{(4)}$
such that $\psi^{(4)} \in \Psi \left( M^{(4)}, V\right)  $ is
a mapping $\psi^{(4)}  : M^{(4)} \to V$
for a complex valued or
${\bf Z}_2$-graded vector space $V$.
Mapping $\chi_A $ satisfies that
 $\chi_A  \left( \psi^{(4)} \right) (x) 
=  \psi^{(4)}  (x) $ if $x\in A$
and that
$\chi_A  \left( \psi^{(4)} \right) (x) 
=  
0 $ if $x\not\in A$, 
and $D(\psi^{(4)})$
gives  the support of $\psi^{(4)}$:
$D(\psi^{(4)}) = supp (\psi^{(4)})$.

In addition,
let us consider
the set ${\cal D} (\tilde M)$ of 
all the differentiable mapping from $\tilde M$ to itself
and the set ${\cal D} ( M^{(4)})$ of 
all the diffeomorphisms of spacetime $M^{(4)}$.
In particle theories,
set ${\cal D}(\tilde M )$ will be regarded as set ${\cal D}(M^{(4)})$;
and,
in field theories, it is the set 
of all the linear transformations of
a field
such that
$\Phi \left( \psi^{(4)}\right) = \psi^{(4)}+ \phi^{(4)} $.

Now,
let us assume
that an object
has its own internal-time 
relative to a domain of the spacetime.

\begin{lw}\label{Variation A}
For every domain $A\in \tilde {\cal O}$,
the mapping
 $\tilde o_A : 
 {\tilde M}  \to S^1$  has an action $S _A :  \tilde M  \to {\bf R} $
and equips  an object $X\in {\tilde M}$
with the internal-time
$\tilde o_A(X) $:
\begin{equation}\label{original Feynamm's rule}
\tilde o _A \left(  X \right) = e^{i 
S_A ( X ) }  .
\end{equation}
\end{lw}
For particle theories,
a one-dimensional submanifold or a curve
$l\subset M^{(4)}$
 represents
the nonrelativistic motion for a particle 
 such that
$\left( t, x(t) \right)  \in l 
 $ for $t\in T$,
 where
$M^{(4)}$ is the Newtonian spacetime $M^{(4)}= T\times 
M^{(3)}$ for the
Newtonian time $T\subset {\bf R}$
and the three-dimensional Euclidean space $M^{(3)}$;
thereby, it 
has
the following action 
for the ordinary Lagrangian $L :TM\to {\bf R}$:
\begin{equation}\label{nonrelativistic action}
S _A\left( l 
\right)  = {\bar h}^{-1} \int_ {l\cap A} d t  \ L \left( x(t)
, {{dx(t)}\over {dt}}
\right)  ,
\end{equation}
where 
$\bar h= h/ {4\pi } $ or
$=\hbar / 2$ for Planck's constant $h$ ($\hbar =h/ 2\pi $).
The
relativistic motion
of a free particle whose mass is $m$
has the following action for the proper-time $\tau \in {\bf R}$:
\begin{equation}\label{relativistic action}
S _A\left( l 
\right)  = {\bar h}^{-1} \int_ {l\cap A} d \tau  \ mc^2  .
\end{equation}
 For field theories,
field variable $X =\psi^{(4)} $ over spacetime $M^{(4)}$
 has the 
following action
for the Lagrangian density ${\cal L}_M$ of matters:
\begin{equation}\label{standard action}
S _{A}\left( \psi^{(4)}
\right)   ={1\over {\bar h c}} \int_{A}
  d v^{(4)}   \left( y  \right) \ {\cal L}_M  \left(  \psi^{(4)}  (y),
d\psi^{(4)}  (y)
\right)   ,
\end{equation}
where $ v^{(4)} $ is the volume measure of $M^{(4)}$.
In the standard field theory,
 $ \psi ^{(4)}$ is a set of ${\bf Z}_2$-graded fields
over spacetime $M^{(4)}$,
the Dirac field for fermions,
 the Yang-Mills field for gauge bosons
and other field under consideration.
For  the
Einstein gravity,
the
Hilbert action
includes
the metric tensor $g$ on $M^{(4)}$ 
with
a cosmological constant $\Lambda \in {\bf R}$:
\begin{eqnarray}
\label{Hilbert action}\nonumber
S _{A}\left(  \psi^{(4)}, g
\right)   &=&{1\over {\bar h c}} \int_{A}
  d v_g^{(4)}   \left( y  \right) \ {\cal L}_M\left(  
\psi  ^{(4)}(y),
d\psi  ^{(4)}(y)
\right) \\
& \ & \ \ \ -  {1\over {\bar h c}} \int_{A}
  d v_g^{(4)} \left( {{c^4} \over {16 \pi G}} R_g +\Lambda \right)
 - {2\over {\bar h c}} \int_{\partial A}
  d v_g^{(3)} {{c^4} \over {16 \pi G}} K_g  ,
\end{eqnarray}
where $R_g$ and $K_g$ are the
four-dimensional and the extrinsic three-dimensional  scalar curvatures on 
domain $A$ and on its boundary ${\partial A}$;
and $G$ is the Newton's constant of gravity.
The last
term of (\ref{Hilbert action}) is necessary to
produce the correct Einstein equation for gravity \cite{Wald}.

Let us now consider
the subset ${\cal D}_A(\tilde M)$ of set ${\cal D}(\tilde M)$
such that
every element $ \Phi\in {\cal D}_A(\tilde M)$
satisfies
$ \chi_{D(X)\setminus A}(\Phi (X) ) = X $,
and assume it as a infinite-dimensional Lie group.
In particle theories,
set ${\cal D}_A(\tilde M)$ is the set ${\cal D}_A(M)$
of all the diffeomorphisms
of $M$ such that
$\Phi (l) \setminus A = l \setminus A $;
and,
in filed theories,, it is the set 
of all the linear transformations of
a field
such that
$\Phi \left( \psi^{(4)}\right) = \psi^{(4)}+ \phi^{(4)} $ 
for an element $\phi^{(4)}\in \Psi \left( M^{(4)}, V\right) $
and that $ \phi^{(4)}(x)=0$  if $x\not\in A $.
Mapping  $\tilde o_A $ may have the symmetry
under a transformation $\Phi\in
{\cal D} (\tilde M)$  such that
it satisfies 
the following relation for every pair $(A,X)$:
\begin{equation}\label{sym}
\tilde o_{ A}\left( \Phi (X)\right) = \tilde o_A (  X) .
\end{equation}
Such symmetry 
verifies the existence of the conserved charge.

Object $X$ and all the rest but $X$
composes
the universe $U$.
The internal-time $\Pi_A (U)$ of 
universe $U$ relative to domain $A$
would be separated into two parts:
\begin{equation}
\Pi_A (U) = \tilde o_A (X) \cdot \tilde o^*_A ( X) .
\end{equation}
Let us call $\tilde o^*_A ( X)\in S^1$ as the external-time
of $X$ relative to $A$.
Thus, the external-time of universe $U$
would always be unity: $\Pi^*_A (U) =1$.

\begin{lw}\label{Variation A2}
For every domain $A\in \tilde {\cal O}$,
the mapping
 $\tilde o_A^* : 
 {\tilde M}  \to S^1$  has an action $S^* _A :  \tilde M  \to {\bf R} $
and equips  an object $X\in {\tilde M}$
with the external-time
$\tilde o^*_A(X) $:
\begin{equation}\label{original Feynamm's rule}
\tilde o ^* _A\left(  X \right) = e^{i 
S_A^*  ( X ) }  .
\end{equation}
\end{lw}
Let us also introduce the mapping $\tilde s_A  \left(  \tilde o  \right)
:
\tilde M  \to S^1$
that relates mappings $\tilde o^*_A $ and $\tilde o_A$:
\begin{equation}\label{original Emergencer}
\tilde o^*_{A} (X) 
= \tilde o _A\left( X \right)
\cdot \tilde s_A  \left(   \tilde o  \right) (X) .
\end{equation}
It has a function $R_A\left( \tilde o \right) $ such that
\begin{equation}
\tilde s_A\left(   \tilde o \right) (X)
= e^{i R_A\left(  \tilde o \right) (X) } .
\end{equation}
There is also the mapping $ \tilde s_{A }^*  \left(  \tilde o ^* \right) :
\tilde {\cal O}   \to S^1$:
\begin{equation}
\tilde o^*_{A} (X)
\cdot \tilde s_{A }^* \left( \tilde o^* \right) (X)
 = \tilde o _A\left( X \right) .
\end{equation}
Mapping  $\tilde \eta ^*_A$ may have the symmetry
under a transformation  $\Phi\in
{\cal D} (\tilde M)$ 
 such that
it satisfies 
the following relation for every pair $(A,X)$:
\begin{equation}
\tilde o^*_{A}\left( \Phi (X)\right) = \tilde o^*_A (  X) .
\end{equation}
If mapping  $\tilde \eta_A$ also
has symmetry (\ref{sym}) for the same transformation $\Phi $,
they must satisfy the following invariance:
\begin{equation}
\tilde s_A\left( \tilde o\right) \left( \Phi (X)\right) = 
\tilde s_A \left( \tilde o\right)  (  X) 
\ \ \ , \ \ \ \ 
\tilde s^*_A \left(
\tilde o^*\right) \left( \Phi (X)\right) = 
\tilde s^*_A \left(\tilde o^*\right)  (  X) .
\end{equation}

The following law further
supplies
the condition that an
object 
has the actual existence on a domain of the spacetime.

\begin{lw}\label{internal-external}
Object $X\in \tilde M$ 
has actual existence on domain $A\in \tilde {\cal O} $
when the internal-time coincides with the external-time:
\begin{equation}\label{compa}
\tilde o^*_{A} (X) = \tilde o _A\left( X \right) .
\end{equation}
\end{lw}
Relation (\ref{compa})
requires the following quantization
condition:
\begin{equation}\label{uncer}
\tilde s_A \left(  \tilde o \right) (X) =1,
\end{equation}
or equivalently,
\begin{equation}
\tilde s_{A}^* \left(  \tilde o^* \right) (X)=1,
\end{equation}
which 
quantizes spacetime $M^{(4)}$
for 
an object $X\in \tilde M$.

For the space $ d_A(\tilde M)$ of all the infinitesimal generators
of ${\cal D}_A(\tilde M)$,
let us consider
an arbitrary element
$\Phi_{\epsilon }\in {\cal D}_A(\tilde M)$,
differentiable by parameter $\epsilon \in {\bf R}$:
\begin{equation}
\lim_{\epsilon \to 0} {{d\Phi_{\epsilon }}\over {d\epsilon }}
 \circ \Phi_{\epsilon }^{-1} 
= \xi \in d_A(X) .
\end{equation}
Thus, we can
introduce
the variation $\delta $ 
as follows:
\begin{eqnarray}
\left\langle 
i \tilde o_A \left(  X \right)  ^{-1 } 
\delta \tilde o_A \left(  X \right) , \xi\right\rangle
&=& i \tilde o_A \left(  X \right) ^{-1 } 
\left. {d\over {d\epsilon }}\right\vert_{\epsilon =0}
\tilde o_A \left(  \Phi _{\epsilon }(X) \right) ,\\
\left\langle 
i \tilde o_A^* \left(  X \right)  ^{-1 } 
\delta \tilde o_A^* \left(  X \right) , \xi\right\rangle
&=& i \tilde o_A^* \left(  X \right) ^{-1 } 
\left. {d\over {d\epsilon }}\right\vert_{\epsilon =0}
\tilde o_A^* \left(  \Phi _{\epsilon }(X) \right) 
\end{eqnarray}
where $\langle \ \ , \ \ \rangle :
 d^*_A(\tilde M) \times  d_A(\tilde M) \to {\bf R}$
is the natural pairing for the dual space $ d^*_A(\tilde M)$
of $ d_A(\tilde M)$.
This variation satisfies the variational principle
of the following law.

\begin{lw}\label{variational}
Object $X\in \tilde M$ must
 satisfy the variational principle
for every domain $A\in \tilde {\cal O} $:
\begin{equation}
 \label{v-pr}
\delta \tilde o _A (X)  = 0
\ \ \ , \ \ \ \
\delta \tilde o^*_{A } (X) = 0 .
\end{equation}
\end{lw}
Thus, Law 4  keeps
Law 3 under the above variation,
and also has the following
expression:
\begin{equation}
 \label{v-pr2}
\delta \tilde s _A\left( \tilde o \right) (X)  = 0
\ \ \ , \ \ \ \
\delta \tilde s ^*_A\left( \tilde o^* \right) (X) = 0 .
\end{equation}

Now, we will
consider 
the mapping
${\cal P} : T \to \tilde {\cal O} $
for 
the time $T\subset {\bf R}$ of an observer's clock $T$.
Domain
${\cal P}(t)$ and
its boundary $\partial  {\cal P} (t )
= \overline{{\cal P} (t )}\setminus {\cal P} (t )$
represent
 the {\it past} and the {\it present}
at time $t \in T$,
where $\overline{A}$ is the closure of $A\in \tilde {\cal O}$;
and it satisfies the following conditions:
\begin{enumerate}
\item
for every $X\in \tilde M$,
$
 t_1 < t_2  \in T \ \ \Rightarrow \ \
{\cal P} (t_1 )  \cap D (X )  \subset {\cal P} (t_2 )  \cap D (X ) $
 (ordering);
\item 
for every $X\in \tilde M$,
the present $
\partial  {\cal P} (t ) \cap D (X ) 
$ is a spacelike hypersurface in $M^{(4)}$
for every time $t\in T$ (causality).
\end{enumerate}
From Law 3, object $X$  emerges into the world  at
time $t \in T$
when it satisfies
\begin{equation}\label{unit}
 \tilde s_{{\cal P}(t)} \left( \tilde o \right) (X)=1\ .
\end{equation}
This condition
of the emergence
determines when object $X$ interacts with all the 
rest in the world,
and discretizes time $T$
in Whitehead's philosophy 
\cite{Whitehead}. In other words,
what a particle or a field $ X$ 
gains actual existence or
emerges into the world, here,
means that it becomes exposed to or has the possibility to interact 
with the other elements or with the ambient world
excluded from the description.
Such occasional influences from the unknown factors
can break the deterministic feature of 
the above description; and it would cause the irreversibility
in general as considered in Subsection 8.3.
The emergence further allows
the observation
of a particle or a field
through an experiment even if the device or its environment is
included in the description as shown in Subsection 8.2.
Besides, the variational principle of Law 4
produces the equation of motion and
the conservation
of the frequency of such emergence in the next section.

\section{Foundation of Protomechanics}

Let us consider
the development of present 
$\partial {\cal P}(t)$ for short time $T=(t_i,t_f) \subset {\bf R}$,
keeping the following description
without the appearance of singularity;
and suppose 
that the time interval extends long enough
to keep the continuity of time
beyond the discretization in the previous section,
where such discretization
would only affect the property of the emergence-measure,
defined below, corresponding to the density matrices
in quantum mechanics.
Assume that present
$\partial {\cal P}(t)$ is diffeomorphic to
a three dimensional manifold $M^{(3)}$
by a diffeomorphism
$\sigma_t :M^{(3)} \to \partial {\cal P}(t)  $
for every $t\in T$.
It induces a corresponding mapping
$\tilde \sigma_t : \tilde M \to M $
for the space $M$
that is three-dimensional
physical space $M^{(3)}$ for particle theories
or the space $M= \Psi (M^{(3)},V)$ 
of all the $C^{\infty }$-fields 
over $M^{(3)}$ for field theories.
For particle theories, mapping
$ \tilde \sigma_t  $ is defined as
$\tilde \sigma_t (l) =
\sigma_t ^{-1} \left( l\cap  \sigma_t (M^{(3)}) \right) $
for a curve $l\subset M^{(4)}$;
for field theories, it is defined as
$ \tilde \sigma_t (\psi^{(4)})  =
\psi^{(4)} \circ \sigma_t $ for a
field $\psi^{(4)} $.

Let us assume that space $M$ is a $C^{\infty }$ manifold
endowed with an appropriate
topology and the induced topological $\sigma $-algebra.\footnote{
$M $ is assumed as an ILH-manifold modeled by the Hilbert space
endowed with an inverse-limit topology (consult \cite{Omori}).}
We will denote the tangent space as $TM$
and the cotangent space $T^*M$;
and we shall consider
the space of all the vector fields over $M$ as $X(M)$
and that 
of all the 1-forms over $M$ as $\Lambda^1(M)$.
To
add a one-dimensional cyclic freedom $S^1$ at
each point of $M$ introduces
the trivial
$S^1$-fiber bundle $E (  M ) $ over $M$.\footnote{The 
introduced freedom would
{\it not} represent what is corresponding to the
{\it local clock} in Weyl's sense
or the {\it fifth-dimension}
in Kaluza's sense \cite{Kaluza} for the 
four-dimensional spacetime  $M^{(4)}$.
To consider such freedom,
 $M^{(4)}$ would be extended to the 
principal fiber-bundle
over $M^{(4)}$ with a $N$-dimensional 
special unitary group
$SU(N)$.}
Fiber $S^1$ represents an 
intrinsic clock of a particle or a field,
which is located at every point on $M$.
For
the space $\Gamma  [ E (  M ) ] $
of all the  global sections
of $E (  M ) $,
every element $ 
\eta \in \Gamma  [ E ( M)   ] $
now represents the system
that a particle or a field
belongs to and carries with,
and a synchronization of every two clocks
located at different points in space $ M$.

For past ${\cal P}(t)$ such that  
$\partial {\cal P}(t) = \sigma_t (M^{(3)})$,
there is an mapping $o_t : TM \to {\bf R}$ such that
every initial position $(x_0, \dot x_0 )\in TM$ 
has an object $X\in \tilde M_{{\cal P}}$
satisfying the following relation
for $x_t=\tilde \sigma_t (X)$:
\begin{equation}\label{def eta}
o_t \left( x_t,  \dot x_t \right) = 
\tilde o_{{\cal P}(t)} \left( X \right)  .
\end{equation}
For the velocity field $v_t \in X(M)$ 
such that $v_t \left( x_t\right) = 
 {{dx_t}\over {dt}} $, we will introduce a
section $\eta_t \in \Gamma \left[ E(M)\right] $
and call it {\it synchronicity} over $M$:
\begin{equation}\label{def eta}
\eta_t (x)=o_t \left( x, v_t (x) \right)   .
\end{equation}
The Lagrangian $L_t^{TM}: TM \to {\bf R}$ 
 characterizes
the speed of the internal-time:
\begin{equation}\label{internal velocity}
L^{TM}_t\left( x_t, {{dx_t}\over {dt}} \right) 
 = -i\bar h   o_t \left( x_t,
{{dx_t}\over {dt}} \right)^{-1} {d\over {dt}}o_t \left( x_t,
{{dx_t}\over {dt}} \right) .
\end{equation}
Since relation (\ref{internal velocity})
is valid for every initial conditions of position $
(x_t,\dot x_t) \in TM$,
it determines the time-development of synchronicity $\eta_t $
in
the following way
for the Lie derivative ${\cal L}_{v_t}$
by velocity field
$v_t \in X(M)$:
\begin{equation}\label{internal Lie velocity}
L_t^{TM}\left( x, v_t (x)\right) =
-i {\bar h}  \eta_t   (x) ^{-1}\left( {\partial \over {\partial t }}
+ {\cal L}_{v_t} \right) \eta_t   (x)  .
\end{equation}

Let us now consider
the mapping $ p :
\Gamma [ E (  M )   ]
\to \Lambda^1(M)
$ satisfying  the following relation:
\begin{equation}
p \left( \eta_t \right) =
- i \bar h  \eta_t ^{ -1}
d \eta_t .
\end{equation}
If 
the energy
$ E_t \left( \eta_t \right) :TM \to {\bf R}$
is defined as 
\begin{equation}
E_t \left( \eta_t \right)\left( x \right) =  i\bar h
\eta_t   (x)   ^{-1 }
{{ \partial   }\over {\partial t }} 
\eta_t   (x) ,
\end{equation}
condition (\ref{internal Lie velocity})
satisfies the following relation:
\begin{equation}
\label{H-L rel.}
E_t \left( \eta_t \right)
 (x)  
=
 v_t  (x) \cdot p \left( \eta_t \right) (x)   - 
L^{TM}_t \left( x, v_t(x)  \right) .
\end{equation}

Attention to
the following calculation by definition (\ref{internal Lie velocity}):
\begin{equation}\label{calc-1}
-i\bar h {\partial \over {\partial v }}
\left\{ o_t \left( x, v_t( x)   \right)^{-1}
\left( {\partial \over {\partial t }}
+ {\cal L}_{v_t} \right) o_t \left( x, v_t( x)  \right) \right\} =
{{\partial L^{TM}_t}\over {\partial v}}\left( x,
v_t  (x)
\right) .
\end{equation}
Since variational principle
(\ref{v-pr}) in
Law 1 implies that $ o_t \left( x, \dot x  \right) $
is invariant under the variation
of $\dot x $ at every point $(x, \dot x)$, i.e.,
\begin{equation}\label{variation>>motion}
{\partial \over {\partial \dot x }}o_t \left( x, \dot x  \right)  = 0
\ \ \ \ \Longleftrightarrow  \ \ \ \ 
{\partial \over {\partial v }}o_t \left( x, v_t (x) \right)    = 0
\end{equation}
then formula (\ref{calc-1})
has the following different expression:
\begin{eqnarray}
\nonumber
-i\bar h {\partial \over {\partial v }} \left\{
o_t \left( x, v_t( x)   \right)^{-1}
\left( 
{{\partial } \over {\partial t }}
+ {\cal L}_{v_t} 
\right) 
o_t \left( x, v_t( x) \right) \right\}
 &=&
{{\partial } \over {\partial v }} \left\{
v_t(x)\cdot  p \left( \eta_t   \right) (x) \right\}
\\ \label{calc-2}
&=&
p \left( \eta_t   \right) (x ) .
\end{eqnarray}
Equations (\ref{calc-1}) and (\ref{calc-2})
leads to the {\it modified}
Einstein-de Broglie relation, that was
$p=h / \lambda $ for Planck's constant $h = 2\pi \hbar $
and wave number $\lambda $
in quantum mechanics:
\begin{equation}\label{p-lagrange}
 p \left( \eta_t   \right) (x) = 
{{\partial L^{TM}_t}\over {\partial v}}\left( x,
v_t  (x)
\right) .
\end{equation}
Notice that this
relation (\ref{p-lagrange})
produces the Euler-Lagrange equation
resulting from
the classical least action principle:
\begin{eqnarray}
\label{E-L}
d L^{TM}_t  \left( x,
v_t ( x)
\right)
- \left( {{\partial }\over {\partial t}}+{\cal L}_{v_t} \right)
{{\partial L^{TM}_t}\over {\partial v}}\left( x,
v_t ( x)
\right) &=&0 \\
\label{E-L4}
\Longleftrightarrow \ \ \ \ \ 
{ { \partial L^{TM}_t  }\over {\partial x^j}}\left( x_t,
\dot x_t
\right)
- {{d}\over {dt}}
{{\partial L^{TM}_t}\over {\partial \dot x^j}}\left( x_t,
\dot x_t
\right) &=&0 \ ;
\end{eqnarray}
thereby, relation (\ref{p-lagrange})
is stronger condition than the classical relation  (\ref{E-L4}).

Under the modified Einstein-de Brogie relation (\ref{p-lagrange}),
relation (\ref{H-L rel.})
gives the Legendre transformation and
introduces Hamiltonian $H_t^{T^*M}$ 
as a real function on cotangent space $T^*M$
such that 
\begin{equation}
E_t \left( \eta_t   \right)
 (x)   =
H_t^{T^*M}\left( x ,p \left( \eta_t   \right) (x)  \right) .
\end{equation}
This satisfies the first equation of Hamilton's canonical
equations of motion:
\begin{equation}\label{velocity-H}
v_t (x)= {{\partial H^{T^*M}_t}\over {\partial p}}\left( x,
p\left( \eta_t  \right) (x)
\right) .
 \end{equation}
Solvability
$\left[ {\partial \over {\partial t}}, d \right] =0$ further
leads to
 the second equation
of Hamilton's canonical equations
of motion:
\begin{equation}\label{solvability}
{\partial \over {\partial t}} p \left( \eta_t   \right) (x)  = 
- d H^{T^*M}_t   \left( x,  p \left(  \eta_t   \right) (x) \right)  ,
\end{equation}
which is equivalent to
equation (\ref{E-L}) of motion under condition (\ref{p-lagrange}).
If Lagrangian $L_t^{TM} $ satisfies
\begin{equation}\label{cons-T}
{{\partial L_t^{TM} }\over {\partial t}}=0 \ ,
\end{equation}
then
equations (\ref{velocity-H}) and (\ref{solvability})
of motion 
prove the conservation
of energy:
\begin{equation}\label{conservation of energy}
\left( {{\partial }\over {\partial t}}+{\cal L}_{v_t}\right) H_t^{T^*M}
\left( x,  p \left(  \eta_t   \right) (x) \right)  =0 .
\end{equation}

On the other hand,
the mapping $ \tilde s_{{\cal P}(t)}\left( \tilde o\right) $
induces a
mapping
$s_t ( o_t   ) :TM \to S^1 $
such that every initial position $(x_0,\dot x_0) \in TM$
has an object $X\in \tilde M_{{\cal P}}$
satisfying the following relation:
\begin{equation}
s_t (  o_t )  \left( x_t, {{dx_t}\over {dt}}\right) = \tilde s_{{\cal P}(t)}
\left(  \tilde o\right) (X).
\end{equation}
For velocity field $v_t$,
we can define the following section
$\varsigma_t \left( \eta_t   \right)  \in \Gamma \left[ E(M)\right] $
and call it {\it shadow} over $M$:
\begin{equation}
\varsigma_t \left( \eta_t   \right) (x) =
s_t (  o_t )  \left( x , v_t(x)\right) .
\end{equation}
Condition (\ref{unit}) of emergence now
has the following form:
\begin{equation}
s_t\left(  o_t\right)  \left( x_t, {{dx_t}\over {dt}}\right) =1
\ \ \ \ \Longleftrightarrow \ \ \ \
\varsigma_t \left( \eta_t   \right) (x) =1,
\end{equation}
when synchronicity
$\eta_t $ comes across the 
section $\eta_t^*=\eta_t
\cdot \varsigma_t\left( \eta_t\right)  $ at position $ x \in M$.
Let us introduce 
the function $T_t( o_t )^{TM}: TM \to {\bf R}$ such that
\begin{equation}\label{internal velocity 2}
T_t(o_t )^{TM}\left( x_t, {{dx_t}\over {dt}} \right)  
= -i  \bar h
 s_t(o_t )\left( x_t, {{dx_t}\over {dt}} \right) ^{-1}
{d\over {dt}}s_t(o_t ) \left( x_t, {{dx_t}\over {dt}} \right) 
.
\end{equation}
Since relation (\ref{internal velocity 2})
is valid for every initial conditions of position $x_t \in M$,
it determines the time-development of shadow $\varsigma_t\left( \eta_t\right)  $
in
the following way
for the Lie derivative ${\cal L}_{v_t}$
by the velocity field
$v_t \in X(M)$ such that $v_t \left( x_t\right) =  {{dx_t}\over {dt}} $:
\begin{equation}\label{s-Lie}
T_t(o_t )^{TM}\left( x, v_t(x) \right)  =
 -i  \bar h \varsigma_t \left( \eta_t   \right) ^{-1}
\left\{ 
{{\partial }\over {\partial t}}
+ {\cal L}_{v_t}\right\} \varsigma_t \left( \eta_t   \right) \ .
\end{equation}
In stead of Hamiltonian
for a synchronicity,
we will consider
the emergence-frequency $f_t \left( \eta_t \right)
:  M \to {\bf R}$ for a shadow such that
\begin{equation}
 2 \pi \bar h    f_t
\left( \eta_t \right) (x)  = i 
\bar h \varsigma_t \left( \eta_t   \right) (x)  ^{-1} {{
\partial  }\over {\partial t}} \varsigma_t \left( \eta_t   \right) (x) ,
\end{equation}
which represents
the frequency that a particle or a field
emerges into the world.
Condition (\ref{s-Lie})
satisfies the following relation:
\begin{equation}
\label{H-L rel.2}
  2 \pi \bar h     f_t \left( \eta_t \right)
(x)=  
v_t  (x) \cdot p \left( s_t \left( \eta_t \right) \right) (x)    
- T_t (o_t )^{TM}\left( x, v_t(x)  \right) .
\end{equation}

Variational principle
(\ref{v-pr2}) from
Law 4 implies that $ s_t(o_t ) \left( x, \dot x  \right) $
is invariant under the variation
of $\dot x $ at every point $(x, \dot x)$, i.e.,
\begin{equation}\label{variation>>emergence}
{\partial \over {\partial \dot x }}s_t (o_t ) \left( x, \dot x  \right)  = 0
\ \ \ \ \Longleftrightarrow \ \ \ \
{\partial \over {\partial v }}
s_t (o_t ) \left( x, v_t( x ) \right)   = 0,
\end{equation}
which
leads to the following relation
corresponding to the {\it modified}
Einstein-de Broglie relation for synchronicity $\eta_t$:
\begin{equation}\label{p-lagrange: emergence}
 p \left( \varsigma _t\left( \eta_t \right)  \right) (x) = 
{{\partial T^{TM}_t (o_t)}\over {\partial v}}\left( x,
v_t  (x)
\right) .
\end{equation}
Relation (\ref{p-lagrange: emergence})
proves the conservation
of emergence-frequency
in the same way as 
relation (\ref{p-lagrange}) proved that of energy
(\ref{conservation of energy}):
\begin{equation}
\left( {{\partial }\over {\partial t}}+{\cal L}_{v_t}\right) f_t
\left( \eta_t\right) (x)  =0 .
\end{equation}
Notice that emergence-frequency $ f_t\left( \eta_t \right)
 $ can
be negative as well as positive,
and that it 
produces a similar property of
the Wigner function
for a
wave function in quantum mechanics as discussed in Section 6.

In addition,
the probability measure $\tilde \nu $
on $\tilde M $
induces
 the probability measure $\nu_t $ 
on $M$ at time $t\in T$
such that
\begin{equation}
 d\nu_t \left( x_t,
{{dx_t}\over {d t}}
\right)   =  d\tilde \nu (X),
\end{equation}
that represents the ignorance of the initial
position in $M$;
thereby it satisfies the  conservation law:
\begin{equation}
{{d}\over {d t}}d\nu_t  \left( x_t,
{{dx_t}\over {d t}}
\right)   =0 .
\end{equation}
This relation can be described by using the Lie derivative 
${\cal L}_{v_t}$  as
\begin{equation}
\left( {{\partial }\over {\partial t}}
+{\cal L}_{v_t}\right) d\nu_t \left( x, v_t(x) \right)  =0 .
\end{equation}
Since
the velocity field $v_t$
has relation (\ref{velocity-H}) with
synchronicity $\eta_t$,
we can define the {\it emergence-measure}
 $\mu_t \left( \eta_t \right) $
as the product of  the probability measure
with the emergence-frequency:
\begin{equation}
d\mu_t \left( \eta_t \right) (x)= d\nu_t \left( x, v_t(x) \right)  \cdot 
f_t
\left( \eta_t \right)  (x)  .
\end{equation}
Thus, we will obtain the following equation of motion
for emergence-measure 
$d\mu_t(\eta_t )$:
\begin{equation}\label{quotient}
\left( {{\partial }\over {\partial t}}
+{\cal L}_{v_t}\right) d\mu_t \left( \eta_t \right) = 0.
\end{equation}

Let me summarize
the obtained mechanics or protomechanics
based on equations (\ref{internal Lie velocity})
and (\ref{quotient}) of motion 
with relation (\ref{velocity-H})
 in the following theorem
that this section proved.

\begin{tm}(Protomechanics)
Hamiltonian $H^{T^*M}_t : T^*M \to {\bf R}$
defines the velocity field $v_t \in {\cal X}(M)$
and Lagrangian $L^{TM}_t: TM\to  {\bf R}$ as follows:
\begin{eqnarray}
v_t &=& {{\partial H^{T^*M}_t}\over {\partial p}}\left( p\left( \eta_t \right)
\right) \\
L^{TM}_t \left( x,
v(x)  \right)
&=&  v  (x) \cdot p \left(\eta _t \right) (x)   -  
H^{T^*M}_t\left( x, p\left(\eta _t \right) (x)  \right) ,
\end{eqnarray}
where mapping
$p: \Gamma [E(M)] \to \Lambda^1 (M)$ satisfies
the modified Einstein-de Broglie relation:
\begin{equation}
p \left(\eta _t \right) =
- i \bar h   \eta_t  ^{ -1}
d \eta_t .
\end{equation}
The equation of motion
is the set of the following equations:
\begin{eqnarray}
\left( {{\partial }\over {\partial t}}
+{\cal L}_{v_t} \right) \eta_t (x)&=&  -i {\bar h}^{-1}
L_t^{TM}\left( x, v_t (x)\right)   \eta_t (x) ,\\
\left( {{\partial }\over {\partial t}}
+{\cal L}_{v_t} \right) d\mu_t \left( \eta_t \right) &=& 0.
\end{eqnarray}
\end{tm}

\section{DYNAMICAL CONSTRUCTION OF PROTOMECHANICS}

Let us 
express the introduced protomechanics 
in the statistical way
for the ensemble of all the synchronicities on $M$,
and
construct
the dynamical description
for the 
collective motion
of the sections of $E(M)$.
Such statistical description realizes
the description within a
long-time interval 
through the introduced relabeling process
so as to change the labeling time,
that is the time for the
initial condition
before analytical problems occur.
In addition,
it clarifies the relationship between
classical mechanics
and quantum mechanics.
For mathematical simplicity,
the discussion below
suppose that
$M$ is a $N-$dimensional manifold for a finite 
natural number
$N\in {\bf N}$.

\subsection{Description of Statistical-State}

The derivative operator $D= \hbar dx^{j}\partial_{j} : T_0^m (M) \to T_0^{m+1} (M)$
($m\in {\bf N}$)
for the space $T^n_0 (M)$ of all the $(0,n)$-tensors on $M$
can be described as
\begin{equation}\label{D-der}
 D^n p(x) = \hbar ^n \left( \prod_{k=1}^{n}\partial_{j_k} p_j(x) \right)
dx^j \otimes \left( \otimes_{k =1}^{n} dx^{j_k} \right) .
\end{equation}
By utilizing this derivative operator $D$,
the following Banach norm
endows the space $\Gamma \left[ E(M) \right] $ of 
all the $C^{\infty }$ sections of $E(M)$
 with a norm topology 
for the family ${\cal O}_{\Gamma \left( E(M) \right) }$
of the induced open balls:
\begin{equation}\label{q-norm}
\left\|   p( \eta  ) \right\| = \sup_{M} 
\sum_{\kappa \in {\bf Z}_{\geq 0} } \hbar ^\kappa
\left\vert D^{\kappa } 
p(\eta )(x) \right\vert_g ,
\end{equation} 
where the metric $(2,0)$-tensor $g = g^{i  j }\partial_i \otimes \partial_j
\in T^2_0(M)  $ on M gives 
\begin{equation}
\left\vert D^{\kappa } p(x) \right\vert_g = \hbar^{\kappa }\sqrt{   
g^{i  j }\prod_{k=1 }^{\kappa }g^{i_k j_k}
\left( \partial_{i_k} p(\eta )_{i }   \right)
\left( \partial_{j_k} p(\eta )_{j } \right) (x) }.
\end{equation}

In terms of the 
corresponding norm topology on $\Lambda^1 (M)$,\footnote{
Assume here that $\Lambda^1 (M)$ has the Banach
norm such that
$
\left\|  p  \right\| = \sup_{M} 
\sum_{\kappa \in {\bf Z}_{\geq 0} }\left\vert D^{\kappa } 
p (x) \right\vert_g ,
$
for $p\in \Lambda^1 (M)$.}
we can consider the space
$C^{\infty } \left( \Lambda^1 \left( M \right)   , C^{\infty }(M) \right) $
of all the $C^{\infty }$-differentiable
mapping from $\Lambda^1 \left( M \right) $ to $C^{\infty }(M)=C^{\infty }(M, {\bf R})$
and the subspaces
of the space
$ 
C( \Gamma  [
E ( M  )  ]   ) $ such that
\begin{equation}
C\left( \Gamma \left[
E\left( M \right) \right]  \right) 
= \left\{  \left. p^*F : 
\Gamma \left[
E ( M ) \right] \to C^{\infty }(M) \  \right\vert  
  F \in   
C^{\infty }\left( \Lambda^1  ( M  ) , C^{\infty }(M) \right)
 \right\} .
\end{equation}
Classical mechanics
requires the local 
dependence on
the momentum for functionals,
while quantum mechanics
needs the wider class
of functions
that depends on their derivatives.
The space of the classical
functionals
and
that of the quantum functionals
are defined as
\begin{eqnarray}
C_{cl}\left( \Gamma \left[
E\left( M \right) \right] \right) 
&=& \left\{ p^*F \in C\left( \Gamma \left[
E\left( M \right) \right] \right)
\ \left\vert \
 p^*F  \left( \eta \right) (x) = 
F^{T^*M} \left( x ,    p (\eta ) (x)  \right) \
\right.
\right\} \\
C_{q\ }\left( \Gamma \left[
E\left( M \right) \right]  \right) 
&=& \left\{   p^*F \in C \left( \Gamma \left[
E\left( M \right) \right] \right)
\ \left\vert \
 p^*F  \left( \eta \right) (x) = 
F ^Q\left( x , p (\eta ) (x),  
..., D^{n} p (\eta ) (x) , ... \right) \ 
\right.
\right\} ,
\end{eqnarray}
and related with each other as
\begin{equation}\label{increasing}
C_{cl}\left( \Gamma \left[
E\left( M \right) \right]  \right) 
\subset
C_q\left( \Gamma \left[
E\left( M \right) \right]  \right) 
\subset
C\left( \Gamma \left[
E\left( M \right) \right] \right) .
\end{equation}
In other words,
the classical-limit
indicates
the limit of $\hbar \to 0$ with fixing $\vert p(\eta )(x) \vert $
finite at every $x\in M$, or what
the characteristic length $[x]$ and momentum $[p]$
such that $x/[x] \approx 1 $ and $p/[p] \approx 1 $
satisfies 
\begin{equation}
[p]^{-n-1} D^n p(\eta )(x)  \ll 1 .
\end{equation}
In addition,
the
$n$-th semi-classical system can
have the following functional space:
\begin{equation}
C_{n+1 }\left( \Gamma \left[
E\left( M \right) \right]  \right) 
= \left\{   p^*F \in C \left( \Gamma \left[
E\left( M \right) \right] \right)
\ \left\vert \
 p^*F  \left( \eta \right) (x) = 
F _{<n>}\left( x , p (\eta ) (x),  
..., D^{n} p (\eta ) (x) \right) \ 
\right.
\right\} .
\end{equation}
Thus, there is the increasing series
of subsets as
\begin{equation}
C_1 \left( \Gamma \left[ E(M)\right] \right)
... \subset  C_{n } \left( \Gamma \left[ E(M)\right] \right)
... \subset 
C_{\infty } \left( \Gamma \left[ E(M)\right] \right)
\subset 
C  \left( \Gamma \left[ E(M)\right]  \right) ,
\end{equation}
where $F _{<1>}=F ^{cl}$ and $F _{<\infty >}=F ^{q}$:
\begin{eqnarray}
C_{1\ } \left( \Gamma \left[ E(M)\right] \right)
&=& C_{cl} \left( \Gamma \left[ E(M)\right] \right) \\
C_{\infty } \left( \Gamma \left[ E(M)\right] \right)
&=&
C_{q \ } \left( \Gamma \left[ E(M)\right] \right) .
\end{eqnarray}

On the other hand,
the emergence-measure $\mu (\eta )  $ has
the Radon measure 
$ \tilde \mu (\eta )  $
for section $\eta \in \Gamma [E(M)]$ such that
\begin{equation}
\tilde \mu (\eta ) \ \left( F \left( p ( \eta  ) \right)
  \right)  =
\int_M d\mu (\eta ) (x) 
F \left( p ( \eta  ) \right) (x) .
\end{equation}
The introduced norm topology
on $ \Gamma \left( E(M) \right) $ induces the topological
$\sigma$-algebra
$ {\cal B}\left( {\cal O}_{ \Gamma \left( E(M) \right) }
\right) $;
thereby
manifold $ \Gamma \left( E(M) \right) $
becomes a measure space
having the probability 
measure ${\cal M}$ such that 
\begin{equation}
 {\cal M} \left(   
\Gamma \left( E(M) \right)  \right) =1.
\end{equation}
For a subset $C_n\left( \Gamma \left( E(M) \right)   \right)
\subset C\left( \Gamma \left( E(M) \right)   \right) $,
 an element
$\bar \mu 
\in C_n\left( \Gamma \left( E(M) \right)   \right) ^* $
is a linear functional
 $ \bar \mu : C_n\left( 
\Gamma \left[ E(M) \right] \right) 
 \to {\bf R}   $
such that
\begin{eqnarray}
 \bar  \mu \left( p^* F \right) &=&
\int_{\Gamma \left[ E(M) \right] }d {\cal M} (\eta ) \
\tilde  \mu (\eta ) \ \left( F \left( p ( \eta  ) \right)
  \right)  \\
&=& \int_{\Gamma \left[ E(M) \right]}d {\cal M} (\eta ) \
\int_M dv(x) \
\rho \left(\eta \right) (x)
F  \left( p ( \eta  ) \right) (x)   ,
\end{eqnarray}
where $ d \mu (\eta )   =
dv  \
\rho \left(\eta \right)   $.
Let us call mapping $\rho :\Gamma [E(M)]\to C^{\infty }(M)$
as the {\it emergence-density}.
The dual spaces
make an decreasing series
of subsets (consult \cite{Bogolubov} in the definition of the Gelfand triplet):
\begin{equation}
C_1 \left( \Gamma \left[ E(M)\right] \right) ^*
\supset ... 
C_n \left( \Gamma \left[ E(M)\right] \right) ^*
\supset ... 
C_{\infty } \left( \Gamma \left[ E(M)\right] \right) ^*\supset
C  \left( \Gamma \left[ E(M)\right]  \right) ^* .
\end{equation}
Thus,
relation
(\ref{increasing}) requires
the opposite sequence
for the dual spaces:
\begin{equation}\label{decreasing}
C_{cl}\left( \Gamma \left( E(M) \right)   \right) ^*
\supset
C_q\left( \Gamma \left( E(M) \right)   \right) ^*
\supset
C \left( \Gamma \left( E(M) \right)   \right) ^*.
\end{equation}

Let us summarize
how the present
theory includes both quantum mechanics and
classical mechanics in the following
diagram,
though leaving detail considerations
for Section 5 and Section 6, respectively.

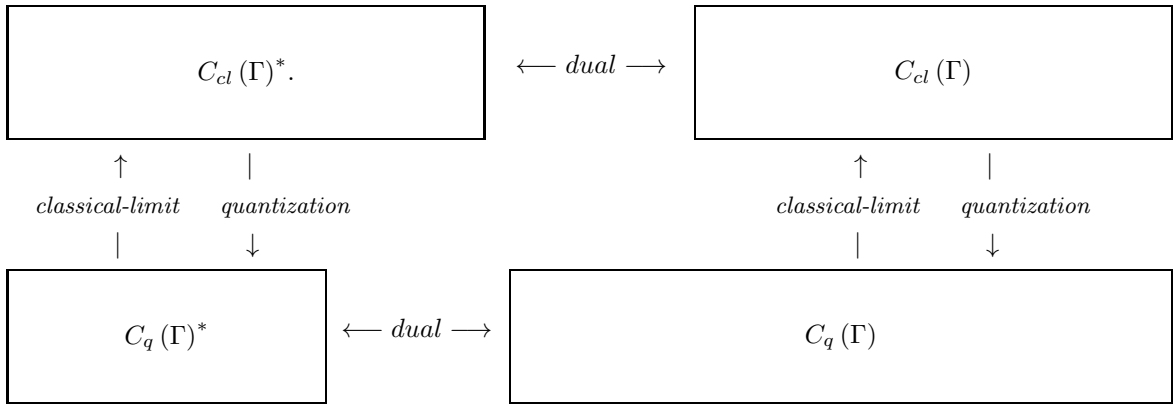
\begin{picture}(380,220)(10,20)
\put(200,50){\framebox(250,50){$C_{q}\left( \Gamma \right) $}}
\put(270,150){\framebox(180,50){$C_{cl}\left( \Gamma \right) $}}
\put(10,50){\framebox(120,50){$C_{q}\left( \Gamma \right) ^*$}}
\put(10,150){\framebox(180,50){$C_{cl}\left( \Gamma \right) ^*$.}}
\put(136,75){$ \longleftarrow   
dual  \longrightarrow $}
\put(202,175){$ \longleftarrow  
dual  \longrightarrow $}
\put(50,137){$ \uparrow $}
\put(20,122){{\small\it classical-limit}}
\put(50,107){$ \vert $}
\put(330,137){$ \uparrow $}
\put(300,122){{\small\it classical-limit}}
\put(330,107){$ \vert $}
\put(100,137){$ \vert $}
\put(90,122){{\small\it quantization}}
\put(100,107){$ \downarrow $}
\put(380,137){$ \vert $}
\put(370,122){{\small\it quantization}}
\put(380,107){$ \downarrow $}
\end{picture}

\subsection{Description of Time-Development}

The group $D(M)$ of
all the
$C^{\infty }$-diffeomorphisms of $M$
and the abelian group
$ C^{\infty }\left( 
M \right) $
 of all the $C^{\infty }$-functions 
on $M$
construct the semidirect product
$ S  ( M ) =
D (M) \times_{semi. } C^{\infty }(M) $
of $D (M)$ with $ C^{\infty }(M) $,
and define
the multiplication $\cdot $
between $ \Phi_1=(\varphi_1,s_1)$ and $\Phi_2=(\varphi_2,s_2)\in   S  ( M  )
$ as  
\begin{equation}\label{(2.1.1)}
\Phi_1\cdot \Phi_2 =(\varphi _1 \circ \varphi _2,(\varphi_2^*
s_1 )\cdot s_2) ,  
\end{equation}
for the pullback $\varphi ^*$ by $\varphi \in
D (M)$.
The
Lie algebra 
$s (M)$
of $ S (M) $ 
has the Lie bracket such that, for   
$ V_1=(v_1 , U_1)$ and $V_2=(v_2 , U_2) \in  s (M)$,
\begin{equation}\label{(2.1.2)}
 [V_1, V_2 ]= \left( [v_1 ,v_2 ], v_1 U_2 - v_2 U_1 
+ \left[ U_1 , U_2 \right] \right) ;  
\end{equation}
 and its dual space $s (M)^*$
is defined by natural pairing $\langle \ , \  \rangle $.
Lie group
$ S  ( M  )  $ now acts on every $C^{\infty }$ section
 of $E ( M   ) $ (consult {\it APPENDIX B}).
We shall further introduce the group $Q(M) 
= Map\left( \Gamma \left[ E(M)\right] , S  (M) \right) $ 
of all the mapping from $ \Gamma \left[ E(M)\right] $ into $S(M) $,
that has
the Lie algebra $q(M) 
= Map\left(  \Gamma \left[ E(M)\right] , s  (M)\right) $ and
its dual space $ 
q(M)^* = Map \left(   \Gamma \left[ E(M)\right] , s  (M)^* \right)  $.

To investigate the 
group structure
of the system considered,
let us further define
the emergence-momentum
$ {\cal J}   
\in   q\left(M\right) ^*$ 
as follows:
\begin{equation}
 {\cal J}   \left( \eta  \right) 
=   d {\cal M} \left( \eta  \right) \
\left( \tilde    \mu \left( \eta  \right)  
 \otimes p (\eta )  ,\tilde    \mu \left( \eta  \right)  
\right)  .
\end{equation}
Thus,
the functional
${\cal F} : q\left(M\right) ^* \to {\bf R} $ 
can always be defined 
as 
\begin{equation}
{\cal F}  \left(  {\cal J} \right)
= \bar \mu  \left( p^*F \right) .
\end{equation}
On the other hand, the derivative
$  {\cal D}_{\rho  } 
F   \left( p \right)   $ can be 
introduced as follows
excepting the point
where the distribution $\rho $
becomes zero:
\begin{equation}
{\cal D}_{\rho  } 
F   \left( p \right)  (x)
=\sum_{ (n_1 ,  ... , n_N) \in {\bf N}^N}
{1\over { \rho  (x)} }
\left\{ \prod_{i}^{N} \left( - \partial_i \right) ^{ n_i }
  \left(  \rho  (x) 
p ( x )
{{\partial F    }\over { \partial \left\{
\left(
 \prod_{i}^{N}
\partial_i^{n_i } \right) p_j \right\}  }}
\right)  \right\}
\partial_j .
\end{equation}
Then,
operator $\hat F \left( \eta \right)  =
{{\partial {\cal F}}
\over {\partial {\cal J}}}  \left( {\cal J}\left( \eta \right)\right) $ 
is defined as
\begin{equation}
\left. {d\over d\epsilon }\right\vert_{\epsilon =0}
{\cal F } \left( {\cal J}  + \epsilon {\cal K}\right) 
= \left\langle  {\cal K} , \hat F   \right\rangle ,
\end{equation}
i.e.,
\begin{equation}
\hat F  \left( \eta \right)
= \left(   {\cal D}_{ \rho (\eta ) } F  
\left(  p  (\eta ) \right) ,
- p  (\eta )  \cdot     {\cal D}_{ \rho (\eta ) } F   
\left(  p  (\eta ) \right)
+ F    \left( p  (\eta )  \right)
\right)  ;
\end{equation}
thereby, the following null-lagrangian relation
 can be obtained:
\begin{equation}
{\cal F}  
 \left( {\cal J}   \right) 
= 
\langle {\cal J}   , \hat F    \rangle  .
\end{equation}

Let us consider the time-development
of the section $\eta ^{\tau }_{t }(\eta ) \in \Gamma [E(M)]$
such that
 the {\it labeling time} $\tau $ satisfies
$\eta ^{\tau }_{\tau }(\eta ) =\eta  $.
It has the momentum
 $p_t^{\tau }(\eta )
=-i \bar h\eta ^{\tau }_t(\eta )^{-1}
d \eta ^{\tau }_t(\eta ) $ and
the emergence-measure $\mu^{\tau }_t(\eta ) $
such that
\begin{equation}\label{q-measure-rel}
d {\cal M} \left( \eta \right) \
  \tilde \mu_t^{\tau } \left( \eta  \right) 
= d {\cal M} \left(  \eta ^{\tau }_t(\eta )\right) \
\tilde \mu_t \left( \eta ^{\tau }_t(\eta )\right) :
\end{equation}
\begin{eqnarray}\label{mu(F)}
\bar  \mu_t \left( p^*F_t\right) 
&=&\int_{\Gamma \left[ E(M) \right]}d {\cal M} (\eta ) \
 \tilde \mu_t (\eta ) \ \left( p^*F_t  (\eta )
  \right)  \\
&=& \int_{\Gamma \left[ E(M) \right]}d {\cal M} \left( \eta \right) \
\tilde \mu^{\tau }_t \left( \eta  \right) \ \left( p^*F  \left( 
\eta ^{\tau }_t(\eta )\right) 
  \right)  \\
&=&
 \int_{\Gamma \left[ E(M) \right]}d {\cal M} \left( \eta \right) \
\int_M dv(x) \ \rho_t^{\tau } (\eta ) (x) 
F_t   \left( p ^{\tau }_t(\eta )\right)  (x)    .
\end{eqnarray}
The introduced labeling time $\tau $
can always be chosen such that $ \eta ^{\tau }_t(\eta ) $
does not have any singularity
within a short time for every $\eta
\in \Gamma \left[ E(M) \right]$.
The emergence-momentum
$ {\cal J}_t^{\tau }  
\in   q\left(M\right) ^*$ 
such that
\begin{eqnarray}
{\cal J}_t^{\tau }  (\eta ) &=&
 {\cal J}_t  \left( \eta ^{\tau }_t(\eta )\right)  \\
&=&   d {\cal M} \left( \eta ^{\tau }_t(\eta )\right) \
\left( \tilde   \mu_t \left( \eta ^{\tau }_t(\eta )\right)  
 \otimes 
p_t^{\tau }(\eta )  ,\tilde   \mu_t \left( \eta ^{\tau }_t(\eta )\right)  \right) \\
&=&  d {\cal M} (\eta ) \
 \left(  \tilde \mu_t ^{\tau }
\left( \eta  \right) \otimes 
p_t^{\tau }(\eta )  ,
\tilde \mu_t ^{\tau }
\left( \eta  \right) \right) \ 
\end{eqnarray}
satisfies the following relation
for the functional
${\cal F}_t   : q\left(M\right) ^* \to {\bf R} $:
\begin{equation}
{\cal F}_t  \left(  {\cal J}^{\tau }_t\right)
=  \mu_t \left( p^* F_t\right) ,
\end{equation}
whose value is independent of
labeling time $\tau $.
The operator $\hat F_t^{\tau } =
{{\partial {\cal F}_t}
\over {\partial {\cal J}}}  \left( {\cal J}^{\tau }_t\right) $ 
is defined as
\begin{equation}
\left. {d\over d\epsilon }\right\vert_{\epsilon =0}
{\cal F }_t  \left( {\cal J}^{\tau }_t + \epsilon {\cal K}\right) 
= \left\langle  {\cal K} , \hat F_t^{\tau }  \right\rangle ,
\end{equation}
i.e.,
\begin{equation}
\hat F_t  ^{\tau }
= \left(   {\cal D}_{ \rho_t^{\tau }(\eta ) } F_t  
\left(  p^{\tau }_t (\eta ) \right) ,
- p_t^{\tau } (\eta )  \cdot     {\cal D}_{ \rho_t^{\tau }(\eta ) } F_t  
\left(  p^{\tau }_t (\eta ) \right)
+ F_t   \left( p^{\tau }_t (\eta )  \right)
\right)  .
\end{equation}
Thus, the following null-lagrangian relation
 can be obtained:
\begin{equation}
{\cal F}_t   
 \left( {\cal J}_t^{\tau }  \right) 
= 
\langle {\cal J}_t^{\tau }  , \hat F _t ^{\tau } \rangle  ,
\end{equation}
while
the normalization condition 
has the following expression:
\begin{equation}\label{q-normalize}
{\cal I}\left( {\cal J}^{\tau }_t \right) =1 
\ \ \ \ \ for \ \ \ \ \ 
{\cal I}\left( {\cal J}^{\tau }_t \right) =
\int_{\Gamma \left[ E(M) \right] }d{\cal M}(\eta )  \
\mu_t (\eta ) (M) .
\end{equation}

\begin{tm}
For Hamiltonian operator $\hat H_t^{\tau }  =
{{\partial {\cal H}_t}\over {\partial {\cal J}}} 
  \left( {\cal J}_t^{\tau } \right)\in   q\left(  M\right) $
corresponding to Hamiltonian
$p^*H_t \left( \eta \right) (x) =
H^{T^*M}_t\left( x, p \left( \eta \right) \right) $,
equations (\ref{internal Lie velocity}) and (\ref{quotient}) of motion
becomes Lie-Poisson equation 
\begin{equation}\label{Hq}
{{\partial  {\cal J} ^{\tau }_t}\over {\partial t}} 
= ad^*_{\hat H_t ^{\tau }     }{\cal J}^{\tau } _t   ,
\end{equation}
which can be expressed as
\begin{equation}
{{\partial  } \over {\partial t}} 
 \rho _t^{\tau }(\eta )(x) 
= -\surd ^{-1}\partial_j \left( 
{ { \partial H_t^{T^*M}  }\over {\partial \ p_j \  }} 
\left( x, p_t^{\tau } \left( \eta \right) (x) \right) 
 \rho  _t^{\tau }(\eta )(x) \surd  \right) ,
\end{equation}
\begin{eqnarray}\nonumber
{{\partial  } \over {\partial t}}
\left( \rho  _t^{\tau }(\eta )(x) 
  p_{tk}^{\tau } (\eta ) (x) \right)
\nonumber
&=&
 - \surd^{-1} \partial_j \left(
{ { \partial H_t^{T^*M}  }\over {\partial \ p_j \  }} 
\left( x, p_t^{\tau } \left( \eta \right) (x) \right) 
\rho  _t^{\tau }(\eta )(x)  p_{tk}^{\tau } (\eta )(x) \surd  
\right) \\
\nonumber
&  &
-  \rho  _t^{\tau }(\eta )(x)  p_{tj}^{\tau } (\eta ) (x) 
\partial_k  
{ { \partial H_t^{T^*M}  }\over {\partial \ p_j \  }} 
\left( x, p_t^{\tau } \left( \eta \right) (x) \right)   \\
&  &
+ \rho  _t^{\tau }(\eta ) (x)  
\partial_k \left(
 p^{\tau }_t (\eta ) (x) \cdot 
{ { \partial H_t^{T^*M}  }\over {\partial \ p \  }} 
\left( x, p_t^{\tau } \left( \eta \right) (x) \right) 
- H_t^{T^*M}   \left( x, p^{\tau }_t (\eta ) (x) \right) 
\right)  .
\end{eqnarray}
\end{tm}
\par\noindent${\it Proof.}\ \ $
{\it
Lie-Poisson equation (\ref{Hq})
 is  calculated 
for ${\cal D} H_t  ^{\tau }(\eta ) 
=  {\cal D}_{ \rho_t^{\tau }(\eta ) } H_t  
\left(  p^{\tau }_t (\eta ) \right) $
as follows:
\begin{equation}
\label{density's-q} 
{{\partial  } \over {\partial t}} 
 \rho _t^{\tau }(\eta )(x) 
= -\surd ^{-1}\partial_j \left(
{\cal D}^j H_t  ^{\tau }(\eta ) (x)
 \rho  _t^{\tau }(\eta )(x) \surd  \right) ,
\end{equation}
\begin{eqnarray}\nonumber
{{\partial  } \over {\partial t}}
\left( \rho  _t^{\tau }(\eta )(x) 
  p_{tk}^{\tau } (\eta ) (x) \right)
\nonumber
&=&
 - \surd^{-1} \partial_j \left(
{\cal D}^j H_t  ^{\tau }(\eta ) (x)
\rho  _t^{\tau }(\eta )(x)  p_{tk}^{\tau } (\eta )(x) \surd  
\right) \\
\nonumber
&  &
-  \rho  _t^{\tau }(\eta )(x)  p_{tj}^{\tau } (\eta ) (x) 
\partial_k  
{\cal D}^j H_t  ^{\tau }(\eta ) 
 (x)  \\
\label{current's-q}
&  &
+ \rho  _t^{\tau }(\eta ) (x)  
\partial_k \left(
 p^{\tau }_t (\eta ) (x) \cdot {\cal D} H_t  ^{\tau }(\eta )  (x)
- H_t   \left( p^{\tau }_t (\eta ) \right) (x)
\right)  ,
\end{eqnarray}
where $ dv= dx^1 \wedge ... dx^N \ \surd $
and
$\surd = \sqrt{det \left\vert g^{jk} \right\vert }$ 
for the
local coordinate ${\bf x}= \left( x^1, x^2, ... , x^N\right) $.
Second equation (\ref{current's-q})
can be rewritten in conjunction with 
the conservation  (\ref{density's-q}) of the
emergence-density
as
\begin{equation}
{{\partial  } \over {\partial t}}
  p_{tk}^{\tau } (\eta ) (x) 
+   
{\cal D}^j H_t  ^{\tau }(\eta ) (x)
 \partial_j   p_{tk}^{\tau } (\eta )(x)  
+     p_{tj}^{\tau } (\eta ) (x) 
\partial_k  
{\cal D}^j H_t  ^{\tau }(\eta ) 
 (x) =  
\partial_k  L^{\tau }_t (\eta ) (x) ,
\end{equation}
where
\begin{equation}
 L^{\tau }_t (\eta ) (x)= p^{\tau }_t (\eta )
 (x) \cdot {\cal D} H_t  ^{\tau }(\eta )  (x)
- H_t   \left( p^{\tau }_t (\eta ) \right) (x) ,
\end{equation}
or, by using Lie derivatives,
\begin{equation}
{\cal L}_{{\cal D} H_t  ^{\tau }(\eta )  }
\  p_{t}^{\tau } (\eta )  =  
d L^{\tau }_t (\eta ) .
\end{equation}
Thus,
we can obtain the 
equation of motion
in the following simpler form
by using Lie derivatives:
\begin{eqnarray}
{\cal L}_{{\cal D} H_t  ^{\tau }(\eta )  }
\  \eta _{t}^{\tau }    &=&  -i\bar h
 L^{\tau }_t (\eta )  \  \eta _{t}^{\tau } \\
{\cal L}_{{\cal D} H_t  ^{\tau }(\eta )  }
\ \rho _t^{\tau }(\eta ) \ dv 
&=& 0 ,
\end{eqnarray}
which is equivalent to the equations  (\ref{internal Lie velocity}) and (\ref{quotient}) 
when $p^*H_t \left( \eta \right) (x) =
H^{T^*M}_t\left( x, p \left( \eta \right) \right) $

}\hspace{\fill} { \fbox {}}\\

\noindent
Equation (\ref{Hq}) will
prove in the following two sections
to include the
Schr\"odinger equation 
in canonical quantum mechanics
and the classical Liouville equations
in classical mechanics.

For
$  {\cal U}_t  ^{\tau }
 \in  Q\left(  M\right) $ such that
$
{{\partial  
{\cal U}_t  ^{\tau } }\over {\partial t}}  
\circ \left( {\cal U}_t  ^{\tau  } \right) ^{-1}= 
\hat H_t  ^{\tau } (\eta ) 
\in q(M)  $,
let us introduce the following operators:
\begin{eqnarray}
  \tilde H_t^{\tau } (\eta )   = Ad^{-1}_{{\cal U}_t  ^{\tau }   }
\hat H_t^{\tau } (\eta )  \ \ \left( =  \hat H_t ^{\tau } (\eta )\right) , \   and \ \ \
 \tilde F_t^{\tau } (\eta )  = Ad^{-1}_{{\cal U}_t  ^{\tau }    }
\hat F_t ^{\tau } (\eta ) .
\end{eqnarray}
It satisfies the following theorem.

\begin{tm}
Lie-Poisson equation (\ref{Hq})
is
equivalent to the following equation:
\begin{equation}\label{general heisenberg}
{{\partial }\over {\partial t}}\tilde F_t  ^{\tau }    = 
\left[ \tilde H _t ^{\tau }    ,\tilde F_t   ^{\tau }   \right] 
+\widetilde{\left( {{\partial F_t  ^{\tau }     }
\over {\partial t}} \right) } .
\end{equation}
\end{tm}
\par\noindent${\it Proof.}\ \ $
{\it
Equation (\ref{Hq}) of motion
concludes the following equation:
\begin{equation}
\left\langle
{{\partial  {\cal J} ^{\tau }_t}\over {\partial t}}  ,\hat F_t   ^{\tau }  
\right\rangle
= \left\langle
ad^*_{\hat H_t ^{\tau }     }{\cal J}^{\tau } _t  ,
\hat F_t   ^{\tau }  
\right\rangle .
\end{equation}
The left hand side can be calculated as
\begin{eqnarray}
L.H.S.
&= & 
{{d  }\over {d t}} {\cal F}_t\left( {\cal J}^{\tau }_t\right)
-{{\partial  {\cal F}_t}\over {\partial t}} \left( {\cal J}^{\tau }_t\right)\\
&= &
\left\langle \left(
{{\partial  }\over {\partial t}}Ad^*_{{\cal U}_t^{\tau}}{\cal J} ^{\tau }_{\tau }  
\right) ,
\hat F_t   ^{\tau }  
\right\rangle  -
\left\langle
Ad^*_{{\cal U}_t^{\tau}}{\cal J} ^{\tau }_{\tau }  ,
\hat {{\partial  F_t^{\tau }}\over {\partial t}} 
\right\rangle \\
&=&
\left\langle
{\cal J} ^{\tau }_{\tau }  ,
{{\partial  }\over {\partial t}} \tilde F_t   ^{\tau }  
\right\rangle
 -
\left\langle
{\cal J} ^{\tau }_{\tau }  ,
\tilde {{\partial  F_t^{\tau }}\over {\partial t}} 
\right\rangle  ;
\end{eqnarray}
and the right hand side becomes
\begin{eqnarray}
R.H.S.
&= & 
\left\langle
ad^*_{\hat H_t ^{\tau }     }Ad^*_{{\cal U}_t ^{\tau }}{\cal J}^{\tau } _t  ,
\hat F_t   ^{\tau }  
\right\rangle \\
&=&
\left\langle
Ad^*_{{\cal U}_t ^{\tau }}ad^*_{\tilde H_t ^{\tau }     }{\cal J}^{\tau } _t  ,
\hat F_t   ^{\tau }  
\right\rangle \\
&=&\left\langle
{\cal J}^{\tau } _t  , \left[
\tilde H_t ^{\tau }    ,
\tilde F_t   ^{\tau }  \right]  
\right\rangle .
\end{eqnarray}
Thus, we can obtain this theorem.
}\hspace{\fill} { \fbox {}}\\

The general theory
for Lie-Poisson systems
certificates that,
 if a  group action of Lie group
$ Q(M)$
keeps the Hamiltonian ${\cal H}_t: q(M)^* \to {\bf R}$  
invariant,
there exists an invariant charge 
functional
$Q : \Gamma \left[ E(M)\right] \to
C(M) $ and the induced
function ${\cal Q}: q(M)^* \to {\bf R}$ such that
\begin{equation}\label{charge-invariant}
\left[ \hat H _t, \hat Q \right]  = 0 ,
\end{equation}
where $\hat Q $ is expressed as
\begin{equation}
\hat Q 
= \left(  {\cal D}_{\rho (\eta )} Q
\left(  p  (\eta ) \right) ,
- p  (\eta )  \cdot     {\cal D}_{\rho (\eta )} Q
\left(  p  (\eta ) \right)
+ Q   \left( p (\eta )  \right)
\right)  .
\end{equation}

\section{DEDUCTION OF CLASSICAL MECHANICS}

In classical Hamiltonian mechanics,
the  state of a particle 
on  manifold $M$
can be
represented as a position  in
the cotangent bundle $T^*M$.
In this section,  we will reproduce
the classical equation of motion
from the general theory
presented in the previous section.
Let us here concentrate ourselves
on the case where $M$ is $N$-dimensional
manifold for simplicity,
though the discussion
below would still be valid if
substituting an appropriate Hilbert space
when
$M$ is infinite-dimensional
ILH-manifold\cite{Omori}.

\subsection{Description of Statistical State}

Now,
we must be concentrated
on the case
where the physical
 functional $F\in C^{\infty }\left( \Lambda^1(M) , C^{\infty }(M) \right) $
does {\it not}
depend  on the derivatives 
of the $C^{\infty }$ 1-form $p\left( 
\eta \right)\in \Lambda^1 (M) $
induced from $\eta \in \Gamma \left[ E(M) \right] $, then it
has the following expression:
\begin{equation}\label{local F}
p^* F  \left( \eta \right) (x) = 
F^{T^*M} \left( x,  p\left( \eta \right) (x) \right)   .
\end{equation}

Let us  choose a  
coordinate system  
$\left( U_{\alpha } , {\bf x}_{\alpha  } 
\right) _{\alpha \in \Lambda_M}$
for  a covering $\left\{ U_{\alpha } 
\right\} _{\alpha \in \Lambda _{M  }} $
over $M$,
i.e.,
$ M = \bigcup_{\alpha \in \Lambda_M }
U_{\alpha }$.
Let us further choose
a reference set $U \subset U_{\alpha }$
such that $v(U) \neq 0$
and consider the set $\Gamma_{U k}\left[ E(M)\right] $ of the
$C^{\infty }$ sections of  $E(M)$
having corresponding momentum
$p\left( \eta \right) $
the
supremum of whose every component
$p_j\left( \eta \right) $ 
in $U$ becomes
 the value $k_j   $ for $ k= (k_1 ,..., k_N) \in {\bf R}^N$:\footnote{
To substitute
$\Gamma_{U k}\left[ E(M)\right] = \left\{ 
\eta \in \Gamma \left[ E(M)\right] \  \left\vert
\ \int_{  U}dv(x) \ p_j\left( \eta
\right) (x) = k_j v\left( U \right) \right.
\right\} $ for
definition (\ref{lambda cl.})
also induces the similar discussion
below,
while there exist a variety of the classification
methods that produce the same result.}
\begin{equation}\label{lambda cl.}
\Gamma_{U k} \left[ E(M)\right] = \left\{ 
\eta \in \Gamma \left[ E(M)\right] \  \left\vert
\ \sup_{  U} p_j\left( \eta
\right) (x) = \hbar k_j \right.
\right\} .
\end{equation}
Thus,
every section $\eta \in \Gamma\left[ E(M)\right] $ has 
some $k\in {\bf R}^N$
such that $\eta =\eta [k]\in \Gamma_{U k} \left[ E(M)\right] $.
Notice that
$\Gamma_{U k} \left[ E(M)\right] $
can be identified with
$\Gamma_{U^{\prime } k} \left[ E(M)\right] $
for every two reference sets $U$ and $U^{\prime }\in M$,
since
there
exists a diffeomorphism $\varphi $
satisfying $\varphi \left( U \right) =U^{\prime } $;
thereby, we will simply denote
$\Gamma_{U k}\left[ E(M)\right] $
as $\Gamma_{ k}\left[ E(M)\right] $.

On the other hand,
let us consider
the space $L\left( T^*M \right) $ of
all the Lagrange foliations,
i.e.,
every element
$\bar p \in L\left( T^*M \right) $
is a mapping $\bar  p :
{\bf R}^N \to \Lambda^1(M) $
such that
each $q  \in T^*M$
has a unique $k \in {\bf R}^N $  as
\begin{equation}
q =\bar  p [ k ] \left( \pi (q) \right) .
\end{equation}
For every $\bar  p 
= p\circ \bar  \eta  \in L\left( T^*M \right) $
such that $\bar  \eta [k] \in \Gamma_{  k}\left[ E(M)\right]  $,
it is possible to
separate an element $\eta [k] \in \Gamma_{  k}\left[ E(M)\right] $
for a
$\xi\in  \Gamma_{ 0}\left[ E(M)\right]  $ as
\begin{equation}
\eta [k]= \bar  \eta [k] \cdot \xi ,
\end{equation}
or to 
separate 
momentum
$p \left( \eta [ k] \right) $ as
\begin{equation}\label{cl-separation}
p \left( \eta [ k] \right) 
= \bar  p    [ k]  
  + p  \left( \xi \right) ;
\end{equation}
thereby, we can express 
the emergence-density $ \rho  
: \Gamma [E(M)] \to C^{\infty }\left( M\right) $ 
in the following form
for the  function $ \varrho \left(  \xi \right)  \in  C^{\infty }
\left( T^*M,{\bf R}\right) $ on $T^*M$:
\begin{equation}\label{rho=rho}
 \rho    \left( \eta [k] \right) 
(x) \surd 
=  
\varrho \left(  \xi \right) \left( x, p \left( \eta [k]\right) (x) \right) .
\end{equation}

We call the set
$ B\left[ E(M)\right]  
= \Gamma_{0}\left[ E(M)\right]  $
the {\bf back ground} of $L\left( T^*M \right) $.
For the Jacobian-determinant
$ \sigma  [ k]  = det 
\left( {{\partial \bar p_{ti} ^{\tau }  [ k]   }
\over {\partial k_j}}\right) $,
we will define the measure
${\cal N}$ on $B\left[ E(M)\right] $
for the $\sigma $-algebra induced from that of $\Gamma \left[
E(M)\right] $:
\begin{equation}
d {\cal M} \left( \eta [k]\right) \
d v (x)
= d^Nk  d {\cal N}  \left(  \xi \right) 
 d v (x) \ \sigma [k](x)\  .
\end{equation}
For  separation (\ref{cl-separation}),
the Radon measure $\tilde  \mu (\eta )$
 induces the measure
$\omega^N $ on $T^*M$
in
the following lemma
such that $\omega^N = \phi_{U_{\alpha }
* } d^Nx \wedge d^Nk$
for
$d^Nx =dx^1 \wedge ... dx^N $
and $d^Nk =dk^1 \wedge ... dk^N $.
\begin{lem}
The following relation holds:
\begin{equation}
\bar \mu \left( p^*F\right)=
\int_{T^*M} \omega^N (q) \
\rho^{T^*M} \left( q \right)  
F^{T^*M} \left( q \right)  ,
\end{equation}
where
\begin{equation}
\rho^{T^*M} \left( q \right)  =
\int_{ B\left[ E(M)\right]  }d {\cal N}  \left( 
\xi \right) \
\varrho \left( \xi \right) \left( q \right)  .
\end{equation}
\end{lem}

\par\noindent${\it Proof.}\ \ $
{\it
The direct calculation
based on separation
(\ref{cl-separation})
shows 
\begin{eqnarray}\nonumber
\bar \mu \left( p^* F\right) &=&
\int_{
\Gamma \left[ E(M)\right] 
}d {\cal M}  \left( \eta  \right)  \
\tilde \mu  \left( \eta  \right)  \left( p^*F  \left( \eta   \right) 
  \right)  \\ \nonumber
&=& \int_{\Gamma \left[ E(M)\right] }d {\cal M} \left( \eta [k] \right)  \
\int_M dv(x) \
\varrho \left( \xi \right)  \left( x, p  \left( \eta [k] \right)  (x)\right)  
F^{T^*M} \left( x,  p  \left( \eta  [k]\right)  (x) \right)   \\ \nonumber
&=& 
\int_{{\bf R}^N} d^Nk
\int_{B \left[ E(M)\right] }
d {\cal N}  \left( 
\xi \right)
\int_M dv(x) \
\sigma   [ k](x) \\ \nonumber
& & \ \ \ \ \ \ \ \ \  \ \ \ \ \ \ \ \ \  \times \
\varrho \left( \xi \right)  \left( x, p \left( 
\eta [k] \right) (x) \right)  
F^{T^*M} \left( x,  p\left( 
\eta [k] \right) (x) \right)  \\ \nonumber
&=& 
\int_{{\bf R}^N} d^Nk
\int_{B \left[ E(M)\right] }
d {\cal N}   \left( \xi\right) 
\int_M dv(x) \
\sigma   [ k](x) \\
& & \ \ \ \ \ \ \ \ \  \ \ \ \ \ \ \ \ \  \times \
\varrho \left( \xi \right) \left( x,\bar  p  [k]   (x) 
+p\left( \xi\right) (x) \right)  
F^{T^*M} \left( x,  \bar  p  [k]   (x) 
+p\left( \xi\right) (x) \right)  \\ \nonumber
&=& 
\int_{B \left[ E(M)\right]  }
d {\cal N}  \left( \xi \right)  \sum_{\alpha \in \Lambda_M }
\int_{\phi_{U_{\alpha }}\left( A_{\alpha }\right) }
d^Nk\wedge d^Nx \
\phi_{U_{\alpha }}^*\varrho \left( \xi \right) \left( x , k \right)  
\phi_{U_{\alpha }}^*F^{T^*M} \left( x,  k\right) 
\\
&=& 
\int_{B \left[ E(M)\right]  }
d {\cal N}  \left( \xi \right) 
\int_{T^*M} \omega^N (q) \
\varrho \left( \xi \right) \left( q\right)  
F^{T^*M} \left( q\right) ,
\end{eqnarray}
where 
$T^*M= \bigcup_{\alpha \in \Lambda_M }
A_{\alpha }$
is  the  disjoint union 
of $A_{\alpha } \in {\cal B}\left( {\cal O}_{T^*M} \right) $
such that
(1) $\pi \left( A_{\alpha }
\right)  \subset U_{\alpha }$ and that (2)
 $A_{\alpha }\cap A_{\beta }
= \emptyset $ for $\alpha \neq \beta \in \Lambda_M $
(consult {\it APPENDIX A}).

If defining the probability function
$\rho^{T^*M} : T^*M \to {\bf R}$ such that
\begin{equation}
\rho^{T^*M} \left( q \right)  =
\int_{B \left[ E(M)\right]  }d {\cal N}  \left( \xi \right) \
\varrho \left( \xi \right) \left( q \right)  ,
\end{equation}
we can obtain this lemma.
}\hspace{\fill} { \fbox {}}\\

\subsection{Description of Time-Development}

Let us consider the time-development
of the functional $\bar 
\mu_t : C^1 \left( \Gamma \left( M\right) ,C\left( M\right)
\right) \to {\bf R}$
for $
p_t^{\tau } \left( \eta [ k] \right) = 
\bar  p _t^{\tau } [k] 
  + p  \left( \xi \right) $.
For the Jacobian-determinant
$ \sigma_t^{\tau }  [ k]  = det 
\left( {{\partial \bar p_{ti} ^{\tau } [ k] }
\over {\partial k_j}}\right) $,
the following relation holds:
\begin{eqnarray}
 \bar \mu_t \left( p^* F_t\right) &=&\int_{T^*M} 
\omega^N (q) \
\rho_t^{T^*M} \left( q \right)  
F^{T^*M} \left( q \right) \\
&=& \int_{{\bf R}^N} d^Nk
\int_M dv(x) \
\bar  \rho_t^{\tau }   [ k] (x) 
F^{T^*M} \left( x, \bar   p_t^{\tau } [k] (x) \right) 
  ,
\end{eqnarray}
where
\begin{equation}\label{classical probability}
\bar  \rho_t^{\tau }   [ k](x)   \surd =
\sigma_t ^{\tau } [k](x) 
\rho^{T^*M} \left( x,  \bar  p_t^{\tau } [k] (x) \right)  .
\end{equation}
The Jacobian-determinant
$ \sigma_t^{\tau }  [ k] $ satisfies
the following relation:
\begin{equation}
 {{d {\cal M}\left( \eta_t^{\tau }(\eta )\right) }\over 
{d {\cal M}(\eta ) }} 
  = {{ \sigma_t^{\tau }  [ k] }\over {\sigma  [ k] }} .
\end{equation}

Thus,
we can define
the reduced emergence-momentum
$ \bar {\cal J}_t  
\in   \bar q\left(M\right) ^*= q\left(M\right)  ^*/
B \left[ E(M)\right]  $ as follows:
\begin{equation}
\bar  {\cal J}_t \left( \bar  \eta [k] \right)  
= \left(  d^Nk \  \wedge dv \
\bar   \rho_t ^{\tau } [k]  \otimes 
\bar  p_t^\tau [k]  , d^Nk\ \wedge 
 dv \ \bar   \rho_t ^{\tau }  [k] \right) ;
\end{equation}
and we can define the functional 
$\bar {\cal F}_t \in C^{\infty }\left( 
  \bar q\left(  M\right) ^*  , {\bf R} \right)  $
as
\begin{eqnarray}
\label{function}
\bar {\cal F}_t  \left(  
\bar {\cal J}_t  \right) 
&=& \bar \mu_t
  \left( p^* F_t   
  \right) \\
&=&
\int_{{\bf R}^N}d^Nk \
\int_M dv(x) \
\bar    \rho_t ^{\tau } [k]  (x) 
F_t^{T^*M} \left( x ,\bar   p_t^{\tau }[  k]  (x) \right) ,
\end{eqnarray}
which is independent of  labeling time $\tau $.

Then, the
operator $\hat F_t^{cl } ={{\partial \bar {\cal F}_t}
\over {\partial \bar {\cal J}}}  \left( \bar {\cal J}_t\right) $ satisfies
\begin{equation}
\hat F_t  ^{cl}
= \left(    {{\partial F_t^{T^*M}  }\over { \partial p \ \ }}
\left(x ,\bar   p_t^{\tau } [k]
(x ) \right) ,
-L^{F_t^{T^*M}} \left(  x , {{\partial F_t^{T^*M}  }\over {\partial p \ \ }} 
\left( x , \bar  p_t^{\tau } [k]
(x ) \right)
  \right) \right)  ,
\end{equation}
where
\begin{equation}\label{F-lagrange}
  L^{F_t^{T^*M}}  \left( { x,{{\partial F_t^{T^*M}  }\over {\partial p \ \ }} 
\left( x, p \right) } \right) =
p \cdot {{\partial F _t^{T^*M}  }\over {\partial p \ \ }}\left( x, p \right)  
-F_t^{T^*M}   \left( x, p  \right) 
\end{equation}
is the Lagrangian if function $F_t$ is Hamiltonian $H_t$.
Thus, the following null-lagrangian relation
 can be obtained:\footnote{
The Lagrangian corresponding to this Lie-Poisson system
is $
\langle \bar {\cal J}_t  , \hat H _t ^{cl} \rangle 
-{\cal H}_t  
 \left( \bar {\cal J}_t  \right) 
$, while
the usual Lagrangian is $L^{H_t^{T^*M}}$. }
\begin{equation}
 \bar {\cal F}_t  
 \left( \bar {\cal J}_t  \right) 
= 
\langle \bar {\cal J}_t  , \hat F _t^{cl}  \rangle  .
\end{equation}
Besides, 
the normalization condition 
becomes 
\begin{equation}\label{normalize}
\bar {\cal I}\left( \bar {\cal J} _t \right) =1 
\ \ \ \ \ for \ \ \ \ \ 
\bar {\cal I}\left(\bar  {\cal J} _t \right) =
\int_{{\bf R}^N}d^Nk  \
\int_M dv(x) \ \bar   \rho_t^{\tau } [k]( x) 
.
\end{equation}

\begin{tm}
For Hamiltonian operator $\hat H_t  =
{{\partial {\cal H}_t}\over {\partial \bar {\cal J}}} 
  \left(\bar {\cal J} _t \right)\in   \bar q\left(  M\right) $,
the equation of motion
becomes Lie-Poisson equation:
\begin{equation}\label{H[X]cl}
{{\partial \bar {\cal J}_t}\over {\partial t}} 
= ad^*_{\hat H_t^{cl}  }\bar {\cal J}_t   ,
\end{equation}
that is  calculated 
as follows:
\begin{equation}
\label{density's}
{{\partial  } \over {\partial t}} 
\bar   \rho _t^{\tau }[k](x) 
= -\surd ^{-1}\partial_j \left(
{{\partial H^{ T^*M }_t}\over {\partial p_j}} 
\left( x, \bar  p_t^{\tau } [k](x)\right)
\bar   \rho  _t^{\tau }[k](x)  \surd  \right) ,
\end{equation}
\begin{eqnarray}\nonumber
{{\partial  } \over {\partial t}}
\left( \bar  \rho  _t^{\tau }[k](x) 
\bar    p_{tk}^{\tau } [k](x) \right)
\nonumber
&=&
 - \surd^{-1} \partial_j \left(
{{\partial H^{ T^*M }_t}\over {\partial p_j}} \left( x, 
\bar  p_t^{\tau } [k](x)\right) 
 \bar  \rho  _t^{\tau }[k](x)  \bar  p_{tk}^{\tau } [k](x)
\surd  \right) \\
\nonumber
&  &
-  \bar  \rho  _t^{\tau }[k](x)  \bar  p_{tj}^{\tau } [k](x) 
\partial_k  
\left( {{\partial H^{ T^*M }_t}\over {\partial p_j}} 
\left( x, \bar  p_t^{\tau } [k](x)\right)  \right) \\
\label{current's}
&  &
+ \bar  \rho  _t^{\tau }[k](x)  
\partial_k  L^{H^{ T^*M }_t } \left( x, \bar  p_t^{\tau } [k](x)\right)  .
\end{eqnarray}
\end{tm}
\par\noindent${\it Proof.}\ \ $
{\it
The above equation can be obtained from
the integration of general equations
(\ref{density's-q}) and (\ref{current's-q})
on the space $\Gamma_{U 0}$;
thereby, it proves the
reduced equation from original Lie-Poisson equation
(\ref{Hq}).
}\hspace{\fill} { \fbox {}}\\

As a most important result,
the following theorem
shows that Lie-Poisson equation
(\ref{H[X]cl}), or the set of
equations (\ref{density's})
and (\ref{current's}),
actually
represents 
the classical Liouville equation.

\begin{tm}\label{canonical equivalent}
Lie-Poisson equation (\ref{H[X]cl}) 
is equivalent to
the classical Liouville equation
for the probability density
function (PDF) $\rho_t^{T^*M} \in C^{\infty }(T^*M,{\bf R})$
of a particle on cotangent space $T^*M$:
\begin{equation}\label{equation for rho}
{{\partial } \over {\partial t }}\rho^{T^*M}_t  =
\{ \rho^{T^*M}_t , H^{T^*M} \} ,
\end{equation}
where the Poisson bracket 
$\{ \ , \ \} $  is defined for
every $A$, $B\in C^{\infty }(M)$ as
\begin{equation}
\{ A , B \} = {{\partial A}\over {\partial p_j}}
{{\partial B}\over {\partial x^j}}
-{{\partial B}\over {\partial p_j}}{{\partial A}\over {\partial x^j}} .
\end{equation}
\end{tm}
\par\noindent${\it Proof.}\ \ $
{\it Classical equation (\ref{equation for rho})
 is equivalent to
 the
canonical equations of motion
through the local expression such that $\phi_{U_{\alpha }}
\left( q_t \right) = \left( x_t , p_t \right) $ for
the bundle mapping $\phi_{U_{\alpha }}: \pi^{-1}(U_{\alpha }) \to 
U_{\alpha } \times {\bf R}^N$:
\begin{equation}\label{classical canonical p:app}
{{dp_{jt}}\over {dt}}= 
-{{\partial  H^{T^*M} }
\over {\partial x^j}}(x_t ,p_t)
\ \ \ \ \ \ \ \ \ \ \ \ \ \ \ \ \ \ \ 
{{dx^j_t}\over {dt}} =
{{\partial H^{T^*M} }\over {\partial p_j}}(x_t ,p_t) .
\end{equation}

If 
$ q_t  
=  \left( x_t ,  \bar  p_{t} ^{\tau }[k]
(x_t ) \right)
$
satisfies
canonical equations  of motion
(\ref{classical canonical p:app}),
the above equation of motion
induces 
\begin{equation}\label{p's eq:app}
{{\partial\bar   p_{tk}^{\tau } } \over {\partial t}}[k](x)
= -{{\partial  H^{T^*M} }\over {\partial x^k }}
 \left( x  ,\bar   p_t^{\tau } [k] (x)
\right) 
-{{\partial H^{T^*M} }\over {\partial p_j}}
 \left( x  ,\bar   p_t^{\tau } [k] (x)
\right)
\partial_j  \bar  p_{tk}^{\tau } [k](x) ,
\end{equation}
then 
relation (\ref{classical probability})
satisfies the following equation:
\begin{eqnarray}\nonumber
{{\partial  } \over {\partial t}} \bar \rho^{\tau } _t[ k](x) 
&=&  \surd ^{-1}  
 \partial_j
\left(  \sigma_t^{\tau } [k] (x)
{{\partial H^{T^*M} }\over {\partial p_j}} \left( x,
\bar   p_t^{\tau } [k](x)\right) 
\right) \rho^{T^*M}_t (x ,\bar  p_t^{\tau }[k] (x) ) \\
\nonumber
&  &
+ \surd ^{-1}  
 \sigma_t^{\tau } [k] (x)
{{\partial   \rho^{T^*M}_t } \over {\partial t}} (x ,\bar  p_t^{\tau }[k] (x) )\\
\nonumber
&  &  
- \surd ^{-1}  \sigma_t^{\tau } [k] (x)
{{\partial H^{T^*M} }\over {\partial x^j}} 
\left(  x, \bar  p_t^{\tau } [ k](x)\right)
  {{\partial  \rho^{T^*M}_t  } \over {\partial p_j}}  (x ,
\bar  p_t^{\tau }[ k] (x) ) \\
\nonumber
&  &
- \surd ^{-1}  \sigma_t^{\tau } [ k] (x)
{{\partial H^{T^*M} }\over {\partial p_j}} \left( x, \bar  p_t^{\tau } [ k](x)\right)
\\
\nonumber
&  &
\times
 \partial_j \bar   p_{tk} [ k](x)
 {{\partial   \rho^{T^*M}_t } \over {\partial p_k}}  (x ,\bar  p_t^{\tau }[ k] (x) )\\
\label{density's eq:app}
&=& - \surd ^{-1} \partial_j   \left(
{{\partial H^{T^*M} }\over {\partial p_j}} \left( x, \bar  p_t^{\tau } [ k](x)\right)
 \rho^{\tau } _t[k](x)  \surd \right) .
\end{eqnarray}
Equations (\ref{p's eq:app}) and (\ref{density's eq:app})
lead to the following equation:
\begin{eqnarray}\nonumber
{{\partial  } \over {\partial t}}
\{ \bar \rho^{\tau } _t[k](x)  \bar  p_{tk}^{\tau }[k](x) \}
&=&   
 -\bar  p_{tk} [k](x)  \surd ^{-1} \partial_j \left(
{{\partial H^{T^*M} }\over {\partial p_j}} \left( x, \bar  p_t^{\tau } 
[k](x)\right)  
 \bar \rho^{\tau } _t[k](x) \surd \right) \\
\nonumber
&  &
 - \bar \rho _t^{\tau }[k](x) {{\partial 
 H^{T^*M} }\over {\partial x^k }}
\left( x, \bar  p_t^{\tau } [k](x) \right) \\
\nonumber
&  &
- \bar \rho^{\tau } _t[k](x) 
{{\partial H^{T^*M} }\over {\partial p_j}} \left( x, \bar  p_t^{\tau } [k](x)\right)
\partial_j  \bar  p_{tk}^{\tau } [k](x) \\
\nonumber
&=&
 -  \surd ^{-1} \partial_j \left(
{{\partial H^{T^*M} }\over {\partial p_j}} \left( x, 
\bar  p_t^{\tau } [k](x)\right)  
\bar  \rho^{\tau } _t[k](x) \bar  p_{tk}^{\tau } [k](x) \surd \right)\\
\nonumber
&  &
- \bar  \rho^{\tau } _t[k](x) 
\left\{ \bar   p_{tj}^{\tau } [k](x) 
\partial_k  
\left( {{\partial  H^{T^*M} }\over {\partial p_j}} 
\left( x, \bar  p_t^{\tau } [k](x)\right)  \right) \right.
\\
\label{current's eq:app}
&  & \left.
+ \partial_k  L^H  \left( x, \bar  p_t^{\tau } [k](x)\right) 
\right\}  .
\end{eqnarray}
Equations (\ref{density's eq:app}) and
(\ref{current's eq:app})
are equivalent to equations (\ref{density's}) and
(\ref{current's});
thereby, canonical equation (\ref{equation for rho})
is equivalent to
Lie-Poisson equation (\ref{H[X]cl}).
}\hspace{\fill} { \fbox {}}\\

The above discussion has
a special example
of the following Hamiltonian:
\begin{equation}\label{special Hamiltonian}
H^{ T^*M }_t   \left( x , p   \right)
= 
g^{ij}  (x)\left( p_{i}   
+ A_i  
\right)
\left( p_{j}   + A_j  
\right)
+U   (x) ,
\end{equation}
where corresponding Hamiltonian
operator $ \hat H_t  $
is calculated 
as 
\begin{equation}
\hat H_t [k]
= \left( 
 g^{ji} \left( \bar  p _{ti}[k]  + A_i   \right)
 \partial_j ,
 -  g^{ji}  \bar  p _{tj}[k] \bar   p _{ti} [k]   
+g^{ji} A_j  A_i   +U  
 \right)  ; 
\end{equation}
thereby, equation (\ref{H[X]cl})
is described for special Hamiltonian (\ref{special Hamiltonian})
 as
\begin{eqnarray}
\nonumber
{ {\partial } \over  {\partial t}  } 
\left( \bar  \rho _t^{\tau }[k] (x) \bar  p_{tj}^{\tau }[k](x) \right)  &=& -
 \surd ^{-1} \partial_i 
\left\{ g^{ik}(x)
\left( \bar  p_{tk}^{\tau }[k](x) 
 +A_k(x) \right)  \bar \rho _t^{\tau } [k](x)
\bar  p_{tj}^{\tau } [k] (x) \surd  \right\} \\
\nonumber
&  &- \bar  \rho _t^{\tau } [k](x) \left( \partial_j 
 g^{ik}(x)  \right) 
\bar  p_{ti}^{\tau }[k] (x)
\bar  p_{tk}^{\tau }[k] (x)\\ \nonumber
&  &-
\left( \partial_j  g^{ik}(x) A_k 
(x) \right) \bar  \rho _t^{\tau }[k] (x) \bar  p_{ti}^{\tau }[k] (x) \\ 
\label{motion equation 1}
&  &  -
\bar  \rho _t^{\tau }[k] (x) \partial_j 
\left\{  U(x) 
+ g^{ik}(x) A_i(x)  A_k (x) \right\} ,
\end{eqnarray}
\begin{equation}\label{motion equation 2}
{ {\partial } \over  {\partial t}  } \bar  \rho _t^{\tau }[k] (x)   = 
- \surd ^{-1} \partial_i 
\left\{ g^{ik}(x)\left( \bar  p_{tk}[k](x) 
+ A_k (x) \right) \bar  \rho _t^{\tau }[k] (x) \surd  \right\} .
\end{equation}

For
$  \bar {\cal U}_t  
 \in \bar Q \left(  M\right) $ such that
$
{{\partial  
\bar  {\cal U}_t  }\over {\partial t}}  
\circ  \bar  {\cal U}_t     ^{-1}= 
\hat H_t^{cl}  
 \in   \bar q\left(  M\right)  $,
let us introduce operators
\begin{eqnarray}
  \tilde H_t^{cl}   =& Ad^{-1}_{\bar {\cal U}_t   }
\hat H_t^{cl}  ,\\
 \tilde F_t ^{cl} =& Ad^{-1}_{\bar {\cal U}_t   }
\hat F_t^{cl} ,
\end{eqnarray}
which induces the following equation
equivalent to
equation (\ref{H[X]cl}):
\begin{equation}
{{\partial }\over {\partial t}}\tilde F_t^{cl}  = 
\left[ \tilde H _t ^{cl},\tilde F_t^{cl}  \right] 
+\widetilde{\left( {{\partial F_t    }
\over {\partial t}} \right) }^{cl} .
\end{equation}
This expression
of the equations of motion
coincides with
the following Poisson equation
because of
Theorem \ref{canonical equivalent}:
\begin{equation}\label{cl. Poisson eq.}
{{d  } \over {d  t }} F_t^{T^*M}
=\left\{ H_t^{T^*M}, F_t^{T^*M} \right\}     +
{{\partial F_t^{T^*M}} \over {\partial  t }}     .
\end{equation}

As discussed in Section 3,
 if a  group action of Lie group
$ Q(M)$
keeps the Hamiltonian $\bar {\cal H}_t: \bar q(M)^* \to {\bf R}$  
invariant,
there exists an invariant charge 
function
$Q ^{T^*M}\in C^{\infty }(T^*M)$ and the induced
function $\bar {\cal Q}: \bar q(M)^* \to {\bf R}$ such that
\begin{equation}\label{charge-invariant(cl)}
\left[ \hat H _t^{cl}, \hat Q^{cl} \right]  = 0 ,
\end{equation}
where $\hat Q^{cl}  $ is expressed as
\begin{equation}
\hat Q ^{cl}
= \left(  {{\partial  Q^{T^*M}}\over {\partial p}}
\left(  x, \bar p_t^{\tau }  [k](x) \right) ,
- p  (\eta )  \cdot    {{\partial  Q^{T^*M}}\over {\partial p}}
\left(  x, \bar p_t^{\tau }  [k](x) \right)
+ Q^{T^*M}   \left(  x, \bar p_t^{\tau }  [k](x) \right)
\right)  .
\end{equation}
Relation (\ref{charge-invariant(cl)})
is equivalent to the following 
convolution relation:
\begin{equation}
\left\{  H _t^{T^*M},   Q^{T^*M} \right\} = 0 .
\end{equation}

In the argument so far on the dynamical construction
of classical mechanics,
the introduced infinite-dimensional freedom
of the background $B\left[ E(M)\right] $
seems to be redundant, while they appear as a natural consequence
of the general theory on protomechanics
discussed in the previous section.
In fact, it is really true that
one can directly induce classical mechanics as the dynamics of the 
Lagrange foliations of $T^*M$ in $L \left( T^*M \right) $.
In the next section, however,
it is observed that we will encounter difficulties
without those freedom
if moving onto the dynamical construction
of quantum mechanics.

\section{DEDUCTION OF QUANTUM MECHANICS}

In canonical quantum mechanics,
the  state of a particle 
on  manifold $M$
can be
represented as a position  in
the Hilbert space ${\cal H}(M)$ 
of all the $L_2$-functions over $M$.
In this section,  we will reproduce
the quantum equation of motion
from the general theory
presented in Section 4.
Let us here concentrate ourselves
on the case where $M$ is $N$-dimensional
manifold for simplicity,
though the discussion
below is still valid if
substituting an appropriate Hilbert space
 when
$M$ is infinite-dimensional
ILH-manifold\cite{Omori}.

\subsection{Description of Statistical-State}

Now,
we must be concentrated
on the case
where the physical
 functional $F\in C^{\infty }\left( \Lambda^1(M) , C^{\infty }(M) \right) $
depends  on the derivatives 
of the 1-form $p\left( 
\eta \right)\in \Lambda^1 (M) $
induced from $\eta \in \Gamma \left[ E(M) \right] $, then it
has the following expression:
\begin{equation}\label{semi-local F}
p^* F  \left( \eta \right) (x) = 
F^{Q} \left( x,  p\left( \eta \right) (x) ,
D p\left( \eta \right) (x) ,
... , D^{n} p\left( \eta \right) (x) ,...\right)   .
\end{equation}

Let us assume that $M$ has a finite
covering $M= \bigcup_{\alpha \in \Lambda_M}U_{\alpha }$
for the mathematical simplicity
such that
$\Lambda_M = \left\{ 1,  2, ... , \Lambda \right\} $
for some $\Lambda \in {\bf R}$,
and choose a  
coordinate system  
$\left( U_{\alpha } , {\bf x}_{\alpha  } 
\right) _{\alpha \in \Lambda_M}$.
Let us further choose
a reference set $U \subset U_{\alpha }$
such that $v(U) \neq 0$
and consider the set $\Gamma^{\hbar }_{U k}\left[ E(M)\right] $ of  the
$C^{\infty }$ sections of $E(M)$
for $ k= (k_1 ,..., k_N) \in {\bf R}^N$
such that\footnote{
As in classical mechanics,
to substitute
$\Gamma^{\hbar }_{U k}\left[ E(M)\right] = \left\{ 
\eta \in \Gamma \left[ E(M)\right] \  \left\vert
\ \int_{  U}dv(x) \ p_j\left( x
\right)  = \hbar k_j v\left( U \right) \right.
\right\} $ for
definition (\ref{gamma})
also induces the similar discussion
below,
while there exist a variety of the classification
methods that produce the same result.
}
\begin{equation}\label{gamma}
\Gamma^{\hbar }_{U k}\left[ E(M)\right]  = 
\left\{ \eta \in \Gamma \left[ E(M)\right] \  \left\vert
\ \sup_{  U} p_j\left( \eta
\right)  (x) = \hbar k_j \right.
\right\} .
\end{equation}
As in classical mechanics,
 we will simply denote
$\Gamma^{\hbar }_{U k}\left[ E(M)\right] $
as $\Gamma^{\hbar }_{ k}\left[ E(M)\right] $,
since 
$\Gamma^{\hbar }_{U k}\left[ E(M)\right] $
can be identified with
$\Gamma^{\hbar }_{U^{\prime } k}\left[ E(M)\right] $
for every two reference sets $U$ and $U^{\prime }\subset M$.

For every $\bar  p 
= p\circ \bar  \eta  \in L\left( T^*M \right) $
such that 
$\bar  \eta [k] \in \Gamma^{\hbar }_{  k}\left[ E(M)\right]  $,
it is further possible to
separate an element $\eta [k] \in \Gamma^{\hbar }_{  k}\left[ E(M)\right] $
for a
$\xi\in  \Gamma^{\hbar }_{ 0}\left[ E(M)\right]  $ as
\begin{equation}
\eta [k]= \bar  \eta [k] \cdot \xi ,
\end{equation}
or to 
separate 
momentum
$p \left( \eta [ k] \right) $ as
\begin{equation}\label{q-separation}
p \left( \eta [ k] \right) 
= \bar  p    [ k]  
  + p  \left( \xi \right) .
\end{equation}
The emergence density $ \rho    \left( \eta [k]\right) 
$ can have  
the same expression as the classical one (\ref{rho=rho})
for the  function $ \varrho  \left( \xi \right)   \in  C^{\infty }
\left( T^*M,{\bf R}\right) $ on $T^*M$
since $C_{q}\left( \Gamma \right) ^*
\subset C_{cl}\left( \Gamma \right) ^*$:
\begin{equation}\label{rho=rho-q}
 \rho    \left( \eta [k]\right) 
(x) \surd 
=  
\varrho  \left( \xi \right)   \left( x, p \left( \eta [k]\right) (x) \right) ,
\end{equation}
which has only the restricted values
if
compared
with  the classical 
emergence  density; it
sometimes causes the discrete spectra of
the wave-function in canonical quantum mechanics.
We call the set
$ B^{\hbar }\left[ E(M)\right]  
= \Gamma^{\hbar }_{0}\left[ E(M)\right]  $ as
the {\bf back ground} of $L\left( T^*M \right) $
for quantum mechanics.
For the  measure
${\cal N}$ on $ B^{\hbar }\left[ E(M)\right]   $
for the $\sigma $-algebra induced from that of $\Gamma \left[ E(M)\right] $:
\begin{equation}
d {\cal M} \left( \eta [k]\right) \
d v (x)
= d^Nk d {\cal N}  \left(  \xi \right) 
 d v (x) \ \sigma [k](x)\  .
\end{equation}

 Let us next consider 
the disjoint union 
$M = \bigcup_{\alpha \in \Lambda_M }
A_{\alpha }$
for $A_{\alpha } \in {\cal B}\left( {\cal O}_{E(M)} \right) $
such that
(1) $\pi \left( A_{\alpha }
\right)  \subset U_{\alpha }$ and that (2)
 $A_{\alpha }\cap A_{\beta }
= \emptyset $ for $\alpha \neq \beta \in \Lambda_M $
(consult {\it APPENDIX A}).
Thus,
every  section $
\eta \in \Gamma \left[ E(M)\right] $ has 
some $k\in {\bf R}^N$
such that $\eta =\eta [k]\in\Gamma^{\hbar }_{ k}\left[ E(M)\right] $; and,
it will be separated
into the product
 of a   $\xi \in B^{\hbar }\left[ E(M)\right] $
and the fixed $\bar \eta [k] =  
e^{2i \{ k_{  j} x^j 
+  \zeta   \} }\in \Gamma_{ k}\left[ E(M)\right] $
 that induces one of the Lagrange foliation $ \bar p
= p \circ \bar \eta \in L\left( T^*M \right) $:
\begin{eqnarray}
\eta \left[ k\right] &=& \sum_{\alpha \in A_{\alpha }}
 \chi_{A_{\alpha }}  \cdot
e^{2i \{ k_{  j} x^j 
+  \zeta   \} } \cdot \xi \\
&=&
\prod_{\alpha \in A_{\alpha }}
\left( 
e^{2i \{ k_{ j} x^j 
+  \zeta  \} } \cdot \xi \right) ^ {\chi_{A_{\alpha }}}
,
\end{eqnarray}
where the test function $\chi_{A_{\alpha }}:M \to {\bf R} $
satisfies 
\begin{equation}
\chi_{A_{\alpha }}(x)
= \left\{ 
{\matrix{1\cr
0\cr }\    \ } \right.
\matrix{{at\  x\in A_{\alpha }}\cr
{at\  x\notin A_{\alpha } }\cr
}  
\end{equation}
and has the projection property 
$ \chi_{A_{\alpha }}^2 = \chi_{A_{\alpha }}$.

If defining
the {\it window mapping} $\chi_{A_{\alpha }}^*:
C^{\infty }(M) \to L^1\left( {\bf R}^N\right) $
for any $f\in C^{\infty }(M)$
such that
\begin{equation}
\chi_{A_{\alpha }}^* f \left( {\bf x}\right) =
\left\{ 
{\matrix{\varphi_{\alpha }^* f \left(
 {\bf x}\right)\cr
0\cr }\    \ } \right.
\matrix{{at\  {\bf x}\in \varphi_{\alpha }\left( A_{\alpha } \right) }\cr
{at\   {\bf x}\notin \varphi_{\alpha }\left( A_{\alpha } \right)  }\cr
}  ,
\end{equation}
we can {\it locally} transform  the function $\rho [k]  \left(
\xi \right)
= \sigma [k] \rho \left( 
\eta \left[  k\right]  \right)   \surd  $
into Fourier coefficients as follows:
\begin{equation}
\chi_{A_{\alpha }}^*
\rho [k] \left(
\xi \right)  \  \left( {\bf x}\right) 
 = 
\int_{{\bf R}^{N }}  d^{N}k^{\prime } \
 \tilde \varrho_{\alpha } \left( \xi \right)
  \left(  {{2k
+ k^{\prime }}\over 2},  {{2k- 
k^{\prime } }\over 2} \right)
e^{ik^{\prime }  {\bf x} ^j } ,
\end{equation}
where introduced function
 $\tilde \varrho_\alpha  $
should satisfies
\begin{equation}
\tilde \varrho_\alpha \left( \xi \right) 
 (k, k^{\prime }) ^* =\tilde \varrho_\alpha 
\left( \xi \right)  ( k^{\prime },k),
\end{equation}
for the value $\rho [k] \left(
\xi \right) (x) $ is real at every $x\in M$;
thereby,
the collective expression
gives
\begin{eqnarray}
\rho   \left[  k\right]   \left(
\xi \right)
 &= & \sum_{\alpha \in A_{\alpha }}
 \chi_{A_{\alpha }}  \cdot
\int_{{\bf R}^{N }}  d^{N}k^{\prime } \
 \tilde \varrho  _\alpha \left( \xi \right)
  \left(  {{2k
+ k^{\prime }}\over 2},  {{2k- 
k^{\prime } }\over 2} \right)
e^{ik^{\prime }x^j } \\
&= &
\int_{
{\bf R}^{N  }} 
 d^{N  } k^{\prime } \
  \tilde \varrho  
\left( \xi \right)  \left(  {{2 k
+  k^{\prime }}\over 2},  {{2 k- 
 k^{\prime }}\over 2} \right) 
\cdot  \eta  \left[ 
 k -{{  k^{\prime }}\over 2}
\right] ^{-{1\over 2}}   \eta      \left[ 
 k+ {{  k^{\prime }}\over 2} 
\right]  ^{ {1\over 2}}  ,
\end{eqnarray}
where
\begin{equation}
  \tilde \varrho  \left( \xi \right)
  \left(  {{2 k
+  k^{\prime }}\over 2},  {{2 k- 
 k^{\prime }}\over 2} \right) 
= \prod_{\alpha \in A_{\alpha }}
\left( 
 \tilde \varrho  _\alpha \left( \xi \right)
  \left(  {{2k
+ k^{\prime }}\over 2},  {{2k- 
k^{\prime }}\over 2} \right) \right)^{\chi_{A_{\alpha }} } .
\end{equation}

Let us introduce
the ketvector $\left\vert   k     \right\rangle $
and bravector $\left\langle   k     \right\vert $
such that
\begin{equation}
  \left\vert   k     \right\rangle 
= \prod_{\alpha \in \Lambda _M }
\left\vert   k   ,\alpha
 \right\rangle    \ \  , \ \ \ \ 
\left\langle   k     \right\vert 
= \prod_{\alpha \in \Lambda _M }
\left\langle   k  ,\alpha   \right\vert ,
\end{equation}
where the local vectors $\left\vert   k   ,\alpha
 \right\rangle $ and $\left\langle   k  ,\alpha   \right\vert $
satisfy
\begin{equation}
\left\langle    x \left\vert  
 k   ,\alpha    \right. \right\rangle 
=
e^{2i \{ k_{ j} x^j 
+  \zeta  \} \chi_{A_\alpha } }  \surd ^{-{1\over 2}} \ \ , \ \ \
\ \left\langle   k, \alpha  \left\vert  
x    \right. \right\rangle 
=
e^{2i \{ -k_{ j} x^j 
+  \zeta  \} \chi_{A_\alpha } } \surd ^{-{1\over 2}}.
\end{equation}
We can define 
the Hilbert space ${\cal H} \left( M\right) $ of  all the vectors
that can be expressed as a linear combination
of vectors $\{ \vert k \rangle \}_{k\in {\bf R}} $.
Now,
let us construct the {\it density matrix}
in the following definition.
\begin{de}\label{q-density}
The {\bf density matrix}  $\hat \rho  $
is an operator
such that
\begin{eqnarray}
\hat \rho 
&=&
\int_{B^{\hbar }\left[ E(M) \right]  }
d{\cal N} (\xi ) 
\int_{
{\bf R}^{N }} d^N  n
\int_{
{\bf R}^{N }} 
d^N n^{\prime } \
\tilde \varrho (\xi )  \left( n,  n^{\prime }  \right) 
\ \xi^{{1\over 2} } 
 \left\vert  n \right\rangle  
\left\langle  n^{\prime } \right\vert    \xi^{-{1\over 2} } \\
&=& \int_{B^{\hbar }\left[ E(M) \right]   }
d{\cal N} (\xi )  \int_{
{\bf R}^{N }} d^N  k \ \hat  \rho [k]  \left( \xi \right) ,
\end{eqnarray}
where
\begin{equation}
\hat \rho  \left[  k\right]  \left( \xi \right) = 
\int_{
{\bf R}^{N }} d^N k^{\prime } \
\tilde \varrho (\xi )  \left( k+  {k^{\prime }\over 2} ,
  k-  {k^{\prime }\over 2}  \right) 
\ \xi^{{1\over 2} } 
 \left\vert   k+  {k^{\prime }\over 2} \right\rangle  
\left\langle   k-  {k^{\prime }\over 2} \right\vert \xi^{-{1\over 2} } .
\end{equation}
\end{de}

Let ${\cal O} \left( M\right) $
be the set of all the hermite operators acting on
Hilbert space ${\cal H} \left( M\right) $,
which has
the bracket $\langle \ \ \rangle : {\cal O} \left( M\right)
\to {\bf R}$ for every hermite operator
$\hat {\bf F}  $ such that
\begin{equation}
\left\langle \hat  {\bf F}    \right\rangle
= \int_{{\bf R}^N}d^Nk \int_M dv(x)\ 
\left\langle  x \left\vert \hat {\bf F}      \right\vert x \right\rangle .
\end{equation}
Set ${\cal O} \left( M\right) $
becomes the algebra
with the  product, scalar product  and addition;
thereby,
we can consider
 the commutation 
and the anticommutaion
between   operators $\hat {\bf A}$, $\hat {\bf B} \in {\cal O} \left( M\right) $:
\begin{equation}
\left[ \hat {\bf A}, \hat {\bf B}\right]_{\pm }
=  \hat  {\bf A}    \hat  {\bf B}   
\pm  \hat  {\bf B}    \hat  {\bf A}  .
\end{equation}
Consider the 
momentum operator $\hat {\bf p} $
that satisfies the following relation
for any $ \left\vert \psi  \right\rangle
\in {\cal H}\left( M \right) $:
\begin{equation}
 \left\langle x \left\vert  \hat {\bf p}  
\right\vert \psi \right\rangle
=-i D
\left\langle x \left\vert \psi \right. \right\rangle ,
\end{equation}
where $D = \hbar dx^j \partial_j$ is the derivative
operator (\ref{D-der}).
Further,
the function operator $ \hat {\bf f}   $
induced from the function $f \in C^{\infty }(M)$
is an operator that satisfies the following relation
for any $ \left\vert \psi  \right\rangle
\in {\cal H}\left( M \right) $:
\begin{equation}
\left\langle x \left\vert  \hat {\bf f} 
\right\vert \psi \right\rangle
=  f(x) 
\left\langle x \left\vert \psi \right. \right\rangle .
\end{equation}
The following commutation relation
holds:
\begin{equation}\label{q-commute}
\left[  \hat {\bf p}_j , \hat {\bf f}   \right] _-
= {\hbar \over i } \widehat{ \partial_j{\bf f}} .
\end{equation}
Those operators $\hat {\bf f} $ and $\hat {\bf p} $
induces a variety of operators
in the form of their polynomials.
\begin{de}
The hermite operator $\hat  {\bf F} $ is
called  an
{\bf observable},
if it 
can be represented as the polynomial
of the momentum operators $\hat p $
weighted
with  function operators $\hat {\bf f}_n^j $
independent of $k$ such that
\begin{equation}
\hat {\bf F}  =  \sum_{n=0}^{\infty } 
\left[ \hat {\bf f}_n^j  ,  \hat {\bf p}^n_{j} \right]_+ .
\end{equation}
\end{de}
The following lemma shows
that every {\it observable} has its own
physical functional.

\begin{lem}\label{q-gen}
Every observable
$\hat  {\bf F}  $ has 
a corresponding functional $F : \Gamma [E] \to C^{\infty }(M)$:
\begin{equation}
\bar \mu \left( p^*F  \right)  =
\left\langle \hat\rho  \ \hat {\bf F} 
\right\rangle .
\end{equation}
\end{lem}
\par\noindent${\it Proof.}\ \ $ 
{\it 
There are corresponding
functionals $g_{nl}^j: \Lambda^1(M) \to C(M)$
($l\in \{ 1, 2, ..., n \} $)
such that
\begin{eqnarray}\nonumber
\left\langle \hat\rho \ \left[ \hat {\bf f}_n^j  ,  \hat {\bf p}^n_{j} \right]_+
\right\rangle 
&=&
\int_{B^{\hbar }\left[ E(M) \right]   }
d{\cal N} (\xi ) 
\int_{{\bf R}^{N }} d^N  n
\int_{{\bf R}^{N }} d^N n^{\prime } \
\tilde  \varrho (\xi )  \left( n,  n^{\prime }  \right) 
\left\langle  n^{\prime } \left\vert   \ \xi^{-{1\over 2}}
\ \left[ \hat {\bf f}_n^j  ,  \hat {\bf p}^n_{j} \right]_+ \
\xi^{{1\over 2}} 
 \right\vert  n \right\rangle \\ \nonumber
&=&
\int_{B^{\hbar }\left[ E(M) \right]  }
d{\cal N} (\xi ) 
\int_{ {\bf R}^{N }}d^N  k
\int_{
{\bf R}^{N }} 
d^N k^{\prime } 
\\ \nonumber
&  & \times
\sum_{\alpha \in \Lambda_M}\int_{U_{\alpha }}  d^Nx \ 
\tilde  \varrho  
 (\xi ) \left( k-{k^{\prime }\over 2}, k+{k^{\prime }\over 2}
\right) e^{ik^{\prime }_jx^j }
  \left\{ 
\sum_{l= 0 }^{n}
g_{nl}^j\left( p \left( \eta [k] \right) \right) (x) k_j^{\prime l}\right\}
\\ \nonumber
&=&
 \int_{B^{\hbar }\left[ E(M) \right] }
d{\cal N} (\xi ) 
\int_{
 {\bf R}^{N }   } d^N  k
\int_{
{\bf R}^{N }} d^N k^{\prime } 
\\ \nonumber
&  & \times
\sum_{\alpha \in \Lambda_M}\int_{U_{\alpha }}  d^Nx \   
\tilde  \varrho   (\xi ) 
  \left( k-{k^{\prime }\over 2}, k+{k^{\prime }\over 2}
\right) 
e^{ik^{\prime }_jx^j } 
 \left\{ 
\sum_{l= 0 }^{n} 
\left( -\hbar {{ \partial }\over {\partial x^j}}\right) ^l
g_{nl}^j\left( p \left( \eta [k] \right)  \right) (x)
\right\}  \\ \nonumber
&=&
 \int_{B^{\hbar }\left[ E(M) \right]  }
d{\cal N}(\xi ) 
\int_{
{\bf R}^{N }} d^N  k
\int_{M}  dv(x)  \ 
\rho \left(  \eta [k]\right) (x) \
p^* F_j^n  \left(  \eta [k]\right) 
 (x)\\ \nonumber
&=&
 \int_{\Gamma \left[ E(M)\right] }
d{\cal M}(\eta ) 
\int_{M }   dv (x) \ 
\rho (\eta ) (x) \  p^* F_j^n (\eta )  (x) \\ 
&=& \bar \mu \left( p^* F_j^n \right)
.\end{eqnarray}
where 
\begin{equation}
p^* F_j^n 
\left(  \eta  [k]
  \right)  (x)  =
\sum_{l= 0 }^{n} \left\{
\left( -\hbar {{ \partial }\over {\partial x^j}}\right) ^l
g_{nl}^j\left(  p\left(  \eta [k]\right)  \right) (x)\right\} .
\end{equation}
}\hspace{\fill} { \fbox {}}\\

\subsection{Description of Time-Development}

Now, 
we can describe a 
$     \eta_t^{\tau } \left( \eta [k] \right)
\in  \Gamma _{Uk}\left[ E(M)\right]  $
as 
\begin{eqnarray}
   \eta_t^{\tau } \left( \eta [k] \right)
&=&  
\sum_{\alpha \in A_{\alpha }}
 \chi_{A_{\alpha }}  \cdot
e^{2i \{ k_{\alpha j} x^j 
+  \zeta_t^{\tau }[k]  \} } \cdot \xi \\
&=&
\prod_{\alpha \in A_{\alpha }}
\left( 
e^{2i \{ k_{\alpha j} x^j 
+  \zeta _t^{\tau } [k]\} } \cdot \xi \right) ^ {\chi_{A_{\alpha }}}
 ,
\end{eqnarray}
where the
function
$ \zeta  _t ^{\tau } 
[  k  ] 
\in C^{\infty }\left( M \right) $
labeled by {\it labeling time}
$\tau  \leq t \in {\bf R}$ satisfies
\begin{equation}
 \zeta _{\tau }^{\tau }  [  k  ] =  \zeta \ \ \ \ : independent\ of\ k   ;
\end{equation}
thereby,
the momentum $  p_t^{\tau }\left( 
\eta   [  k]\right) =
\bar p_t^{\tau } [  k ] + p \left( 
\xi \right) 
\in \Lambda ^1 (M) $ for $\bar p^{\tau }_{t} = 
p^{\tau }_{t} \circ \bar \eta \in L\left( T^*M \right) $
satisfies the  Einstein-de Broglie relation:\footnote{
Relation (\ref{improvement}) is 
the most crucial improvement from
the corresponding relation
in previous letter \cite{Ono}.}
\begin{equation}\label{improvement}
\bar   p_t^{\tau } [   k ] =
-i {\hbar \over 2} 
\bar   \eta_t^{\tau }   [  k ]  ^{-1} d 
\bar   \eta_t ^{\tau }  [   k ]  .
\end{equation}

The   
density operator $\hat \rho_t^{\tau } [k] \left( \xi \right) $
is introduced as
\begin{equation}\label{tau-dep density op}
 \hat \rho_t^{\tau } [k]  \left( \xi \right)  =
\int_{
{\bf R}^{N }} d^N k^{\prime } \
\tilde \varrho_t^{\tau } (\xi )  \left( k+  {k^{\prime }\over 2} ,
  k-  {k^{\prime }\over 2}  \right) 
\ \xi^{{1\over 2} } 
 \left\vert   k+  {k^{\prime }\over 2} \right\rangle  
\left\langle   k-  {k^{\prime }\over 2} \right\vert \xi^{-{1\over 2} } ,
\end{equation}
which satisfies the following lemma.
\begin{lem}
\begin{equation}\label{quantum-sigma-1}
\hat \rho_t          =   \int_{\Gamma _U}d{\cal N}(\xi )
 \int_{{\bf R}^N} d^Nk \
 U^{\tau }_t [k]   \hat \rho_t^{\tau }  [k] \left( \xi \right)
 U^{\tau }_t [k]^{-1}  ,
\end{equation}
where
\begin{equation}
{ U^{\tau }_t} \left[  k\right]  =   e^{i \{ \zeta_t^{\tau } 
\left[  k\right]  - \zeta  \} }   .
\end{equation}
\end{lem}
\par\noindent${\it Proof.}\ \ $
{\it 
The direct calculation shows for 
the observable $\hat {\bf F}_t $
corresponding to every functional $F$
\begin{eqnarray}\nonumber
\left\langle \hat\rho_t 
\ \hat {\bf F}_t  \right\rangle  &=&
\bar \mu_t
\left( p^*F_t    \right) \\ \nonumber
&=& \int_{\Gamma \left[ E(M)\right] }d{\cal M}(\eta ) \
\int_M dv \ \rho_t^{\tau }
\left(  \eta \right)  
(x) \ p^*F_t
\left(  \eta _t^{\tau } \left(\eta \right)  
  \right)   \\ \nonumber
&=&  \int_{B\left[ E(M)\right] }
d{\cal N}\left( \xi \right)  
\int_{ {\bf R}^{N }} d^N  k 
\int_{M}  dv(x)  \ 
\rho_t^{\tau } [k] \left( \xi \right) (x)  \ p^*F
\left(  \eta_t^{\tau }  [k]
  \right) (x)
  \\ \nonumber
&=&
\int_{B \left[ E(M)\right] }
d{\cal N}\left( \xi \right) 
\int_{
{\bf R}^{N }} d^N  k
\int_{M}  dv(x)  \ 
\rho_t^{\tau } [k] \left( \xi \right) (x) \ p^*F
\left( \eta [k] \cdot e^{i \{ \zeta_t^{\tau } 
\left[  k\right]  - \zeta  \} }   .
  \right) (x) 
 \\ \nonumber
&=&
\int_{B \left[ E(M)\right] }
d{\cal N}\left( \xi \right) 
\int_{
{\bf R}^{N }} d^N  k
\int_{M}  dv(x)  \ \left\langle x
\left\vert \
{1\over 2}\left[ \ U^{\tau }_t [k]  \hat
\rho_t^{\tau } [k] \left( \xi \right)   U^{\tau }_t [k]^{-1}  
\ , \ \hat {\bf F}_t \right]_+ \ \right\vert x
\right\rangle \ 
\\
&=&
  \ \left\langle 
\left\{
\int_{B\left[ E(M)\right] }
d{\cal N}\left( \xi \right) 
\int_{
{\bf R}^{N }} d^N  k \
  U^{\tau }_t [k]  \hat
\rho_t^{\tau } [k]\left( \xi \right)   U^{\tau }_t [k]^{-1} \right\} \ 
  \hat {\bf F}_t    
\right\rangle \  .
\end{eqnarray}
}\hspace{\fill} { \fbox {}}\\
Relation (\ref{quantum-sigma-1})
represents
relation (\ref{q-measure-rel}):
\begin{equation}
\tilde \mu_t   (\eta ) 
= {{d{\cal M} (  \eta  )  }
\over {d{\cal M}\left(  \eta_t^{\tau \ -1} (  \eta  ) \right)
}} \cdot \tilde \mu_t^{\tau }\left(
 \eta_t^{\tau \ -1}  \left(   \eta  \right) \right) .
\end{equation}

Emergence-momentum $ {\cal J}_t^{\tau }
={\cal J}\left( \eta _t^{\tau }  \right) \in q(M)^*$
has the following expression:
\begin{eqnarray}
  {\cal J}_t^{\tau } &=& d^Nk
d{\cal N}\left( \xi \right) dv \ \left( \rho_t^{\tau }
[k]\left( \xi \right)    p_t^{\tau }\left( \eta [k] \right) ,
\rho_t^{\tau }
[k]\left( \xi \right) \right) \\
\label{J}
&=& d {\cal N}\left( \xi\right)
 d^N k \wedge dv \
\left( \ {1\over 2}\left\langle 
x \left\vert \left[  \hat \rho _t^{\tau }  [k]  \left( \xi \right)
 ,\hat {\bf p}_t^{\tau } [k]  \right] _+
\right\vert
x \right\rangle \ ,\
 \left\langle 
x \left\vert   \hat \rho_t^{\tau }  [k]  \left( \xi \right)
\right\vert
x \right\rangle \ \right) ,
\end{eqnarray}
where the momentum operator $\hat {\bf p}_t^{\tau } [k]$
satisfies
\begin{equation}
\hat {\bf p}_t^{\tau } [k]  =  U_t^{\tau } [k]^{-1} \ \hat {\bf p}\  U_t^{\tau } [k]  .
\end{equation}
The following calculus
of the fourier basis
for $2k_j = n_j + m_j $ justifies expression (\ref{J}):
\begin{eqnarray}\nonumber
e^{-i\{
n_j {\bf x}^j + \zeta_t^{\tau }[k]   \} } d
e^{+i\{
m_j {\bf x}^j +  \zeta_t^{\tau }[k]    \} } -
e^{+i\{
m_j {\bf x}^j +  \zeta_t^{\tau }[k]      \} } d
e^{-i\{
n_j {\bf x}^j +  \zeta_t^{\tau }[k]    \} }
=\ \ \ \ &  &\\ \nonumber
e^{-i\{
n_j {\bf x}^j +  \zeta_t^{\tau }[k]    \} } d
\left\{  e^{+i\{
(m_j+n_j ) {\bf x}^j +  2\zeta_t^{\tau }[k]    \} } 
\cdot
e^{-i\{
n_j {\bf x}^j +  \zeta_t^{\tau }[k]   \} }
\right\} -
e^{+i\{
m_j {\bf x}^j + \zeta_t^{\tau }[k]     \} } d
e^{-i\{
n_j {\bf x}^j +  \zeta_t^{\tau }[k]     \} }
=\ \ \ \ &  &\\
 e^{-i (
n_j- m _j ) {\bf x}^j   } 
\cdot  e^{-i\{ (
n_j+m_j ) {\bf x}^j + 2  \zeta_t^{\tau }[k]     \} } 
d  e^{ +i\{ (
n_j+m_j ) {\bf x}^j + 2  \zeta_t^{\tau }[k]    \} } .&  &
\end{eqnarray}
For Hamiltonian operator $\hat  H_t ^{\tau } =
{{\partial {\cal H}_t}\over {\partial \bar {\cal J}}} 
  \left(\bar {\cal J} _t^{\tau } \right)\in   q\left( M\right) $,
the equation of motion
is the Lie-Poisson equation 
\begin{equation}\label{H[X]qq}
{{\partial  {\cal J} _t^{\tau }}\over {\partial t}} 
= ad^*_{\hat H_t  }  {\cal J} _t ^{\tau }  ,
\end{equation}
that is  calculated 
as follows:
\begin{equation}
\label{density's-qq}
{{\partial  } \over {\partial t}} 
 \rho _t^{\tau }[k]\left( \xi \right)  (x) 
= -\surd ^{-1}\partial_j \left(
{{\partial H^{ T^*M }_t}\over {\partial p_j}} 
\left( x,  p_t^{\tau }\left( \eta [k] \right) (x)\right)
\rho  _t^{\tau }[k]\left( \xi \right)  (x)  \surd   \right) ,
\end{equation}
\begin{eqnarray}\nonumber
{{\partial  } \over {\partial t}}
\left( \rho  _t^{\tau }[k]\left( \xi \right)  (x) 
 p_{tk}^{\tau }\left( \eta [k] \right) (x) \right)
\nonumber
&=&
 - \surd^{-1} \partial_j \left(
{{\partial H^{ T^*M }_t}\over {\partial p_j}} \left( x, 
p_t^{\tau } \left( \eta [k] \right) (x)\right) 
 \rho  _t^{\tau }[k]\left( \xi \right)  (x)  p_{tk}^{\tau }
 \left( \eta [k] \right) (x) \surd  
\right) \\
\nonumber
&  &
-  \rho  _t^{\tau }[k]\left( \xi \right)  (x)  
p_{tj}^{\tau } \left( \eta [k] \right) (x) 
\partial_k  
\left( {{\partial H^{ T^*M }_t}\over {\partial p_j}} 
\left( x, p_t^{\tau } \left( \eta [k] \right) (x)\right)  \right) \\
\label{current's-qq}
&  &
+ \rho  _t^{\tau }[k]\left( \xi \right)  (x)  
\partial_k  L^{H^{ T^*M }_t } \left( x, p_t^{\tau } 
\left( \eta [k] \right) (x)\right)  .
\end{eqnarray}
Notice that the above expression
is still valid even if
Hamiltonian $H_t^{T^*M}$
has 
the ambiguity of the operator ordering
such as that for the Einstein gravity.

To elucidate the relationship between the present
theory and canonical quantum mechanics,
we will concentrate on the case
of the canonical Hamiltonian
having the following form:
\begin{equation}\label{q-canonical Hamiltonian}
H^{ T^*M }_t   \left( x , p   \right)
= {1\over 2}
h^{ij} \left( p_{i}   
+ A_{ti}  
\right)
\left( p_{j}   + A_{tj}  
\right)
+U  _t (x) ,
\end{equation}
where $d h^{ij} =0$.
Notice that almost all the canonical 
quantum theory
including the standard model
of the quantum field theory,
that have empirically been 
well-established, 
really belong to this class of  Hamiltonian systems.
For Hamiltonian (\ref{q-canonical Hamiltonian}),
we will define the Hamiltonian
operator $\hat {\bf H}_t$ as
\begin{equation}
\hat {\bf H}_t=
 {1\over 2}
\left( \hat p_i + A _{ti}   \right)
h^{ij }  \left( \hat p_j +A _{tj} \right)  + U_t ,
\end{equation}
or
$
\langle x \vert \hat {\bf H}_t
\vert \psi \rangle
= {\cal H} _t\langle x \vert  
 \psi \rangle 
$
where
\begin{equation}
 {\cal H} _t=  {1\over 2}
\left( -i \hbar \partial_i + A _{ti} (x) \right)
h^{ij }  \left( -i  \hbar \partial_j +A _{tj} (x) \right)  + U_t (x) .
\end{equation}
\begin{lem}
Lie-Poisson equation  (\ref{H[X]qq})
for Hamiltonian (\ref{q-canonical Hamiltonian})
induces the following equation:
\begin{eqnarray}\label{q-liouville 1}
i \hbar {\partial 
\over {\partial t}} 
\left\langle x \left\vert
\hat \rho^{\tau } _t 
[k] \left( \xi \right) 
\right\vert x \right\rangle
&=&
- \left\langle x \left\vert
 \left[   \hat \rho^{\tau } _t [k] \left( \xi \right)  ,
\hat {\bf H}^{\tau } _t [k]   \right]_- 
\right\vert x \right\rangle
\\ \label{q-liouville 2}
i \hbar {\partial 
\over {\partial t}}
\left\langle x \left\vert
{1\over 2}
\left[ \hat \rho^{\tau } _t 
[k] \left( \xi \right) ,
 \hat {\bf p}^{\tau } _t [k]   \right] _+
\right\vert x \right\rangle
&=&
- \left\langle x \left\vert
\left[ \ {1\over 2}
\left[ \hat \rho_t ^{\tau }  [k] \left( \xi \right) 
 \ , \ \hat {\bf H}^{\tau } _t [k]  \right]_-  \ , \ \hat {\bf p}^{\tau } _t [k]   \right]_+
\right\vert x \right\rangle \ .
\end{eqnarray}
\end{lem}
\par\noindent${\it Proof.}\ \ $ 
{\it 
If we define the operators:
\begin{eqnarray}
 \hat {\bf H} _{(0)}  &=& \left. {{1 } \over {2  }}
h^{ij}  \hat {\bf p}^{\tau } _{t i}[k]  \hat {\bf p}^{\tau } _{tj} [k] \right/ i\hbar \\
\hat  {\bf H} _{(1)}     &=& \left.
{{1 } \over {2  }}\{  \hat  {\bf A} _i h^{ij}  
\hat {\bf p}^{\tau } _{tj} [k] +   \hat {\bf p}^{\tau } _{ti} [k] h^{ij}  \hat  {\bf A} _j \}\right/
i\hbar 
\\
\hat  {\bf H} _{(2)}   &=& \left. \left( \hat  U  +
{{1} \over {2 }}h^{ij} \hat  {\bf A} _i  \hat  {\bf A} _j  \right) \right/ i\hbar ,
\end{eqnarray}
then Hamiltonian operator $\hat {\bf H} _t $
can be represented as
\begin{eqnarray}
\left. \hat {\bf H} _t\right/  i\hbar  &=& \hat  {\bf H} _{(0)}   
 +\hat  {\bf H} _{(1)}   +\hat  {\bf H} _{(2)}   .
\end{eqnarray}

Thus, for 
density operator $\hat \rho^{\tau } _t 
[k] \left( \xi \right)   $ defined as equation
 (\ref{tau-dep density op}),
\begin{equation}
{{-1}\over {2  i\hbar  } }\left\langle x \left\vert
\left[ \ 
\left[ \hat \rho_t ^{\tau }  [k] \left( \xi \right) 
 \ , \ \hat {\bf H}^{\tau } _t [k]  \right]_-  \ , \ \hat {\bf p}^{\tau } _t [k]   \right]_+
\right\vert x \right\rangle \ 
= term_{(1)}  \left( \hat  {\bf H} _{(0)}   \right) +term_{(1)} \left(\hat  {\bf H} _{(1)}   \right) 
 +term_{(1)} \left(\hat  {\bf H} _{(2)}  \right) ,
\end{equation}
where
\begin{eqnarray}\nonumber
term_{(1)} \left( \hat  {\bf H} _{(0)}   \right) &=& 
{{-1}\over {2i\hbar }} \left\langle x \left\vert
\left[ \ {1\over 2}
\left[ \hat \rho_t ^{\tau }  [k] \left( \xi \right) 
 \ , \  \hat  {\bf H} _{(0)}   \right]_-  \ , \ \hat {\bf p}^{\tau } _t [k]   \right]_+
\right\vert x \right\rangle \ 
\\ \nonumber
term_{(1)} \left(\hat  {\bf H} _{(1)}   \right) &=&
{{-1}\over {2i\hbar }}\left\langle x \left\vert
\left[ \ {1\over 2}
\left[ \hat \rho_t ^{\tau }  [k] \left( \xi \right) 
 \ , \   \hat  {\bf H} _{(1)}   \right]_-  \ , \ \hat {\bf p}^{\tau } _t [k]   \right]_+
\right\vert x \right\rangle \ 
\\ \nonumber
term_{(1)} \left(\hat  {\bf H} _{(2)}   \right) &=& 
{{-1}\over {2i\hbar }}\left\langle x \left\vert
\left[ \ {1\over 2}
\left[ \hat \rho_t ^{\tau }  [k] \left( \xi \right) 
 \ , \  \hat  {\bf H} _{(2)}     \right]_-  \ , \ \hat {\bf p}^{\tau } _t [k]   \right]_+
\right\vert x \right\rangle \ .
\end{eqnarray}

First term 
results 
\begin{eqnarray}
term_{(1)} \left( \hat  {\bf H} _{(0)}  \right) 
 &=&
 -  \partial_j \left\{ h^{ij}  p_{ti}\left( \eta [k ] \right)   
\rho^{\tau } _t [k] \left( \xi \right) p_{tk}\left( \eta [k ] \right)  \right\}
 dx^k
\end{eqnarray}
from
 the following computations:
\begin{eqnarray}
\left\langle x \left\vert \
\hat {\bf p}^{\tau } _{tk} [k]  
\ \hat \rho^{\tau } _t [k] \left( \xi \right)  \
\hat  {\bf H} _{(0)}     \ \right\vert x \right\rangle
 &=& {1\over 2}\left. 
\int_{{\bf R}^N}d^Nk^{\prime } \
\tilde \rho^{\tau } _t  \left( \xi \right) \left(
k+{k^{\prime }\over 2} ,k-{k^{\prime }\over 2}
\right) e ^{ik^{\prime }\cdot x}
 \ \right\{ \\
&   &  \left( p_{tk}\left( \eta [k ] \right)+  \hbar
 {k_k^{\prime }\over 2} \right)
h^{ij}\left( p_{ti}\left( \eta [k ] \right)-  
\hbar {k_i^{\prime }\over 2} \right)
\left( p_{tj}\left( \eta [k ] \right)-   \hbar
{k_j^{\prime }\over 2} \right)\\
&  & \left. 
+i\hbar \left( p_{tk}\left( \eta [k ] \right)+ 
\hbar {k_k^{\prime }\over 2} \right) h^{ij}
\partial_j\left( p_{ti}\left( \eta [k ] \right)- 
\hbar {k_i^{\prime }\over 2} \right)
\right\}  ;
\end{eqnarray}
\begin{eqnarray}
\left\langle x \left\vert
\
\hat  {\bf H} _{(0)}        \
  \hat \rho^{\tau } _t [k] \left( \xi \right) 
 \
\hat {\bf p}^{\tau } _{tk} [k]  \ \right\vert x \right\rangle
 &=& {1\over 2}\left. 
\int_{{\bf R}^N}d^Nk^{\prime } \
\tilde \rho ^{\tau } _t \left( \xi \right) \left(
k+{k^{\prime }\over 2} ,k-{k^{\prime }\over 2}
\right) e ^{ik^{\prime }\cdot x}
 \ \right\{  \\
&   &  \left( p_{tk}\left( \eta [k ] \right)-  \hbar
{k_k^{\prime }\over 2} \right)
h^{ij}\left( p_{ti}\left( \eta [k ] \right)+ \hbar {k_i^{\prime }\over 2} \right)
\left( p_{tj}\left( \eta [k ] \right)+ \hbar {k_j^{\prime }\over 2} \right)\\
&  & \left. 
-i\hbar  \left( p_{tk}\left( \eta [k ] \right)- \hbar {k_k^{\prime }\over 2} \right) h^{ij}
\partial_j\left( p_{ti}\left( \eta [k ] \right)+ \hbar {k_i^{\prime }\over 2} \right)
\right\}  ;
\end{eqnarray}
\begin{eqnarray}
\left\langle x \left\vert\
\hat \rho^{\tau } _t [k] \left( \xi \right)  \
\hat  {\bf H} _{(0)}     \
\hat {\bf p}^{\tau } _{tk} [k] \ \right\vert x \right\rangle 
&=& {1\over 2}\left. 
\int_{{\bf R}^N}d^Nk^{\prime } \
\tilde \rho ^{\tau } _t \left( \xi \right) \left(
k+{k^{\prime }\over 2} ,k-{k^{\prime }\over 2}
\right) e ^{ik^{\prime }\cdot x}
 \ \right\{ \\
&   &   \left( p_{tk}\left( \eta [k ] \right)- \hbar {k_k^{\prime }\over 2} \right)
h^{ij}\left( p_{ti}\left( \eta [k ] \right)- \hbar {k_i^{\prime }\over 2} \right)
\left( p_{tj}\left( \eta [k ] \right)-  \hbar {k_j^{\prime }\over 2} \right)\\
&   & +i\hbar  
\left( p_{tk}\left( \eta [k ] \right)- \hbar  {k_k^{\prime }\over 2} \right)
h^{ij}\partial_i \left( p_{tj}\left( \eta [k ] \right)- \hbar {k_j^{\prime }\over 2} \right)\\
&   & -\hbar^2
h^{ij}\partial_k\partial_i \left( p_{tj}\left( \eta [k ] \right) - \hbar {k_j^{\prime }\over 2} \right)\\
&   &+i\hbar\left.
h^{ij}\partial_k
\left\{ \left( p_{ti}\left( \eta [k ] \right)- \hbar {k_i^{\prime }\over 2} \right)
\left( p_{tj}\left( \eta [k ] \right)-  \hbar {k_j^{\prime }\over 2} \right) \right\}
\right\}  ;
\end{eqnarray}
\begin{eqnarray}
\left\langle x \left\vert \
\hat {\bf p}^{\tau } _{tk} [k]  \ \hat  {\bf H} _{(0)}    
\ \hat \rho^{\tau } _t [k] \left( \xi \right)  \
 \right\vert x \right\rangle 
&=&{1\over 2}\left. 
\int_{{\bf R}^N}d^Nk^{\prime } \
\tilde \rho^{\tau } _t  \left( \xi \right) \left(
k+{k^{\prime }\over 2} ,k-{k^{\prime }\over 2}
\right) e ^{ik^{\prime }\cdot x}
 \ \right\{ \\
&   & +\left( p_{tk}\left( \eta [k ] \right)+\hbar {k_k^{\prime }\over 2} \right)
h^{ij}\left( p_{ti}\left( \eta [k ] \right) +\hbar {k_i^{\prime }\over 2} \right)
\left( p_{tj}\left( \eta [k ] \right)+  \hbar {k_j^{\prime }\over 2} \right)\\
&   & - i\hbar  
\left( p_{tk}\left( \eta [k ] \right)+  \hbar {k_k^{\prime }\over 2} \right)
h^{ij}\partial_i \left( p_{tj}\left( \eta [k ] \right)+ \hbar {k_j^{\prime }\over 2} \right)\\
&   & -\hbar^2
h^{ij}\partial_k\partial_i \left( p_{tj}\left( \eta [k ] \right) + \hbar {k_j^{\prime }\over 2} \right)\\
&   &
-i\hbar \left.
h^{ij}\partial_k
\left\{ \left( p_{ti}\left( \eta [k ] \right)+ \hbar {k_i^{\prime }\over 2} \right)
\left( p_{tj}\left( \eta [k ] \right)+ \hbar  {k_j^{\prime }\over 2} \right) \right\}
\right\} .
\end{eqnarray}

Further,
\begin{eqnarray}\nonumber
term_{(1)} \left(\hat  {\bf H} _{(1)}    \right)  
&=&- \left\{
 \partial_i \left(
h^{ij}    A _j  
\rho^{\tau } _t [k] \left( \xi \right)   p_{tk}\left( \eta [k ] \right)
\right) \right. \\ 
&  &\left. +\rho^{\tau } _t [k] \left( \xi \right) 
 \left( \partial_k h^{ij}  A_j  \right)
 p_{ti}\left( \eta [k ] \right)  \right\}  dx^k;
\\ 
term_{(1)} \left(\hat  {\bf H} _{(2)}    \right)  
&=&
-  \rho^{\tau } _t [k] \left( \xi \right) 
 \partial_k\left( U  +
{{1} \over {2}}h^{ij}  A _i A _j
\right)  dx^k  .
\end{eqnarray}

Thus, second equation
(\ref{q-liouville 2}) in this lemma
becomes 
\begin{eqnarray}
{\partial \over {\partial t}} 
\left\{ \rho^{\tau } _t [k] \left( \xi \right) 
 p_{tk}\left( \eta [k ] \right) \right\}
&=&
-  \partial_j \left\{ h^{ij}  \left( p_{ti}\left( \eta [k ] \right)  + A _j  \right) 
\rho^{\tau } _t [k] \left( \xi \right) p_{tk}\left( \eta [k ] \right)  \right\}\\ 
&  &  +\rho^{\tau } _t [k] \left( \xi \right) 
p_{tj}\left( \eta [k ] \right)   
 \left( \partial_k h^{ij}  A_i  \right)
\\ 
&   & -  \rho^{\tau } _t [k] \left( \xi \right) 
\partial_k \left( U  +
{{1} \over {2}}h^{ij}  A _i A _j
\right)   ,
\end{eqnarray}
which is equivalent to
equation
(\ref{current's-qq}) for
Hamiltonian (\ref{q-canonical Hamiltonian}).

On the other
hand,
\begin{equation}
 {{-1}\over {i\hbar }}
\left\langle x \left\vert
\  
\left[ \hat \rho_t ^{\tau }  [k] \left( \xi \right) 
 \ , \ \hat {\bf H}^{\tau } _t [k]  \right]_-  \  
\right\vert x \right\rangle \ 
= term_{(2)}  \left( \hat  {\bf H} _{(0)}   \right) +term_{(2)} \left(\hat  {\bf H} _{(1)}   \right) 
 +term_{(2)} \left(\hat  {\bf H} _{(2)}  \right) ,
\end{equation}
where
\begin{eqnarray}\nonumber
term_{(2)} \left( \hat  {\bf H} _{(0)}   \right) &=& 
{{-1}\over {i\hbar }}\left\langle x \left\vert
\
\left[ \hat \rho_t ^{\tau }  [k] \left( \xi \right) 
 \ , \  \hat  {\bf H} _{(0)}   \right]_-  \ 
\right\vert x \right\rangle \ 
\\ \nonumber
term_{(2)} \left(\hat  {\bf H} _{(1)}   \right) &=&
{{-1}\over {i\hbar }}\left\langle x \left\vert
\
\left[ \hat \rho_t ^{\tau }  [k] \left( \xi \right) 
 \ , \   \hat  {\bf H} _{(1)}   \right]_-  \ 
\right\vert x \right\rangle \ 
\\ \nonumber
term_{(2)} \left(\hat  {\bf H} _{(2)}   \right) &=& 
{{-1}\over {i\hbar }}\left\langle x \left\vert
\
\left[ \hat \rho_t ^{\tau }  [k] \left( \xi \right) 
 \ , \  \hat  {\bf H} _{(2)}     \right]_-  \ 
\right\vert x \right\rangle \ .
\end{eqnarray}

Each term can be calculated as follows:
\begin{eqnarray}
term_{(2)} \left( \hat  {\bf H} _{(0)}   \right) 
 &=& {{-1}\over {2i\hbar }}\left. 
\int_{{\bf R}^N}d^Nk^{\prime } \
\tilde \rho^{\tau }_t \left( \xi \right) \left(
k+{k^{\prime }\over 2} ,k-{k^{\prime }\over 2}
\right) e ^{ik^{\prime }\cdot x}
 \ \right\{ \\
&   &
h^{ij}\left( p_{ti}\left( \eta [k ] \right)- \hbar {k_i^{\prime }\over 2} \right)
\left( p_{tj}\left( \eta [k ] \right)- \hbar {k_j^{\prime }\over 2} \right)\\
&  &
+i\hbar  h^{ij}
\partial_j\left( p_{ti}\left( \eta [k ] \right)- \hbar  {k_i^{\prime }\over 2} \right)\\
&  &
-h^{ij}\left( p_{ti}\left( \eta [k ] \right)+ \hbar {k_i^{\prime }\over 2} \right)
\left( p_{tj}\left( \eta [k ] \right)+ \hbar  {k_j^{\prime }\over 2} \right)
\\
&  & \left.
-i\hbar  h^{ij}
\partial_j\left( p_{ti}\left( \eta [k ] \right)- \hbar  {k_i^{\prime }\over 2} \right)
\right\} \\
&=&  - \partial_j
\left(  \rho^{\tau } _t [k] \left( \xi \right) 
h^{ij}
  p_{ti}\left( \eta [k ] \right)  \right)
\\ \nonumber
term_{(2)} \left( \hat  {\bf H} _{(1)}   \right) 
&=&
 -  \partial_i h^{ij} 
\left( A _j 
\rho^{\tau } _t [k] \left( \xi \right) 
\right) ;\\
term_{(2)} \left( \hat  {\bf H} _{(2)}   \right) &=&0 .
\end{eqnarray} 

Thus, first equation
(\ref{q-liouville 2}) in this lemma
becomes
\begin{equation}
{\partial \over {\partial t}} \rho^{\tau } _t [k] \left( \xi \right) 
=
-  \partial_j \left\{ h^{ij}  \left( p_{ti}\left( \eta [k ] \right)   + A _j  \right)
\rho^{\tau } _t [k] \left( \xi \right)   \right\}   ,
\end{equation}
which is equivalent to
equation
(\ref{density's-qq}) for
Hamiltonian (\ref{q-canonical Hamiltonian}).

Therefore,
Lie-Poisson equation  (\ref{H[X]qq})
proved to be equivalent to the equation set
(\ref{q-liouville 1}) and (\ref{q-liouville 2}) in this lemma.
}\hspace{\fill} { \fbox {}}\\

The above lemma
leads us to one of the main theorem
in the present paper,
declaring
that
Lie-Poisson equation  (\ref{H[X]qq})
for Hamiltonian (\ref{q-canonical Hamiltonian})
is
equivalent to the quantum Liouville equation.
\begin{tm}
\label{main theorem}
Lie-Poisson equation  (\ref{H[X]qq})
for Hamiltonian (\ref{q-canonical Hamiltonian})
is equivalent to the following quantum Liouville equation:
\begin{eqnarray}
{\partial \over {\partial t}} \hat \rho  _t  
= \left[   \hat \rho  _t  ,
\hat {\bf H} \right]_- /(-i \hbar ) .
\end{eqnarray}
\end{tm}
\par\noindent${\it Proof.}\ \ $
{\it 
The following computation
proves this theorem
based on the previous lemma:
\begin{eqnarray}\nonumber
{\partial \over {\partial t}}
\left\langle \hat \rho_t \ \hat {\bf F}_t \right\rangle
&=&
{\partial \over {\partial t}}
\left\langle \hat \rho_t ^{\tau } \ \hat {\bf F}_t \right\rangle
\\ \nonumber
&=&
\int_{\Gamma _U }
d{\cal N}\left( \xi \right) 
\int_{{\bf R}^N} d^N k
\int_M dv(x) \times \\  \nonumber
&  &
\left\{ \left\langle  x \left\vert  
  \hat {\bf F}^{\tau } _t [k]    \
\hat \rho_t^{\tau } [k] \left( \xi \right)   \ \hat {\bf H} ^{\tau } _t [k] 
\right\vert x \right\rangle
- \left\langle  x \left\vert  \hat {\bf H}^{\tau } _t [k]  \
\hat \rho_t^{\tau } [k] \left( \xi \right)   
\  \hat {\bf F}^{\tau } _t [k] 
\right\vert x \right\rangle \right. \\ \nonumber
&  &  
+ 
\left\langle  x \left\vert
\ \hat \rho_t ^{\tau }  [k] \left( \xi \right) \ 
 \ \right\vert x \right\rangle 
\ { {\partial p_t^{\tau }[k] (x)}\over {\partial t}} 
\cdot
{\cal D} F _t\left( \eta_t^{\tau } [k]  \right)  (x) 
\\ \nonumber
&  &  
\left.
+
\left\langle  x \left\vert
\ \hat \rho_t ^{\tau }  [k]  
 \ \right\vert x \right\rangle 
p^*
{ {\partial F _t }\over {\partial t}} \left( \eta_t^{\tau } 
[k]  \right) (x) \right\} \\ \nonumber
&=&
\int_{\Gamma _U }
d{\cal N}\left( \xi \right) 
\int_{{\bf R}^N} d^N k
\int_M dv(x) \times \\ \nonumber
&  &
 \left\{
\left\langle  x \left\vert  \left[
\hat \rho_t^{\tau } [k] \left( \xi \right)
 ,  \hat {\bf H}^{\tau } _t [k] 
\right]_-
\right\vert x \right\rangle \
 p^* F_t \left( \eta_t^{\tau } [k] \right) (x) 
\right.\\ \nonumber
&  & 
+ 
\left( 
{\partial \over {\partial t}} 
\left\langle  x \left\vert \
{1\over 2}
\left[
\ \hat \rho_t ^{\tau }  [k] \left( \xi \right) \ 
,  \hat {\bf p}^{\tau } _t [k]  \right] _+ \ \right\vert x \right\rangle 
\right)
\cdot
{\cal D} F_t \left( \eta_t^{\tau } [k]  \right)  (x)
\\ \nonumber
&  & 
- \left\langle  x \left\vert \
{{\partial \hat \rho_t ^{\tau }  [k] \left( \xi \right) }
\over {\partial t}}\ 
 \right\vert x    \right\rangle 
\  p_t^{\tau } [k] (x) 
\cdot
{\cal D} F_t \left( \eta_t^{\tau } [k]  \right)   (x)
\\ \nonumber
&  & \left.
+
\left\langle  x \left\vert
\ \hat \rho_t ^{\tau }  [k]  
 \ \right\vert x \right\rangle 
p^*
{ {\partial F _t }\over {\partial t}} \left( \eta_t^{\tau } 
[k]  \right)
 (x) \right\}  \\ \nonumber
&=&
\int_{\Gamma _U }
d{\cal N}\left( \xi \right) 
\int_{{\bf R}^N} d^N k
\int_M dv(x) \times \\ \nonumber
&  & 
 \left\{
\left\langle  x \left\vert
\left[
\ {1\over 2}
\left[
\ \hat \rho_t ^{\tau }  [k] \left( \xi \right) \ 
,  \hat {\bf p}^{\tau } _t [k] \right] _+ \ ,\ \hat {\bf H}^{\tau } _t [k] 
\right]_- \right\vert x \right\rangle 
\cdot
{\cal D} F_t \left( \eta_t^{\tau } [k] \right)  (x)
\right. \\ \nonumber
&  &  
\left\langle  x \left\vert  \left[
\hat \rho_t^{\tau } [k] \left( \xi \right)
\  , \ \hat {\bf H}^{\tau } _t [k] 
\right]_-
\right\vert x \right\rangle \
\left\{
 p^* F_t \left( \eta_t^{\tau } [k] \right) (x) 
-   p_t^{\tau } [k] (x) 
\cdot
{\cal D} F_t \left( \eta_t^{\tau } [k]
\right)  (x) \right\}
\\ \nonumber
&  & 
\left.
+
\left\langle  x \left\vert
\ \hat \rho_t ^{\tau }  [k]  
 \ \right\vert x \right\rangle 
p^*
{ {\partial F _t }\over {\partial t}} \left( \eta_t^{\tau } 
[k]  \right) (x) \right\} 
\\ 
&=&
\left\langle ad^*_{\hat H_t^{\tau }}{\cal J}_t^{\tau }
, \hat F_t^{\tau }
\right\rangle
+\left\langle {\cal J}_t
, { {\partial \hat F _t }\over {\partial t}}
\right\rangle \
.
\end{eqnarray}
}\hspace{\fill} { \fbox {}}\\

Now, 
the density matrix $\hat \rho_t  $
becomes the summation of
the pure sates $ \left\vert \psi_t^{(l ; \pm )}\right\rangle
\left\langle  \psi_t^{(l ; \pm )} \right\vert   $
for the set $ 
\left\{ \left\vert \psi_t^{(l ; \pm )}
\right\rangle \right\}_{l \in {\bf R}^{N}} $
of the orthonormal wave vectors  
such that 
$\left\langle  \left. \psi_t^{(l^{\prime };s^{\prime })}
\right\vert  \psi_t^{(l; s)}
\right\rangle  
=\delta (l^{\prime }-l ) \delta_{s, s^{\prime }}$:
\begin{equation}
\hat \rho_t  = \int_{\Lambda } 
d P_{+} (l)  
\left\vert \psi_t^{(l ; + )}
 \right\rangle
\left\langle  \psi_t^{(l ; + )}
 \right\vert 
-
\int_{\Lambda } 
d P_{-} (l)  
\left\vert \psi_t^{(l ; - )}
 \right\rangle
\left\langle  \psi_t^{(l ; - )}
 \right\vert  ,
\end{equation}
where  $ P_{\pm }   $
is a corresponding probability
measure 
on the space $\Lambda $ of 
a spectrum
and the employed integral
is the Stieltjes integral \cite{Neumann}.
If the system is open and has the continuous spectrum,
then it admits 
$\Lambda $ be the continuous superselection rules (CSRs).
The induced wave function has the following expression
for a $L^2$-function $\psi_t^{(l ; \pm )}  =\left\langle  x  \left\vert \psi_t^{(l ; - )} \right.
 \right\rangle \in L^2(M)$:
\begin{equation}
\chi_{\alpha }^*\psi_t^{(l ; \pm )} (x )=
\int_{{\bf R}^N}d^Nk \
\tilde  \psi_{\alpha \ t}^{(l ; \pm )} (k) e^{i\{k_j{\bf x}^j +
\zeta_t (x) \} } .
\end{equation}
The existence of the probability
measure $P_-$ would be corresponding to
the existence of the  antiparticle for the elementary
quantum mechanics.

For example,
the motion of the particle
on a N-dimensional rectangle box  $[0, \pi ]^N $
needs the following boundary condition
on the verge of the box:
\begin{quote}
if $x_j = 0$ or $ \pi $  for some $j \in \{ 1,..., N\} $,
then $ \left\langle x \left\vert
\hat \rho_t   \right\vert x \right\rangle = 0 $,
\end{quote}
Density matrix $\hat \rho_t $
is the summation
of  integer-labeled pure states:
\begin{equation}
\hat \rho_t =  
 \sum_{  ( n ,  n^{\prime } ) \in  {\bf Z}^{2N}  }
\ 
\tilde \rho^{t} _{ t} ( n^{\prime }, n )
\left\vert    n  ; t \right\rangle
\left\langle    n^{\prime } ;  t  \right\vert .
\end{equation}

Let us now concentrate on the case
where $\hat \rho_t $ is a pure state in the following form:
\begin{equation}
\hat \rho_t =  \left\vert \psi_t \right\rangle
\left\langle  \psi_t \right\vert  ;
\end{equation}
there exists a wave function $\psi_t  \in L^2 (M)$
\begin{equation}
\psi_t (x)  =
\int_{{\bf R}^N} d^Nk  \
\tilde \psi_t   (k) e^{i\{
k_j {\bf x}^j + \zeta_t (x)  \} }  ,
\end{equation}
where
\begin{equation}
\tilde \rho_t^t (k, k^{\prime }) 
=  
\tilde  \psi_t (k)^* \tilde  \psi_t ( k^{\prime })
.
\end{equation}
Theorem \ref{main theorem}
introduces the Schr\"odinger equation as the following collorary.

\begin{cl}
\label{main theorem for psi}
Lie-Poisson equation  (\ref{H[X]qq})
for Hamiltonian (\ref{q-canonical Hamiltonian})
becomes the following
Schr\"odinger equation:
\begin{equation}\label{Srodinger}
 i \hbar \partial_t \psi_t     
=   {\cal H}
\psi_t   ,
\end{equation}
where
\begin{equation}
 {\cal H} = {1\over {2m}}\surd ^{-1}   
\left( -i \hbar \partial_i +  A _{ti}(x)\right)
g^{ij }(x)\surd  \left( -i  \hbar \partial_j + A _{tj}(x)\right)  +U_t (x) .
\end{equation}
\end{cl}
Therefore,
the presented theory induces not only
canonical, nonrelativistic quantum mechanics
but also the canonical, relativistic or nonrelativistic quantum
field theory if proliferated for the grassmanian field variables.
In addition,
Section 7 will discuss how the present theory 
also justifies
the regularization procedure
in the appropriate renormalization.

On the other hand,
if introducing the unitary transformation $\hat U_{t} = e^{it\hat {\bf H}_t }$,
Theorem \ref{main theorem} 
obtains the Heisenberg equation
for Heisenberg's representations 
$ \tilde {\bf H} _t = \hat U_{t} \hat {\bf H} _t \hat U_{t} ^{-1}$
and $ \tilde {\bf F} _t = \hat U_{t} \hat {\bf F} _t \hat U_{t} ^{-1}$:
\begin{eqnarray}\label{Heisenberg}
 {\partial \over {\partial t}} \tilde {\bf F} _t
= \left[   \tilde {\bf H} _t  ,
\tilde {\bf F} _t\right]_-  / (-i \hbar )+\widetilde{\left( {{\partial {\bf F} _t    }
\over {\partial t}} \right) },
\end{eqnarray}
since $ \hat \rho  _t =\hat U_t^{-1} \hat \rho  _0 \hat U_t  $.

As discussed in Section 3,
 if a  group action of Lie group
$ Q(M)$
keeps the Hamiltonian ${\cal H}_t: q(M)^* \to {\bf R}$  
invariant,
there exists an invariant charge 
functional
$Q : \Gamma \left[ E(M)\right] \to
C(M) $ and the induced
function ${\cal Q}: q(M)^* \to {\bf R}$ such that
\begin{equation}\label{charge-invariant(q)}
\left[ \hat H _t, \hat Q \right]  = 0 ,
\end{equation}
where $\hat Q $ is expressed as
\begin{equation}
\hat Q 
= \left(  {\cal D}_{\rho (\eta )} Q
\left(  p  (\eta ) \right) ,
- p  (\eta )  \cdot     {\cal D}_{\rho (\eta )} Q
\left(  p  (\eta ) \right)
+ Q   \left( p (\eta )  \right)
\right)  .
\end{equation}
Suppose that functional $p^*Q:\Gamma \left[E(M)\right]  \to C(M)$
has the canonical form such that
\begin{equation}
Q ^{T^*M}(x,p)= A^{ij} p_i p_j +  B(x)_i p_j + C(x) ,
\end{equation}
then the corresponding generator
is equivalent to the observable:
\begin{equation}
\hat {\bf Q} = A^{ij} \hat {\bf p}_i  \hat {\bf p}_j +  
\hat {\bf B} _i  \hat {\bf p}_j +  \hat {\bf p}_j \hat {\bf B} _i 
+ \hat  {\bf C} .
\end{equation}
In this case, relation (\ref{charge-invariant(q)})
has the canonical expression:
\begin{equation}\label{charge-invariant(q) op}
\left[ \hat {\bf H} _t, \hat {\bf Q} \right]  = 0 .
\end{equation}
Those operators can have the eigen values at the same time.

As shown so far,
protomechanics successfully deduced quantum mechanics
for the canonical Hamiltonians
that have no problem in the operator ordering,
and proves still valid
for the noncanonical Hamiltonian
that have the ambiguity of the operator ordering
in the ordinary quantum mechanics.
In the latter case,
the  infinitesimal
generator $\hat {\bf F}_t^{tr}$ corresponding to $\hat F
\in q(M)$
is not always equal to
observable $\hat {\bf F}_t$:
\begin{equation}
\hat {\bf F}_t\neq \hat {\bf F}_t^{tr} .
\end{equation}
If one tries to quantize the Einstein gravity,
he or she can proliferate the present theory
in a direct way by utilizing
Lie-Poisson equation (\ref{H[X]qq}).
But,
some calculation method
should be developed for this purpose
elsewhere.

\subsection{Interpretation of Spin}

It has been believed
that a half-integer spin in quantum mechanics
does not have
any classical analogies well-established in classical
mechanics.
Such belief may prevent 
 quantum  mechanics from the realistic interpretation.
This section shows that
the present theory
allows such an classical analogy with a half-integer spin.

Now,
let us consider the
particle motion on space $M^{(3)}$
with the polar coordinate
${\bf x}=(r, \theta ,\phi ) \in [0 , +\infty ) \times
[0 , 2\pi ) \times (0, \pi )$.
The three-dimensional orthogonal group $SO\left( 3,{\bf R}^N\right)
$ acts on ${\cal J}_t= \left(\rho _t^{\tau } p^{\tau }_t
 , \rho_t^{\tau }   \right) $ by the coadjoint
action.
The infinitesimal
generator $ M = M^j  \hat L_j 
\in so\left( 3,{\bf R}^N\right) 
\subset q\left(M \right) $  ($M_j \in {\bf R}$, $ j\in \{1,2,3\}$) 
has an corresponding operator $ \hat {\bf M}=
 M^j \hat {\bf L}_j   \in su(2,{\bf C }) $
that satisfies
\begin{equation}
\left\langle  ad^*_{\hat M} {\cal J}_t ,
 \hat F \right\rangle   =  -i\hbar ^{-1}
  \left\langle    \left[ \hat \rho_t ,  \hat {\bf M}\right] _{-}
\ \hat  {\bf F}  \right\rangle  
 .
\end{equation}
Infinitesimal generator
 $ \hat L_j  $
has the following expression:
\begin{eqnarray}
\hat L_1 & =&    -sin \phi {\partial \over {\partial \theta }}
-\cot\theta \cos\phi   {\partial \over {\partial \phi }}   \
, \\
\hat L_2 & =&   \cos \phi {\partial \over {\partial \theta }}
-\cot\theta \sin\phi   {\partial \over {\partial \phi }}  
\ ,\\
\hat L_3 & =&   {\partial \over {\partial \phi }}   \ .
\end{eqnarray}
while corresponding operator $ \hat {\bf L}_j $ satisfies
\begin{eqnarray}
\left\langle \theta , \phi
\left\vert \hat {\bf L}_1 \right\vert \psi \right\rangle 
& =&    \left\{ - {\hbar \over i}\sin \phi {\partial \over {\partial \theta }}
- {\hbar \over i} \cot\theta \cos\phi   {\partial \over {\partial \phi }}  \right\}  \
\left\langle \theta , \phi
\left\vert  \psi \right. \right\rangle  , \\
\left\langle \theta , \phi
\left\vert \hat {\bf L}_2 \right\vert \psi \right\rangle  & =& 
\left\{ {\hbar \over i} \cos \phi {\partial \over {\partial \theta }}
- {\hbar \over i} \cot\theta \sin\phi   {\partial \over {\partial \phi }}   \right\} 
\ \left\langle \theta , \phi
\left\vert  \psi \right. \right\rangle  ,\\
\left\langle \theta , \phi
\left\vert \hat {\bf L}_3 \right\vert \psi \right\rangle  & =&
 {\hbar \over i}  {\partial \over {\partial \phi }}  \ 
\left\langle \theta , \phi
\left\vert  \psi \right. \right\rangle  .
\end{eqnarray}
Notice that these operators are 
hermite or self-conjugate,
$
\hat {\bf L}_j^{\dag }   = 
\hat {\bf L}_j $,
and induces
the angular momentum or
the integer spin
of the particle:
\begin{eqnarray}
\left\vert   \psi_t   \right\rangle
&=&  
\sum_{m =  -l }^{  l}
\ 
c_m^l(t)
\left\vert    l,m   \right\rangle\\
&  &for  \ \ \ \ \ 
 \left\langle \theta , \phi \right. 
\left\vert l; m\right\rangle
= Y^m_l(\theta , \phi ) ,
\end{eqnarray}
where
\begin{equation}
\left.
 \hat {\bf L}\cdot  \hat {\bf L}
\left\vert    l,m   \right.\right\rangle
= \hbar^2 l(l+1) 
\left\vert l; m\right\rangle \ \ , \ \ \ \
\left.
 \hat {\bf L}_3
\left\vert    l,m   \right.\right\rangle
=   \hbar m 
\left\vert l; m\right\rangle .
\end{equation}

If the Hamiltonian for
the motion in the three-dimensional Euclid space
has the following form
in a central field of force, it is invariant under the rotation about z-axis:
\begin{equation}\label{spin-hamiltonian}
H \left( x, p \right) = p^2 +x\cdot \left( p\times B \right) + U(r) ,
\end{equation}
where $r=\sqrt{x^2+y^2+z^2}\neq 0$.
Since this Hamiltonian has the canonical form,
the corresponding infinitesimal generator is
equivalent to the following quantum observable \cite{Dirac}:
\begin{equation}
\hat {\bf H} = \hat {\bf P_r}^2 +
 {{\hat {\bf L}\cdot \hat {\bf L}  }\over {r^2}}+ 
{1\over 2}
\left\{ \hat {\bf L}\cdot B +B\cdot \hat {\bf L} \right\} + U(r),
\end{equation}
where
\begin{equation}
\left\langle \theta , \phi , r
\left\vert \hat {\bf P_r} \right\vert \psi \right\rangle 
= -{\hbar \over {ir}}{\partial \over {\partial r} } r\left\langle \theta ,
 \phi , r \left\vert \psi \right.\right\rangle .
\end{equation}

On the other hand,
the infinitesimal generators
for the half-integer spin are different from those discussed
in the above for the angular momentum and integer spin:
\begin{eqnarray}
\hat S_1 & =& \left( \  \hat L_1 \  , \
{\hbar \over 2}\cdot {{\cos \phi }\over {\sin \theta }} +
\hat L_1  s \ \right) \
, \\
\hat S_2 & =& \left( \ \hat L_2 \ , \
{\hbar \over 2}\cdot {{\sin \phi }\over {\sin \theta }} 
+\hat L_2  s \ \right)
\ ,\\
\hat S_3 & =&\left( \  \hat L_3 \ ,\ \hat L_3  s  \ \right) \ ,
\end{eqnarray}
where function  $s:M^{(3)}\to
{\bf R}$ represents the gage freedom in electromagnetism.
The corresponding generators in quantum mechanics
becomes
\begin{eqnarray}
\left\langle \theta , \phi
\left\vert \hat {\bf S}_1 \right\vert \psi \right\rangle 
& =&   \left\{ {\hbar \over i}\hat L_1 +
{\hbar \over 2}\cdot {{\cos \phi }\over {\sin \theta }} +
 \left( \hat L_1 s \right)   \right\} 
\ \left\langle \theta , \phi
\left\vert  \psi \right. \right\rangle
, \\
\left\langle \theta , \phi
\left\vert \hat {\bf S}_2 \right\vert \psi \right\rangle  & =& 
 \left\{ {\hbar \over i}\hat L_2
+ {\hbar \over 2}\cdot {{\sin \phi }\over {\sin \theta }} 
+ \left( \hat L_2 s \right)   
\right\} 
\ \left\langle \theta , \phi
\left\vert  \psi \right. \right\rangle ,\\
\left\langle \theta , \phi
\left\vert \hat {\bf S}_3 \right\vert \psi \right\rangle  & =&
\left\{ 
{\hbar \over i}\hat L_3 + \left( \hat L_3 s \right)   \right\}
\ \left\langle \theta , \phi
\left\vert  \psi \right. \right\rangle .
\end{eqnarray}
These operators induce the
half-spin:
\begin{equation}
\left\vert  \psi_t \right\rangle 
=
c_+(t)
\left\vert +  \right\rangle +
c_-(t)
\left\vert -  \right\rangle ,
\end{equation}
where the eigenstates have the following expression:
\begin{equation}
 \left\langle \theta , \phi 
\left\vert  +
\right. 
\right\rangle
=  {1 \over {\sqrt{ 2\pi } } } e^{-i s  } 
e^{i{\phi  \over 2 } }
\cos  {\theta  \over 2 }   \ \ , \ \ \ \
 \left\langle \theta , \phi 
\left\vert  -
\right. \right\rangle
=  {1 \over {\sqrt{ 2\pi } } }  e^{-i s  }
e^{-i {\phi  \over 2 }  }
\sin   {\theta  \over 2 }  .
\end{equation}
They satisfy
\begin{equation}
\left.
 \hat {\bf S}\cdot  \hat {\bf S}
\left\vert    \pm  \right.\right\rangle
=   {3\over 4}\hbar^2
\left\vert \pm \right\rangle  \ \ , \ \ \ \
\left.
 \hat {\bf S}_3
\left\vert   \pm   \right.\right\rangle
=   \pm {\hbar \over 2} 
\left\vert \pm \right\rangle .
\end{equation}
If these ketvectors
are denoted as
\begin{equation}
\left\vert  +
\right\rangle  =\left( {\matrix{ 1\cr
0\cr
}} \right)  \ \ , \ \ \ \
\left\vert  -
\right\rangle  =\left( {\matrix{ 0\cr
1\cr
}} \right) ,
\end{equation}
then,
$  \hat {\bf S}_j =    {\hbar \over 2} \sigma _j $
for the Pauli matrices:
\begin{equation}
\sigma _1 =\left( {\matrix{0&1\cr
1&0\cr
}} \right)
  \ \ , \ \ \ \
\sigma _2 =\left( {\matrix{0&-i\cr
i&0\cr
}} \right) 
 \ \ and  \ \ \ \
\sigma _3 =\left( {\matrix{1&0\cr
0&-1\cr
}} \right)  .
\end{equation}
A general  state 
of
the half-integer spin
of a particle
has the following expression:
\begin{equation}
\left\vert   \psi_t   \right\rangle
= 
\sum_{m =  -l-1}^{  l}
\ 
c_m^{l+1/2}(t)
\left\vert    l+1/2,m +1/2  \right\rangle ,
\end{equation}
where, for the normalization constant $N_{l+1/2}^{m+1/2}$,
\begin{eqnarray}
 \left\langle \theta , \phi \right. 
\left\vert l+1/2; m +1/2 \right\rangle
&=&   N_{l+1/2}^{m+1/2} \sqrt{{l+m+1}\over {2l+1}}
e^{-i s  } 
e^{i{\phi  \over 2 } }
\cos  {\theta  \over 2 } \  Y^{m}_l(\theta , \phi ) \\
& \ & \ \ \
 + N_{l+1/2}^{m+1/2}  \sqrt{{l-m}\over {2l+1}} e^{-i s  }
e^{-i {\phi  \over 2 }  }
\sin   {\theta  \over 2 } \
 Y^{m+1}_l(\theta , \phi ) ;
\end{eqnarray}
and the eigen states satisfy 
\begin{eqnarray}
\left.
 \hat {\bf S}\cdot  \hat {\bf S}
\left\vert    l+1/2,m +1/2    \right.\right\rangle
&=& \hbar^2 (l+1/2)(l+3/2) 
\left\vert  l+1/2; m +1/2 \right\rangle \ \ , \\
\left.
 \hat {\bf S}_3
\left\vert    l+1/2,m +1/2    \right.\right\rangle
&=&   \hbar (m+1/2) 
\left\vert  l+1/2; m +1/2 \right\rangle .
\end{eqnarray}

For Hamiltonian (\ref{spin-hamiltonian}),
the infinitesimal generator  of motion is
equivalent to the following observable:
\begin{equation}
\hat {\bf H} = \hat {\bf P_r}^2 +
 {{\hat {\bf S}\cdot \hat {\bf S}  }\over {r^2}}+
{1\over 2}
\left\{ \hat {\bf S}\cdot (B-C) +
 (B-C)\cdot \hat {\bf S} \right\} + U(r) -C\cdot (B-C)
,
\end{equation}
where
\begin{equation}
C={\hbar \over 2}\left( {x \over {2 \left( x^2 +y^2\right) }}  ,
{y \over {2 \left( x^2 +y^2\right) }} , 0\right) +x\times \nabla s .
\end{equation}
Now, we can investigate the internal structure of such
a
half-integer spin particle, an  quark
or lepton as an
electron or a constituted particle
as a nucleus,
which would have the following spin 
for the internal three-dimensional Euclid space:
\begin{equation}\label{internal spin}
S (x,p)=    x
\times \left( p + \nabla s \right)
 + {\hbar \over 2}\left( {x \over {2 \left( x^2 +y^2\right) }}  ,
{y \over {2 \left( x^2 +y^2\right) }} , 0\right) .
\end{equation}
Such an interpretation
of half-integer spin
allows us to describe the Dirac
equation as the equation of the motion
for the following Hamiltonian:
\begin{equation}
H (x, p,\alpha ,\beta )=   \alpha_1\beta \cdot  \left( p -  {e\over c}A \right) 
+mc^2  \alpha_3
- {e}A_0  ,
\end{equation}
where
$\alpha $ 
and $\beta $
are the internal spins
expressed as relation (\ref{internal spin}).
Since the obtained Hamiltonian
is also canonical as discussed in the previous subsection,
it has the following  infinitesimal generator:
\begin{equation}
\hat {\bf H} =  \left( \hat \gamma_ j
\left( \hat {\bf p}^j  -  {e\over c}A \right) 
+mc^2 \right) \hat \gamma_0
- {e}A_0 ,
\end{equation}
where $ \hat \gamma $ is the Dirac matrices.
In the same way,
the internal freedom like the isospins
of a particle
can be expressed as the invariance of motion,
if its Lie group is a subset
of the infinite-dimensional semidirect-product group
$S(M)$.
More detailed consideration
on the relativistic quantum mechanics
will be held elsewhere.

\section{COMPARISON WITH QUANTIZATION METHODS}

As touched on in Introduction,
 canonical quantum mechanics
has the difficulty for the arbitrariness
on the operator ordering in itself.
The present theory proved so far to
solve this structural difficulty, but
is not the first attempt to overcome it;
some alternative
quantization methods have
tried to resolve it and
helped the birth of the present theory.
Several aspects 
they own are still alive within protomechanics
introduced in this paper.

\subsection{Group-theoretic Property}

For the classical Hamiltonian $H^{T^*M}$ 
on the cotangent space $T^*M$ of a 
 N-dimensional oriented manifold $M$,
we can describe the classical motion of
the
particle having position $ \left( x_t,
p_t\right) \in T^*M $ for the function $F^{T^*M}_t $
on $T^*M$
by using Poisson bracket 
$\{ \ , \ \} $ as
\begin{equation}\label{cl. Poisson eq.rev.}
{{d  } \over {d  t }} F_t^{T^*M}
=\left\{ H_t^{T^*M}, F_t^{T^*M} \right\}     +
{{\partial F_t^{T^*M}} \over {\partial  t }}     .
\end{equation}
In canonical
quantum mechanics,
the corresponding equation of motion is
that 
for associated self-adjoint (or hermit) 
operators
$\tilde  {\bf F}_t $ and $\tilde  {\bf H}_t $
in the Heisenberg representation,
which
 acts on the Hilbert space
with the commutator $[\ ,\ ]$:
\begin{equation}
{{\partial } \over {\partial  t }}\tilde {\bf F}_t 
=\left[ \tilde {\bf H}_t , \tilde {\bf F}_t  \right] /(- i\hbar ) +
\widetilde{ {{\partial {\bf F}_t } \over {\partial  t }} },
\end{equation}
where $i=\sqrt{-1}$.
It should be noticed that
this program 
guarantees the existence of such operators,
but not the possibility of
the concrete expression
for all of them.

The Dirac rule 
that transfers 
position observable $x_j $
and momentum observable $p_j $
into operators 
$\hat {\bf x}_j \to x_j $ and
$\hat {\bf p}_j \to 
-i\hbar  {\partial \over {\partial x_j}}$ 
in the Schr\"odinger representation
usually determines
the correspondence
between functions  $H_t^{T^*M}$, $F_t^{T^*M}$
and operators $\tilde  {\bf F}_t $, $\tilde  {\bf H} _t $;
however, as proved by Groenwald \cite{Groenwald} and van Hove 
\cite{van Hove}, it
can not fully work if one considers
the self-adjoint operator 
$\widehat{ {\bf x}_i^n 
{\bf p}_j^m}$ corresponding to the
classical observable $x_i^n p_j^m$ for 
integers $n>1$ and $m\geq 2$;
the position and the momentum operator
must act with {\it infinite} multiplicity \cite{A&M}.
We can  classify some of the quantizations
into the following  two types,
that avoid such difficulty and that
determine the operator ordering.

\begin{enumerate}
\item deformation:
\begin{quote}
deformation  (Moyal, Bayen, {\it et.al.}) \cite{Moyal,B&F&F&L&D}/
path-integral (Feynmann) \cite{Feynmann} and 
stochastic  (Nelson, Parisi-Wu) \cite{P&W}/
etc.. 
\end{quote}
\item  homomorphism:
\begin{quote}
canonical  (Schr\"odinger, Heisenberg, {\it et.al.}) \cite{Heisenberg,Schrodinger} and 
canonical group (Mackey) \cite{Mackey}/
geometric  (Soriau, Kostant) \cite{Soriau,Kostant}/ etc.;
\end{quote}
\end{enumerate}

The quantizations
in the first category
deform  the Poisson algebra  in classical mechanics
into Moyal's algebra or the generalized one
for quantum mechanics;
Moyal's theory,
as well as the
path-integral and the
stochastic quantization,
can obtain 
observable $ \widehat{ {\bf x}_i^n {\bf p}_j^m}$
as
the Weyl product
$\{ \hat {\bf x}_i ^n \hat {\bf p}_j ^m\} _W$
of operators $\hat {\bf x}_i $ and $ \hat {\bf p}_j $:
\begin{equation}\label{Weyl product}
\widehat{ {\bf x}_i^n {\bf p}_j^m} = \{ \hat {\bf x}_i ^n \hat {\bf p}_j ^m\} _W,
\end{equation}
where, for the set
${\bf Z}^+$ of all non-negative
integers,
\begin{equation}
exp (\alpha \hat {\bf x}_i  + \beta \hat {\bf p}_j )
= \Sigma_{(n ,m )\in {\bf Z}^+
\times {\bf Z}^+  }  {1\over
{n !  \cdot m! }}\alpha ^n \beta  ^m \{ \hat {\bf x}_i ^n \hat {\bf p}_j ^m\} _W,
\end{equation}
These theories can also regard observable $\{ \hat {\bf x}_i ^n \hat {\bf p}_j ^m\} _W$
 as the infinitesimal generator
induced by  function $ x_i^n p_j^m$; thereby,
they have no problem of the ambiguity in the operator ordering,
while they produce the same result for canonical Hamiltonians
with the methods in the second category.
The present theory does {\it not} attribute an infinitesimal generator
to the
Weyl product, though
it has strong similarity with the path-integral method
as discussed in Section 2.

On the other hand, 
the second category
indicates that  each quantization method
belonging to it
bases itself on the homomorphism as a Lie algebra
between the Poisson algebra in classical mechanics
and the operator algebra with commutation relation
in quantum mechanics.
As shown so far,
the present theory safely belongs to the first group 
and postulates that
\begin{quote}
a quantum 
system shares the same group structure
with
the corresponding classical system.
\end{quote}

Sharing the same motivation with the present theory,
Kostant and Soriau \cite{Soriau,Kostant} proposed the
geometric quantization
to overcome the structural difficulty in the canonical theory,
which
succeeded
in constructing a Lie algebra isomorphism between the Lie algebra
for classical observables
and
that for quantum observables
by means of prequantization (on the first step).
They considered a $S^1$-bundle 
(or complex-line bundle) $L  =E\left( P , S^1\right) $
over the {\it symplectic manifold} $P$ with a
symplectic structure $\omega $
where $P=T^*M$ for canonical Hamiltonian systems,
a connection $\hat \nabla $ on $L$ whose
curvature form is $\omega $,
and a metric $\left( \ , \ \right) $ invariant
under the parallel transport.
If $L$ is quantizable when the above
objects are globally well-defined  on $L$,
then the quantization map from a classical
function $F^P$ on $P$
to an operator $\tilde F^P $ on the space $\Gamma \left[
L \right] $ 
of all smooth sections of $L$ is obtained
as follows:
\begin{eqnarray}\label{geometric operator}
\tilde F^P &=& -i\hbar \hat \nabla _{X_{F^P}} + 
F^P\\ \label{GQ}
&=& -i\hbar {X_{F^{T^*M}}} +
 \left( F^{T^*M} -p \cdot {{\partial F^{T^*M}} 
\over {\partial p}}\right) ,
\end{eqnarray}
where the second line (\ref{GQ})
is satisfied when $P=T^*M$.
The operators described as (\ref{GQ})
generate 
the semidirect product group
$S\left( P,{S^1}\right) $
of the space of all diffeomorphisms over $M$
with the space of all smooth functions
on $M$. 
Their quantization method completed itself
by means of polarization (on the second step)
(consult \cite{Kirillov} and \cite{Woodhouse})
and succeeded
in quantizing concretely the important class of classical systems.

If considered as a "quantization method",
the present theory
also overcomes that difficulty in the canonical theory,
and  quantizes
all the classical Hamiltonian systems in a concrete manner,
while 
requiring
no additional correction as
metaplectic correction \cite{R&R} in
the geometric quantization.
In addition, it
considers the $S^1$-fiber bundle
$E = E\left( M , S^1\right) $ 
instead of $L = E\left( T^*M , S^1\right) $,
and newly adds the infinite-dimensional freedom
to  the Hilbert space of all the $L_2$-functions
on $M$
unlike the geometric quantization.
As shown in Section 4,
we obtained the operator $\tilde  F_t$,
corresponding to the observable $F_t^{T^*M}$
on $T^*M$ as an element of Lie algebra $q(M) $:
\begin{equation}\label{new operator}
\tilde  F_t  = \left(
 {{\partial F_t^{T^*M} } 
\over {\partial p}} ,
 F_t^{T^*M} -p \cdot {{\partial F_t^{T^*M}} 
\over {\partial p}}  \right) .
\end{equation}
This form is similar to 
that of the geometric quantization
(\ref{geometric operator})
since both utilize their similar semidirect product
groups.
The induced equation of motion (\ref{general heisenberg})
had the following form for the
operators $\tilde F _t  $ and $\tilde H _t  $:
\begin{equation}
{{\partial }\over {\partial t}}\tilde F_t       = 
\left[ \tilde H _t     ,\tilde F_t      \right] 
+\widetilde{\left( {{\partial F_t    }
\over {\partial t}} \right) } .
\end{equation}
Section 3 elucidated
the difference between classical mechanics and
quantum mechanics
as that of their function spaces,
being the space 
$C_{cl}\left( \Gamma \right) $
of the extended classical observables and the space
$C_{q}\left( \Gamma \right) $ of the extended quantum observables
such that
\begin{equation}
C_{cl}\left( \Gamma \right) 
\subset 
C_{q}\left( \Gamma \right) .
\end{equation}
The integration of the additional infinite-dimensional
freedom was
indispensable not only to deduce from this
equation
the Heisenberg
equation in quantum mechanics
for the canonical Hamilonians,
but also the Poisson equation in classical mechanics
through the classical-limit.\\

\subsection{Statistical Property}

As discussed above, 
classical mechanics and the
quantum mechanics
 basically share a 
group structure,
or have a Lie algebra homomorphism between
their own algebras
not only in the present theory
but also in the quantization methods
belonging to the first group.
The difference between 
those mechanics
comes from
what they act on or
how they are represented.

The Poisson
equation (\ref{cl. Poisson eq.rev.})
is equivalent to the classical Liouville equation
for the probability density
function (PDF) $\rho_t^{T^*M} \in C^{\infty }(T^*M,{\bf R})$
of a particle:
\begin{equation}\label{equation for rho rev.}
{{\partial } \over {\partial t }}\rho^{T^*M}_t  =\{ \rho^{T^*M}_t , H^{T^*M} \} .
\end{equation}
In canonical quantum mechanics,
 the corresponding  equation  of motion 
is  the following
quantum Liouville equation
for the density  matrix 
$ \hat \rho_t =\vert \psi_t \rangle \langle \psi_t \vert$:
\begin{equation}\label{Sro-eq}
{{\partial } \over {\partial t }} \hat \rho_t
=[ \hat \rho_t , \hat {\bf H}_t ] /(-i\hbar )  ,
\end{equation}
which is equivalent to the Schr\"odinger  equation:
\begin{equation}
i\hbar {{\partial }\over {\partial t}}\vert \psi_t \rangle =
\hat {\bf H}_t\vert \psi_t \rangle . 
\end{equation}
A wave function $\psi_t $ for quantum mechanics
 has often been compared
with a delta  function or a point on $T^*M$ for classical mechanics,
while it is not so in the protomechanics. 
We may classify
the quantization methods
in the following way.

\begin{enumerate}
\item  wave function:
\begin{quote}
canonical/
geometric  (Soriau, Kostant) \cite{Soriau,Kostant}/ 
path-integral (Feynmann) \cite{Feynmann} and
stochastic  (Nelson, Parisi-Wu) \cite{P&W}/ etc.; 
\end{quote}
\item density function:
\begin{quote}
hydrodynamic (Mandelung) \cite{Mandelung, Takabayashi}/
etc..
\end{quote}
\item density matrix:
\begin{quote}
 $C^*$-algebraic (Segal) \cite{Segal}/
phase-space  (Wigner, Moyal) \cite{Moyal}/etc.; 
\end{quote}
\end{enumerate}

The quantization methods
belonging to the first category
consider that the unitary group or the corresponding
semigroup
acts on
the $L^2$-space of
wave functions over physical space $M$. 
Besides,
 the hydrodynamic description
of quantum mechanics in  the third category
utilizes the analogy between 
the Schr\"odinger equation and the Euler equation
for the  classical fluid motion,
where  the diffeomrophism group acts on
the  
space  of the density function and the velocity field
over $M$ (consult {\it APPENDIX B} for the group theory
of the classical fluid motion).
On the other hand,
the methods in  the second category
assume
 that the unitary group or the 
deformed group acts on the representation space of the density
matrices.
The quantization 
by density function and those by density matrix
seem based on a belief
as remarked by Moyal \cite{Moyal}:
\begin{quote}
"\ldots\   the fundamental entities
would be the statistical varieties
representing the dynamical parameters of each system;
the operators, matrices and the wave functions
of quantum theory would no longer
be considered as having an intrinsic meaning, but
would appear rather as aids to the calculation of 
statistical averages and distributions."
\end{quote}

As shown so far,
the present theory  belongs to non of the above classification,
but shares the same belief as referred by  Moyal's words,
and postulates that
\begin{quote} 
a quantum 
system shares the same statistical structure
with
the corresponding classical system.
\end{quote}
\noindent
This statistical property
can be different from the "classical" one, but
includes it.
On top of that,
 it is close to the entries in the third category
since it
induces a quantization method belonging to  this class 
for the canonical
Hamiltonian; and 
it inherits the hydrodynamic analogy in the second category
between quantum mechanics and classical mechanics.

Mandelung  \cite{Mandelung}
rewrote (\ref{Sro-eq}) into a hydrodynamics
equation
 that the de Broglie-Bohm theory \cite{Bohm} utilized,
and considered the difference between
classical and quantum mechanics,
and is now summarized
in a different manner from the usual explanation.
For the Hamiltonian system with the
Hamiltonian 
\begin{equation}\label{Ham1}
 H^{T^*M}\left( x_t , p_t \right) 
={1 \over {2m}}
\left( p_t-{{e}\over c}A _t (x_t) \right) ^2 +U(x_t ),
\end{equation}
equation (\ref{equation for rho rev.}) 
induces 
the following 
hydrodynamics equation:
\begin{eqnarray}\label{current equation2}
{{\partial  }\over {\partial t}} \left( \bar \rho_t \bar p_{tj}\right)
+  \partial_i  \left( \bar \rho_t \bar v_{t}^i  \bar p_{tj} \right) 
+ \bar \rho_t \partial^j U 
+{{e}\over c}\bar \rho_t \bar v_{t}^i \partial_j A_{ti} &=&
\partial_i T^i_{tj}
\\ 
\label{density equation2}
{{\partial  }\over {\partial t}}\bar \rho_t  +
 \partial_i \left( \bar \rho_t v_t^i  \right) &=&0 ,
\end{eqnarray}
where averaged
PDF $\bar \rho_t$,  averaged momentum
$  \bar p_{tj}$,  averaged velocity $\bar v_t^j$
and stress tensor
$
T_{tj}^i$ are all defined as 
\begin{eqnarray}
\bar \rho_t(x) &=& \int_{{\bf R}^N} 
d^Np\  \rho_t^{T^*M} \left( x,p \right) ,\ \ \
 \bar p_{tj}(x) = 
\bar \rho_t(x) ^{-1}\int_{{\bf R}^N} d^Np \ \rho_t^{T^*M} 
 \left(x,p \right) p_j ,\\
 \bar v_t^j(x)  &=& 
 \bar \rho_t(x) ^{-1} \int_{{\bf R}^N} d^Np\ \rho_t ^{T^*M}
 \left( x, p \right) 
{{\partial H^{T^*M} }\over {\partial p_j} } 
\left( x,p \right) ,\\
T_{tj}^i(x) &=& -
\int_{{\bf R}^N} 
d^N{\bf p} 
 \left\{ \left(
\bar v_t^i(x) - 
{{\partial  H^{T^*M} }\over {\partial p_i} } \left(
x,p \right)
 \right) \rho_t ^{T^*M}\left(  x,p \right)
(\bar p_{tj} (x) - p_j ) \right\} .
\end{eqnarray}
On the other hand, he transformed
the
wave function $\psi_t (x) = R_t(x) e^{iS_t (x)} $
that
satisfies
the
Schr\"odinger equation
(\ref{Sro-eq}) for canonical Hamiltonian (\ref{Ham1})
into the following
 variables:
\begin{eqnarray}
\bar \rho_t (x) &=& R_t(x)^2 ,\ \ \
\bar p_{tj}(x) = \hbar \partial_j S_t (x)  ,\\
\bar v_{tj}(x) &=& {1\over m}\bar p_{tj}(x) - {{e }\over c}A_{tj}(x),\\
\label{s-tensor}
 T^{qi}_{tj}(x) &=&
 \left(
- 
{{\hbar ^2 }\over {2m}} 
\left\{ \bar \rho (x )^{-1}
\left( \partial^i \bar \rho _t (x ) \right) - \partial^i \right\}
\partial_j \bar \rho _t (x ) \right) ,
\end{eqnarray}
and rewrite the
Schr\"odinger equation
in the following
form:
\begin{eqnarray}\label{CE}
{ {\partial }\over {\partial t}}
\left(  \bar \rho_t \bar p_{tj} \right)
+  \partial_i  \left( \bar \rho_t \bar v_{t}^{i} \bar p_{tj} \right)
+ \bar \rho_t \partial_j U  
+{{e}\over c}\bar \rho_t \bar v_{t}^{i} \partial_j A_{ti} 
&=&
\partial_i T^{qi}_{tj}
\\ 
\label{DE}
{{\partial   }\over {\partial t}}\bar\rho_t +
 \partial_i \left( \bar \rho_t  \bar v_t^{qi} \right) &=&0 .
\end{eqnarray}
If the pressure term in R.H.S.,
so-called the quantum effect,
of equation (\ref{CE})
can be taken to be  equivalent to that in R.H.S.
of equation (\ref{current equation2}),
or if 
$
\partial_i  T^{qi}_{tj}
= \partial_i  T^i_{tj}  
$, 
equations 
(\ref{current equation2})
and (\ref{density equation2})
reduces to equations (\ref{CE})
and (\ref{DE});
but their statistical relationship
seems rather obscure if
one asks what stress tensor $T^{qi}_{tj}$
corresponding to the so-called quantum potential 
is all about.

In order to understand the
mechanism making the difference between the  stress tensors
$T^{i}_{tj}$ and $T^{qi}_{tj}$,
one has to consider the information
of the probability
on the phase space $T^*M$  at least, since
equations (\ref{current equation2}) and (\ref{density equation2})
do not include full information
of the classical equation of motion.
Wigner \cite{Wigner} considered
the 
so-called Wigner function $\rho_t ^W :
{\bf R}^{2N} \to {\bf R}$ defined
for the Fourier transformation $\psi_t (x) = \int_{{\bf R}^N}
dk \ \tilde \psi_t (k) e^{ik\cdot x}$ as
\begin{equation}
\rho_t^W (x , k )= \int_{{\bf R}^N}
d^Nk^{\prime } \ \tilde\psi_t \left( k +{k^{\prime }\over 2}\right) ^* 
\tilde \psi_t \left( k
-{k^{\prime }\over 2}\right)  e^{ik^{\prime }\cdot x} ,
\end{equation}
and compared it with
classical PDF $\rho_t^{T^*M}  $ on $T^*M \simeq {\bf R}^{2N}$:
\begin{equation}\label{Moyal's Wigner}
\rho_t ^W\left(x, {p\over \hbar } \right) = \hbar^N \rho_t^{T^*M} (x, p ) .
\end{equation}
Based on this statistical analogy 
 in conjunction with the hydrodynamic analogy,
the previous letter \cite{Ono}
tried to
reconstruct the quantum Liouville equation with keeping
the Lie algebraic structure
unlike Moyal's theory;
and it proved possible, but not  natural.
 
As already discussed in Section 3 to 6,
the present theory further introduced the infinite-dimensional freedom
to the previous attempt \cite{Ono},
and considers the Lie-Poisson equation
for the extended Lie group $Q\left( M  \right) 
 $:
\begin{equation}   
{{\partial  {\cal J} ^{\tau }_t}\over {\partial t}} 
= ad^*_{\hat H_t  }{\cal J}^{\tau } _t   ,
\end{equation}
whose concrete expressions (\ref{density's-q}) and (\ref{current's-q})
were very similar   not only
to equations (\ref{current equation2})
and (\ref{density equation2}) in classical mechanics
but also to
 equations (\ref{CE})
and (\ref{DE}) in quantum mechanics
without the pressure term.
Section 3 elucidated
the difference between classical mechanics and
quantum mechanics
as that of the dual spaces for function spaces
$C_{cl} \left( \Gamma \right) $ and 
$C_{q} \left( \Gamma \right) $,
being the space $C_{cl}^*\left( \Gamma \right) $
of the classical emergence measures
and the space $C_{q}^*\left( \Gamma \right) $
of the quantum emergence measures:
\begin{equation}
C_{cl}^*\left( \Gamma \right) 
\supset 
C_{q}^*\left( \Gamma \right) .
\end{equation}
The integration of the additional infinite-dimensional
freedom was
indispensable not only to deduce
the Schr\"odinger 
equation 
or the quantum Liouville equation
in quantum mechanics
for the canonical Hamilonians,
but also the classical Liouville equation in classical mechanics
through the classical-limit discussed in the previous section.
The introduced emergence density function $\rho [k]\left( \xi \right) $
for wave number $k$ and additional freedom $\xi $,
further,
produces the Wigner function
after such integration:
\begin{equation}
\rho_t^W \left( x, k \right) = \int_{B\left[ E(M)\right] } 
d{\cal N}\left( \xi \right) \varrho_t\left( \xi \right) [k] (x),
\end{equation}
which satisfies relation (\ref{Moyal's Wigner}) 
with the classical probability density
through classical-limit.
As Moyal remarked,  we
can hardly understand it as a kind of ordinary PDF
by itself,
since they must
generally take
negative as well as positive values
on the phase space $T^*M$.
For this reason,
the present theory
introduced the concept of the
emergence measure in Section 3,
that can have the negative values.

\subsection{Semantics of Regularization}

The present theory
introduces the energy-cutoff $\Lambda_0 $
as the superior of the emergence-frequency $f$:
\begin{equation}
\sup \left\vert f  \left(  \eta   \right) \left( x \right)  \right\vert
= \bar h^{-1}\Lambda_0 .
\end{equation}
In elementary quantum mechanics for the motion
of a particle,
$\Lambda_0 $ is large enough
and almost irrelevant for its formulation,
while,
 in the quantum field theory
where $x$ stands for the value of a field variable
in the above formula,
it justifies the
 regularization procedure
in the renormalization method
that Tomonaga \cite{Tomonaga} and Schwinger \cite{Schwinger}
introduced
and that Feynmann, Dyson \cite{Dyson} and their followers
completed.

Since a  field variable in the field theory can
emerge 
and the created particles interact
with one another
at the vertex in Feynmann's Diagram
when external time $t$ has
countable numbers in the period of $T= 1/f $,
the integration with the high
wave number $k\geq c^{-1}\Lambda  $ 
has nonsense if $\Lambda  \geq \Lambda_0$.
In the standard field theory,
some
one-particle-irreducible Feynmann's diagrams
contain the logarithmic divergences in $\Lambda $.
The energy-cutoff $\Lambda $ and
the constants $g=g(g_0,\Lambda )$
such as the masses,
the coupling constants depending on $\Lambda $
first describes such a theory,
while
every calculated observable
$K=K(g_0,\Lambda ) $ should be independent of $\Lambda $:
$
K=K(g)
$.

For the renormalization parameter
$\mu \leq\Lambda_0 $,
it has the following relation with the dimensional regularization
introduced by 't Hooft and Veltman \cite{tH&V}
that decreases the dimension of the spacetime
as
$4 \to 4-\epsilon $ for small $\epsilon $
and that keeps the Lagrangian density for
the relativistic quantum theory invariant under the Lorentz transformation:
\begin{equation}
\epsilon^{-1} = \ln \left( {\Lambda^2  \over \mu^2 }\right) .
\end{equation}
The minimum subtraction
represents the invariance of the theory
under the variation by $\Lambda $.
In addition, Weinberg
\cite{Weinberg} 
proved that all the standard theories of the
elementary particles
are renormalizable.
Thus,
the present theory can provide such theory with 
the semantics of the regularization,
while the detailed study
in its application will be developed elsewhere.

\subsection{Quantization of Phenomenology}

In addition,
the present theory
can quantize several 
phenomenological systems
possibly having dissipation and/or stochasticity,
since it does {\it not} directly rely on
the Poisson nor the symplectic  structure
on a classical phase space
but only on its group structure;
it can rely on
 the ambient semigroup structure through the generalization.
If we can interpret a phenomenological classical system
as that deduced from the following system
for an operator
$ {\cal L}_t :   q(M)^* \to  q(M)^*$ as in Section 5,
we will make the corresponding quantum system 
under the method discussed in Section 6:
\begin{equation}\label{LL}
{{\partial {\cal J}_t}\over {\partial t}} 
= {\cal L}_t \left( {\cal J}_t \right)   .
\end{equation}
This may be one of the most remarkable features
for its application,
that has not be seen in the other theories.

\section{REALIZATION OF SELF-CONSISTENCY}

Born \cite{Born} 
interpreted the square amplitude
of
a Schr\"odinger's wave function
a probability density function (PDF).
Heisenberg  \cite{Heisenberg2}  further
discovered the uncertainty relations
as a peculiar nature of quantum mechanics:
\begin{equation}\label{uncertainty}
\Delta x \cdot \Delta p   \geq \bar h ,
\end{equation}
where $ \Delta x $ and  $ \Delta p $ are 
the accuracy of the position  variable and the momentum
of a particle.
Such relation showed it impossible
to determine how a particle exists in the sense
of the classical mechanics,
and indicated that
we must give up such an idea of existence
or that of causality.
The present theory
provides a new idea of existence,
and explains how the wave-reduction occurs
in an experiment.

\subsection{Interpretation}

The problem
to provide  a
self-consistent interpretation of reality
has still been 
open 
under the hypothesis that quantum mechanics
is a universal theory.
Some theories 
tried to interpret
quantum mechanics
as the ontic theory
referring object systems,
others
 as the  epistemic
 theory
referring measurement outcomes.
Let me classify some of them as follows \cite{BLM,Bub,Mittelstaedt}:

\begin{enumerate}
\item epistemology: 
\begin{quote}
the Copenhagen (Bohr, Heisenberg) \cite{Bohr}/
orthodox (von-Neumann, Wigner) \cite{Neumann}/ many-worlds (Everett, DeWitt)
 \cite{DG}/ etc..;
\end{quote}
\item ontology: 
\begin{quote}
causal  (de Broglie, Bohm) \cite{Bohm,Holland}/
consistent or decoherence  history (Griffiths, Omn\'es)/
other modal interpretations (van Fraassen) \cite{Fraassen}.\\
\end{quote}
\end{enumerate}

The most important interpretation that
has been
supporting the physics in this century
has been the Copenhagen interpretation classified
in the first category.
Bohr considered that
the referents of quantum mechanics
are observed phenomena,
and that the notion of an individual microsystem 
is meaningful for a human being 
only within the context of the whole macroscopic experimental setup
that should be undoubtedly suffers the classical description
\cite{Bohr}.
This view of {\it complementarity}  becomes a self-consistent 
 epistemic idea
once one admits the following postulates:
\begin{enumerate}
\item the impossibility to understand quantum mechanics
by using "the classical description"
and 
\item the possibility to understand the measuring
devices or the compound system with the object system 
by using "the classical description".
\end{enumerate}
The $C^*$-algebraic theory of quantum mechanics
(refer to monograph \cite{Haag}) 
has developed this interpretation
in a rigorous way,
where the classical property
of the measuring devices can be described by
the continuous superselection rules (CSRs)
in the similar way for the measurement theory as 
many-Hilbert-space theory discussed in the following subsection.
These theories all
identify the objectification with the wave-reduction,
and
need the limit such that  the number of the particles constituting the devices
and that
the 
time spent for the experiment are infinite.
If so, however,
there remain
one
question on this theory
\cite{BLM,Espagnat}:
\begin{quote}
"Can such objectification
allow  some
approximation
or  limiting process for itself?"
\end{quote}
If one believes that
an object can exist {\it approximately},
he or she has to face the problem of the
metaphysics to
understand what it means.

von Neumann 
did not accept Bohr's view
on the second postulate listed above
on the possibility of  the classical description 
for the measuring
devices,
and assumed \cite{Neumann} that 
quantum mechanics
is a universally valid theory which applies equally well to the description
of macroscopic measuring devices as to microscopic atomic objects,
and he faced the problem 
that the object system
and the measuring apparatus had to be separated 
though it is impossible
within quantum mechanics,
and solved that dilemma by
introducing the projection postulate
that the final termination of any measuring process
is the conscious observer.
Many-worlds interpretation \cite{DG}
 evaded this dilemma
to suppose that the quantum theory in Hilbert space
describes some reality
which is composed of many distinct worlds,
and that the observer is aware of himself in only one of these worlds.
Thus, such epistemological theory 
leaves the mind of a human-being
enigmatic thing
{\it beyond} quantum mechanics.
In other words,
these attempts apparently failed
to provide the universality of quantum mechanics
that should describe the human mind,
while the recent development in the neuroscience
would show that the brain seems constituted of  the 
neural networks obeying some physical laws.
If they  admit
this criticism
and conclude that the reality occurs not in the "objective mind"
of the other persons
but
in "Ich" or the subjectivity itself,
they allow themselves to give up 
the illustration itself as some physical problem.

Now,
we can doubt whether or not
{\it the classical description}  represents that of {classical mechanics}.
Apparently, as pointed by Bohm \cite{Bohm},
these words are not equivalent to each other.
The present theory
shall give
the first postulate for Bohr's interpretation
more explicit expression by
substituting the word "mechanics" for "description",
and assumes 
\begin{quote}
the impossibility to understand quantum mechanics
by using "classical mechanics".
\end{quote}
In
Section 2,
we interpreted our mathematical
formalism
in terms of  the 
self-creation or
"self-objectification," 
which can be classical description
in the sense that we can understand
it by using the ordinary language,
but that is not the description
by classical mechanics.
In this sense,
the present interpretation
is  close to 
Bohmian mechanics 
in the second category,
that 
assumes that a particle in quantum mechanics
can exist objectively 
and has a position as its preferred variable 
at every time
even
{\it before} the measurement;
and it
 is a variant of the modal interpretations
\cite{Fraassen}.
Thus, it takes into account that
only the position of a particle can be
directly measured \cite{Bohm},
in other words, 
that the observables such as
the momentum, the angular momentum and the spin
of a particle
are always indirectly observed 
by measuring its position;
and it postulates that
\begin{quote}
a quantum 
mechanics shares
the same ontology with classical mechanics.
\end{quote}
Unlike Bohmian mechanics,
the ontology in the present theory was not the same as 
the traditional one in  classical mechanics
that  each particle has its trajectory
as a complete line in the spacetime,
but the new one that
it appears as "quantized events" in the spacetime;
basically,
this interpretation 
would not change
also
in the quantum field theory
that substitute the value of a field variable
for the position of a particle
in quantum mechanics.

The present
interpretation follows,  in a sense,
the thought
by
Plato in the ancient Greece,
that is
based on the
 distinction between an {\it ideate} 
belonging to "the world of immutable being"
out of our universe,
and a {\it phenomenon}, the appearance of an ideate,
belonging to "the world of generation" within our universe,
and that is
sophisticated by Whitehead 
under some resent
knowledge on the general relativity
and the elementary quantum mechanics 
in his "Philosophy of Organism" or
so-called "Process Philosophy"\cite{Whitehead}
(see \cite{K,M} for some brief summary of Whitehead's
philosophy). 
He considered the actuality
or the existence of 
what he called actual entities
as the process of their self-creation
or the "throb of experience":
they  do not endure in time
and flash in and out of existence
in spacetime.
The present theory supports his philosophy in this
sense, and
assumes that
quantum mechanics
and then classical mechanics
deal with such actual existence.
Whitehead \cite{Whitehead} also dispelled and unified
the distinction between
the subjective and the objective 
that would have
sustained the Western culture,
and, as he indicated,
shared the similar 
idea with the philosophies related with Budism
in the
Eastern culture.\footnote{This may be one reason why
I could easily feel sympathy
with the philosophy of organism
when I knew it
after finishing almost all the mathematical
formulation of the protomechanics, and
why I was
deeply inspired to complete its semantics.}
His philosophy is the inversion of Kant's
philosophy \cite{K,Whitehead}:
\begin{quote}
"for Kant, the world emerges from the subject;
for the philosophy of organism, the subject
emerges from the world."
\end{quote}
Every thing of the world including us
shares the subjectivity with one another
through the individual experience,
being the emergence from  the objectivity.
Protomechanics 
can rely on Whitehead's philosophy,
while the quantum theories have rested on Kant's
in twentieth century.

\subsection{Measurement Process}

In the present theory,
the emergence of a particle
does {\it not} represent the wave reduction
itself,
since the density matrix or the wave function
represents merely a statistical state
of the emergence-momentum.
The wave-reduction should   occur 
through the measurement process
independently of 
the objectification problem;
and it means 
the transformation
of the information 
stored in the object system
to the external system, that sometimes includes
 observers, through the measurement process;
thereby,
it does not sense the objectification itself
nor need
the complete wave-reduction
for such purpose.

There have already exist 
several
theories of the
measurement in
quantum mechanics mainly
in the relation with the objectification problem,
which would be classified in what
the wave-reduction represents \cite{BLM,Bub}:
\begin{enumerate}
\item projection:
\begin{quote}
orthodox (von-Neumann, Wigner, Wheeler)/
relative-state (Everett)/
etc.; 
\end{quote}
\item wave-collapse:
\begin{quote}
nonlinear hidden-variable (Bohm, Bub)/
unified dynamics  (Ghirardi, {\it et.al.})/
ergodic-environment (Daneri, Loinger, Prosperi)/
 etc.;
\end{quote}
\item decoherence:
\begin{quote}
environment  (Zeh, Zurek)/
many-Hibert-space  (Machida, Namiki) \cite{M&N}/ 
algebraic (Hepp, {\it et.al.})/
etc.; \\
\end{quote}
\end{enumerate}

The projection postulates
in the first category
assume in some axiomatic sense
that the wave reduction occurs
in the human mind or abstract "Ich"
who can be aware of the universe where they are living,
as discussed in the previous subsection:
\begin{equation}
\left\vert \psi \right\rangle
\to c _j \left\vert j \right\rangle .
\end{equation}
Thus, they would never explain 
 the wave-reduction as the consequence of
the measurement process.
On the other hand,
the entities in
the second category
attempted to obtain the wave-reduction
as the wave-collapse of a wave function
into new wave functions
by introducing the additional nonlinear effects
into quantum mechanics
or by assuming the irreversible effects
from the environments
(consult \cite{BLM}).
They require some additional
postulates {\it beyond} quantum mechanics.

The present theory
prefers
the entities in the third categories
that considers the wave-reduction as
the decoherence that
the density matrix loses their 
nonorthogonal parts 
after the interaction with 
the measuring apparatus and/or its environment:
\begin{equation}
\hat \rho^{in} =\sum_{j,k}  c_j  c_k^*
\vert j \rangle \langle  k \vert 
 \  \ \ \ \ \to  \ \ \ \ \ \hat \rho^{f} =\sum_{j}  c_j  c_j^*
\vert j \rangle \langle  j \vert ,
\end{equation}
where $ \vert j \rangle $ represents
an eigen vector
for $ \hat F $ with the eigen value $c_j\in {\bf R}$;
and it postulates that
\begin{quote} 
the wave-reduction mechanism
should be explained
within the present frame work.
\end{quote} 

Let us
assume
that the following three processes constitute
the measurement process 
that completes
the measurement of an observable
$\hat F $  through that of some position observables.  
\begin{enumerate}
\item the preparing process
to select an initial  state,
\item
the scattering process
to decompose a spectrum, and
\item
 the detecting process to detect a particle.
\end{enumerate}
They always substitute
the measurement of the position of an observed particle
or a radiated particle like a photon
not only for that of the position itself
but also for that
of the spin, the momentum, or the energy.

Suppose that
the initial wave function
is prepared as $\left\vert \psi^{in} \right\rangle 
= \sum_{j}c_j\vert j \rangle \otimes 
\vert  \phi \rangle  $:
\begin{equation}
\hat \rho^{in}  = \sum_{j,k}  c_j  c_k^*
\vert j \rangle \langle  k \vert \otimes
\vert  \phi \rangle \langle \phi \vert \in \Omega^P 
\end{equation}
where  $
\vert  \phi \rangle $ represents
the wave
function  
for the motion of an observed particle
such that its emergence-frequency (EF) is non-negative
at everywhere:
\begin{equation}\label{preparation}
\rho_{\phi } \left( \eta \right) (x) \geq 0;
\end{equation}
thereby, the
 Wigner function (WF) is non-negative at everywhere
from the discussion in the previous section:
\begin{equation}
\int_{{\bf R}^3}d^3k  \left\langle  \phi \left\vert 
k-{k^{\prime }\over 2}  \right.
\right\rangle 
e^{ik^{\prime }\cdot x}
\left\langle \left.
k+{k^{\prime }\over 2} \right\vert  \phi \right\rangle \geq 0,
\end{equation}
which is the  appropriate
condition considered by Moyal \cite{Moyal} for the prepared Wigner function.
If $ \phi   $ stands for the envelope function of the wave packet,
it will satisfy
this condition.
In addition, the measuring process conserves the positive nature of EF and WF
because of the conservation law of EF
discussed in Section 3.
If EF does {\it not} satisfy
such non-negative property,
not only a particle but also an antiparticle
can appear since
the negative  emergence frequency for a particle  
can be translated as
the positive frequency for
an antiparticle within quantum mechanics,
which will serve an appropriate interpretation
of the relativistic quantum mechanics
without proceeding to the quantum field theory.

 Then,
the spectral decomposition
will change it into
\begin{equation}
\hat \rho^{ex}  = \sum_{j,k}  c_j  c_k^*
\vert j 
\rangle \langle  k \vert
\otimes
\vert  \phi_j \rangle \langle \phi_k \vert \in \Omega^P ,
\end{equation}
where 
$\vert  \phi_j \rangle $ represents
the spatial wave function moving toward the j-th detector.
For the eigen state $\left\vert x^{(j)}\right\rangle $ of the position
of the j-th detector,
the present theory immediately
describes 
the emergence density that the particle appears at position $x= x^{(j)}$
as
\begin{eqnarray}\label{objectification}
\rho^{ex} \left( x^{(j)}\right) &=& \left\langle  x^{(j)} \left\vert   \hat \rho^{ex}
\right\vert  x^{(j)}\right\rangle \\
&=&  \vert c_j \vert^2,
\end{eqnarray}
since $\left\vert  x^{(j)}\right\rangle  = 
\left\vert  \phi_j \right\rangle 
$
by definition.
Notice
that relation (\ref{objectification})
does {\it not} represent the wave-reduction itself.

Machida
and Namiki \cite{M&N}
consider that a macroscopic device 
is an open system that interacts
with the external environment,
and
describes the
state
of the measuring 
apparatus 
by introducing the
continuous super-selection rules (CSRs) 
for 
Hilbert spaces:
 the state of the j-th detector  
is described for
continuous measure $P$ 
on the region $L\subset M$ occupied
with considerable number of atoms
constituting the detector:
\begin{equation}
 \hat \rho^{(j)}   = 
\int_{L } d P (l ) \
\hat \rho (l ) ^{(j)}
\in \Omega^M .
\end{equation}
The present theory admits CSRs within its formalism,
and then justifies their consideration.
Thus, the state of the total system
after the spectral decomposition
is $ \hat \rho^I  =  \hat \rho^{ex}  \otimes 
\prod_{k}  \hat \rho^{(k)} $.
They further utilized the Riemann-Lebesgue
Lemma to induce
the decoherence of the density matrix
 $ \hat \rho^I $
or makes all the
off-diagonal
part zero through the interaction
between the particle and the detector
(consult \cite{M&N,N&P} for the detail illustration):
\begin{equation}\label{reduction}
\hat \rho^I  \to  
\hat \rho^F =
e^{-i t \hat H_0 } \
\hat \rho^{f} \otimes
 \prod_{k}  \hat \rho^{(k)} \
e^{ i t \hat H_0 } 
\end{equation}
where $\hat H_0 $ is the total free Hamiltonian operator
after the interaction.

As shown as above,
the present theory allows the
many-Hilbert-space
theory
successfully to
induce
the wave reduction
in a self-consistent manner.
In addition,
the present theory justifies not only
CSRs indispensable for the proof of the wave reduction
(\ref{reduction})
but also
the utilized
approximation or limiting process
that takes
the particle number consisting
the detector as infinite,
since the wave-reduction in itself 
is independent of the objectification of a particle or a field.

\subsection{Thermodynamic Irreversibility}

In the previous subsection, the decoherence
decreases $H$-function:
\begin{equation}
\left\langle \hat \rho^{in}  \ln   \hat \rho^{in}    \right\rangle = 0
\ \ \ \ \to \ \ \ \ \left\langle \hat \rho^{f}  \ln   \hat \rho^{f}    \right\rangle =
\sum_j \vert c_j \vert^2 \ln  \left( \vert c_j \vert^2 \right) \leq 0.
\end{equation}
If the observer who describes the system
obtains the information where the particle appears
in probability
(\ref{objectification}) through the 
measurement process,
he will know the new initial state of the particle:
\begin{equation}
 \hat \rho^{f}   \ \to \  \vert j\rangle \langle j\vert
\otimes \vert \phi_j\rangle \langle \phi_j \vert ;
\end{equation}
thereby, the entropy becomes zero again.
In this sense,
the entropy represents the incompleteness of the information
for the deterministic description,
and would always increase itself
and cause the irreversibility
 through the interaction
between a macroscopic or open system
and a microscopic system after 
the instability as the spectrum decomposition.

As in the generalized measuring process,
 gas molecules
interact with the macroscopic wall 
constituting the box in which they move,
or with
an open system surrounding the considered area,
 after the {\it instability}
caused by the interaction or the collision among molecules,
that would be expressed as the  {\it nonlinear} terms
in the interaction Hamiltonian in the field theory.
In this case,
the thermodynamic 
irreversibility occurs through the following tree steps:
\begin{enumerate}
\item the knowledge of the initial condition (preparing),
\item 
the instability including nonlinearity (scattering) and
\item
the influence from an open system (detecting).
\end{enumerate}
In the final stage, the wave reduction increases the entropy
without 
the information of all new initial conditions.

In the equilibrium,
the maximum entropy requires that
the infinitesimal variation of the following
thermodynamics potential $\Omega =-pV$
becomes zero
for the grand canonical system:
\begin{equation}\label{potential}
\Omega \left( \beta , \mu , V\right) \left\langle  \hat \rho     \right\rangle 
=
\beta^{-1} \left\langle \hat \rho  \ln   \hat \rho   \right\rangle 
+ \left\langle \hat \rho\   \hat {\bf H}   \right\rangle
- \mu  \left\langle \hat \rho\   \hat {\bf N}   \right\rangle 
+  \beta^{-1}  \left(  \left\langle \hat \rho\     \right\rangle  -1\right) ,
\end{equation}
where 
$\beta ^{-1} $ and
$\mu  $ are  the Lagrange coefficients.
Suppose that  the canonical
Hamiltonian $\hat {\bf H} $ and particle number $\hat {\bf N}   $
have
an
eigen  vector $\left\vert N, E \right\rangle $
for
the eigen values $E$ and $N$:
\begin{equation}
\hat \rho  = \sum_{E,N} \varrho_{E,N}\left\vert E,N \right\rangle
\left\langle E,N \right\vert ,
\end{equation}
the variation of potential $\Omega $
for the coefficients $ \varrho_{E,N}$
induces the following:
\begin{equation}\label{BE,FD}
\varrho_{E,N}    =  e^{\beta \Omega  \left( \beta , \mu , V\right) 
} e^{-\beta \left( E-\mu N \right) }.
\end{equation}
Relation (\ref{BE,FD}) concludes
the  Bose-Einstein and the Fermi-Dirac statistics
for bosons and fermions, respectively,
and also
the Maxwell-Boltzmann statistics
in the high temperature approximation.

If a  Maxwell's devil obtains the full information
of the system to describe the system in a deterministic way, 
he must find
a new initial condition to keep his description whenever only 
one among $10^{23}$ molecules
interacts with the macroscopic wall open to the external
area.
Protomechanics would support such an interpretation
for
the second law of the thermodynamics,
while the detailed consideration
should be held elsewhere.

\subsection{Compatibility with Causality}

Now, the introduced
interpretation of density matrices
would enable us to understand
the causality in quantum mechanics.
On the EPR gedanken experiment \cite{EPR},
the violation
of Bell's inequality \cite{Bell}
does not necessarily contradict with objectivity nor
with 
causality in the present theory,
since
this inequality
relies on the {\it positiveness}
of classical probability density functions.

Consider for example
the system of two spin-${1\over 2}$ particles
that are prepared to move 
in different directions towards two measuring apparatuses
$A$ and $B$
that measure the spin component
along the directions $\alpha $ and $\beta $
respectively.
If  there exists
the 
initial PDF 
depending on the hidden variables
for given quantum mechanical state,
the results of the measurement
at the measuring apparatuses $A=\pm 1$ and $B=\pm 1$
do not depend 
respectively on $\beta $ and $\alpha $
under the locality requirement.
For the probability measure 
$P $ on the space 
$\Lambda $ of all the concerned
hidden variables 
including  that contained
in the apparatus themselves,
the correlation function
$P(\alpha , \beta )$
is defined for a PDF $\rho :
\Lambda \to {\bf R}^+$ 
for the set ${\bf R}^+$ of all non-negative real values as
\begin{equation}
P(\alpha , \beta ) = \int_{\Lambda } dP(\lambda ) \
\rho (\lambda ) A( \alpha , \lambda  ) 
B ( \alpha , \lambda ) .
\end{equation}
Alternative settings $\alpha ^{\prime }$
and $ \beta ^{\prime }$ of the 
measuring apparatuses satisfies
Bell's inequality:
\begin{equation}\label{Bell}
\left\vert P(\alpha , \beta ) - P(\alpha , \beta ^{\prime })\right\vert
+\left\vert P(\alpha ^{\prime } , \beta ^
{\prime }) + P(\alpha ^{\prime }, \beta )\right\vert
\leq 2 ,
\end{equation}
whose proof needs the positiveness of
PDF $\rho $.

In quantum mechanics,
$\hat {\bf A}( \alpha   )
= \alpha^j \sigma_j \otimes 1 $ and $\hat {\bf B}( \beta  )
= 1 \otimes \beta^j \sigma_j $ 
for Pauli matrices $\sigma_j $ ($j=1,2,3$) 
are spin observables
corresponding to
classical ones $ A( \beta , \lambda  )$ and
$ B( \beta , \lambda  )$;
and the
probability operator $\hat \rho $
corresponding to PDF $\rho $
is described as
\begin{equation}
\hat \rho   =  
\vert A \rangle  
\langle A  \vert \otimes \vert
B \rangle  
\langle B  \vert ,
\end{equation}
where $\vert A  \rangle  $
and $\vert B  \rangle  $
are initial wave vectors.
Thus, the correlation function
$P(\alpha , \beta )$ has the following form
in this case:
\begin{equation}
P(\alpha , \beta ) =  
\langle
\hat \rho \ \hat {\bf A}( \alpha    ) \
\hat {\bf B} ( \alpha )  \rangle .
\end{equation}
For this correlation function,
relation (\ref{Bell}) can be violated,
since  
probability operator $\hat \rho $
does not
have such property of the positiveness.

In the present paper, however,
we could interpret such probability operators
as the 
emergence-momentum.
Thus, 
we conclude that
the violation of Bell's inequality
does not deny neither the objectivity
nor the causality in the present context,
since the considered emergence density
can have negative value.
On top of that,
the measurement of 
the spin of a particle
is completed as
that of the corresponding position
after the spectrum decomposition
as discussed in Subsection 8.2;
thereby, we can always the positive emergence
density as the probability
density for the observed values
under preparation condition
(\ref{preparation}).
The same
consideration would
prove that the present theory
 has no contradiction
with any delayed-choice experiments.

\section{CONCLUSION}
The present paper
attempted to reveal the structure behind mechanics,
and proposed a basic theory of
time realizing Whitehead's philosophy.
It induced 
protomechanics that
deepened Hamiltonian mechanics
under the modified  Einstein-de Broglie relation,
and that solved the problem of the ambiguity in the operator ordering
in quantum mechanics.
It further provided a self-consistent interpretation for
quantum mechanics
and examined what is the measurement process.
In addition, the
introduced paradigm
produced conjectures on the following subjects:
\begin{enumerate}
\item interpretation of spin (Subsection 6.3),
\item semantics of regularization (Subsection 7.3),
\item quantization of phenomenological system  (Subsection 7.4),
\item origin of irreversibility (Subsection 8.3) and
\item compatibility with causality (Subsection 8.4).
\end{enumerate}

Needless to say,
the first task will be to apply the present theory
to investigate the behavior of the gravity 
in Planck's scale,
since it solved the operator-ordering problem in quantum mechanics.
On the other hand,
the basic theory presented in Section
2 has nonconstructive nature and is valid whatever the considered scale is,
as discussed in Introduction.
The author considers that such a theory 
will appear through the nonlinearity
of a macroscopic system
and appeal to some experiments in future.
In addition,
the present theory may supply an appropriate
description for the motion of a biological system.
It needs the continuous study
 how to apply the present theory to such systems
and how to check it in experimental ways.

\section*{Acknowledgment}

Sincere
gratitude is extended 
to
Professor Tsutomu Kambe
(University of Tokyo),
Professor Miki Wadati
(University of Tokyo),
Professor 
Tsutomu Yanagida (University of Tokyo),
Professor Ken
Sekimoto (Yukawa Institute for Theoretical
Physics),
Professor Toshiharu Kawai (Osaka City University),
Professor Yasuhide Fukumoto (Kyushu University),
Professor Akira Shudo (Tokyo Metropolitan University),
and Dr. Yasunori Yasui (Osaka City University)
for their discussions;
to Professor Akira Saito
(Tokyo Metropolitan University)
on Moyal's  quantization,
Dr. Katsuhiro Suzuki
(Tokyo Agricultural and Technological College)
on the stochastic quantization,
Professor Tudor
Ratiu (University of California
Santa Cruz)
on  Lie-Poisson mechanics,
Dr. Yasushi Mino (Kyoto University)
on the property of a Wigner function,
Dr. Shigeki Matsutani
on the operator ordering problem, 
Mr. Takashi Nozawa 
on the recent developement
in the neuroscience,
Dr. Atsushi Tanaka (Tsukuba University)
on the measurement problem,
Professor Tomoko Yamaguchi (Tokyo Metropolitan 
Industrial College)
on the divergence in the field theory,
and
Dr. Kyo Yoshida (University of Tokyo)
on  the semantics of the probability
for their advice and discussions.

\setcounter{equation}{0}
\renewcommand{\theequation}{A\arabic{equation}}
\section*{APPENDIX A: INTEGRATION ON MANIFOLD}\label{on Manifold}

Let us here determine  the properties of the manifold $ M$
that is
 the three-dimensional
physical space
for the particle motion
in classical or quantum mechanics,
or the space of graded field variables
for the field motion
in classical or the quantum field theory
(consult \cite{K&N} for more detail information
on manifold theory).

Let  $ \left(  M,  
{\cal O}_M\right) $ be
a Hausdorff space
for the family ${\cal O}_M$ of its open subsets,
and also
a
N-dimensional oriented
$C^{\infty }$
manifold that is
modeled
by the N-dimensional Euclid space ${\bf R}^N$
and thus it
 has an atlas
$\left( U_{\alpha } ,
\varphi_{\alpha }
\right)_{\alpha \in \Lambda_M} $
(the set of a local chart of $M$)
for some  countable set $\Lambda_M$ such that\\
\begin{enumerate}
\item  $M= \bigcup_{\alpha \in \Lambda_M}  U_{\alpha } 
 $,
\item  $\varphi_{\alpha }: U_{\alpha } \to  V_{\alpha }$
is a $C^{\infty }$ diffeomorphism
for some $V_{\alpha }\subset {\bf R}^N$ and
\item  if $ U_{\alpha } \cap  U_{\beta } \neq \emptyset $,
then $ g^{\varphi }_{\alpha \beta }=  
\varphi_{\beta }\circ \varphi_{\alpha }^{-1}  :
 V_{\alpha } \cap  V_{\beta }  \to
 V_{\alpha } \cap  V_{\beta } 
$ is a $C^{\infty }$ diffeomorphism.
\end{enumerate}
The above definition would
be extended
to include that of the infinite-dimensional manifolds
called ILH-manifolds.
A ILH-manifold
that is modeled by the infinite-dimensional  Hilbert space
having an inverse-limit topology instead of ${\bf R}^N$
\cite{Omori}.
We will, however, concentrate ourselves on
the finite-dimensional cases for simplicity.
Let us further assume that
$M$ has no boundary
$\partial M = \emptyset $
for the smoothness of
the $C^{\infty }$ diffeomorphism group $D(M)$
over $M$, i.e.,
in order to consider the 
mechanics on a manifold $N$
that has the boundary $\partial N \neq \emptyset $,
we shall substitute the doubling of $N$ for $M$:
$M=   N\cup \partial   N \cup  N $.

Now, manifold
 $ M $ is
the
topological measure space 
 $  M = ( M , {\cal B}\left( {\cal O}_M \right)
,  {\it vol}) $
that
has the volume measure ${\it vol}$
for the topological $\sigma $-algebra
${\cal B}\left( {\cal O}_M \right) $.
For the Riemannian  manifold $M$,
the (psudo-)Riemannian structure induces
the volume measure ${\it vol}$.

Second, we assume that
the particle moves on manifold $M$ and
has
its internal freedom represented by
a oriented manifold
$F= (F, {\cal O}_F) $,
where ${\cal O}_F$ is the family of
open subsets of $F$.
Let $F= (F, {\cal B}\left( {\cal O}_F\right) ,
m_F) $ be the topological measure space 
with the invariant measure $m_F$
under the group transformation
$G_F$: $\tilde g_* m_F  =m_F $ for 
$\tilde g \in G_F$
where
$\tilde g_* m_F\left(
\tilde g(A)\right)  =m_F (A)$ for $A\in {\cal B}\left(
{\cal O}_F\right) $.
In this case, 
the  state of the particle 
can be
represented as a position  on
the 
locally trivial,
oriented
fiber bundle
$E  = ( E , M  , F , \pi ) $  
with fiber $F$
over $M $
with a canonical projection 
$\pi  :E  \to M $,
i.e., for every
$x \in M$,
there is an open neighborhood $  U (x) $ and a
$C^{\infty }$ diffeomorphism
$\phi_{U } : \pi^{-1}\left( U(x)\right) \to U(x) \times F $ such that
$ \pi = \pi_U \circ \phi_{U } $
for $\pi_U : U(x)\times F \to U (x) :(x, s) \to x $.
Let $G_F$ be
the structure group of fiber bundle $E$:
the mapping 
$\tilde g_{\alpha \beta }=\phi_{U_{\alpha }}
\circ\phi_{U_{\beta }}^{-1}:
U_{\alpha }\cap U_{\beta } \times
F \to U_{\alpha }\cap U_{\beta } \times
F $ satisfies 
$\tilde g_{\alpha \beta }(x,s)  \in G_F$
for $(x,s) \in U_{\alpha } \cap \ U_{\beta } \times F$
and
 the cocycle condition:
\begin{equation}\label{cocycle}
\tilde g_{\alpha \beta } ( x,s )
\cdot \tilde g_{\beta \gamma } ( x,s ) =
\tilde g_{\alpha \gamma } ( x,s )  
\ \ \ \ \ \ \ \
for \ \ ( x,s ) \in U_{\alpha }\cap U_{\beta }\cap 
U_{\gamma } \times F,
\end{equation}
where $\alpha , \ \beta , \ \gamma \in \Lambda_M$;
and
condition (\ref{cocycle}) includes the following relations:
\begin{equation}
\tilde g_{\alpha \alpha } (x,s)  = id. \ \
\ \  for \ \  x\in U_{\alpha }, \ \ \ and \ \ \ \
\tilde  g_{\alpha \beta } (x,s)  =\tilde  g_{\beta \alpha }(x,s) ^{-1} 
\ \ \ \ 
for \ \ (x,s) \in U_{\alpha }\cap U_{\beta }\times F.
\end{equation}
Thus, $\left( E ,
{\cal O}_E\right) $ is the Hausdorff space
for the family ${\cal O}_E$ of the open
subsets of $E$ such that
$ \tilde U \in {\cal O}_E $
satisfies $
\phi_{U_{\alpha } } \left(
\tilde U \right) =
U_{\alpha }  \times 
U_{\alpha }^{\prime } $
for some $U_{\alpha } $ ($\alpha \in \Lambda_M$) and 
$U_{\alpha }^{\prime } \in {\cal O}_F$.

Now, $\left( E ,{\cal B}\left(
{\cal O}_E\right) , m_E \right) $ becomes
the topological measure space
with the measure $m_E$ 
induced by the measures
${\it vol} $ and $m_F $ as follows.
For $A \in
{\cal B}\left( {\cal O}_E \right) $,
there exists the following disjoint union
corresponding to the covering $M=
\bigcup_{\alpha \in \Lambda_M } U_{\alpha } $
such that
\begin{enumerate}
\item $A = \bigcup_{
\alpha \in \Lambda_M }A_{\alpha } $ 
where \ $\pi \left( A_{\alpha }
\right)  \subset U_{\alpha }$, and
\item $A_{\alpha }\cap A_{\beta }
= \emptyset $ for $\alpha \neq \beta $.
\end{enumerate}

Thus, the measure $m_E $
can be defined as
\begin{equation}
m_E (A) = \sum_{\alpha \in \Lambda_M  }
\left( {\it vol} \otimes m_F \right)
\circ \phi_{U_{\alpha }} (A_{\alpha }) 
.
\end{equation}
Notice that the above definition of
$m_E$ is independent of the choice of $\left\{ 
A_{\alpha }\right\}_{\alpha \in \Lambda_M} $
such that $A = \bigcup_{
\alpha \in \Lambda_M }A_{\alpha } $ 
is a disjoint union
since $m_F$ is the invariant measure 
on $F$ for the group transformation of $G_F$.

Let us introduce the
space ${\cal M}\left(E \right) $ of all
the possible {\it probability Radon
measures} for
the particle positions on $E$ 
defined as follows:
\begin{enumerate}
\item every $\nu \in {\cal M}\left(E \right) $
is the linear
mapping $\nu : C^{\infty }(E)\oplus {\bf M} \to {\bf R}$
such that
$\nu (F ) < +\infty  $
for $F \in C^{\infty }(E)$, and
\item for every $\nu \in {\cal M}\left(E \right) $,
there exists a 
$\sigma$-additive positive measure $P$
such that
\begin{equation}
\nu  (F) = \int_{E} dP(y ) \left( F (y )  \right)
\end{equation}
and
 that $ P(M) =1 $, i.e.,
$\nu \left( 1 \right) = 1  $.
\end{enumerate}
For every $\nu \in {\cal M}\left(E \right)$,
the probability density
function
(PDF) $\rho \in L^1 \left(
E, {\cal B}({\cal O}_E)\right)
$ 
is 
the positive-definite, 
and satisfies
\begin{eqnarray}
\label{measure's relation 0}
\nu \left( F    \right)  
&=&\int_{E= \cup_{\alpha \in \Lambda_M}
A_{\alpha } } dm_E ( y )  \
\rho   ( y )  \left(
F  
( y  )   \right) \\
&=& \sum_{\alpha \in \Lambda_M}
\int_{ \phi_{U_{\alpha }}(A_{\alpha })} d{\it vol} (x)\
dm_F (\vartheta )  \
\rho \circ  \phi_{U_{\alpha }}^{-1} ( x , \vartheta )  \left(
F  \circ  \phi_{U_{\alpha }}^{-1} 
( x , \vartheta  )   \right)  ,
\end{eqnarray}
where $dP = dm_E \otimes \rho   $.

\setcounter{equation}{0}
\renewcommand{\theequation}{B\arabic{equation}}
\section*{APPENDIX B: LIE-POISSON MECHANICS}\label{Lie-Poisson Mechanics}

Over a century ago, 
in an effort to elucidate the relationship 
between Lie group theory and classical mechanics,
Lie \cite{Lie} introduced the
 {\it Lie-Poisson system}, being
a Hamiltonian system 
on the dual space of an arbitrary finite-dimensional Lie algebra.
Several years later, as a generalization of
the
Euler equation of a rigid body, Poincar\'e \cite{Poincare} 
applied the standard variational principle on the  
tangent space of an arbitrary finite-dimensional Lie group and
independently
obtained the {\it Euler-Poincar\'e equation}
on the Lie algebra,
being
equivalent to
the {\it Lie-Poisson equation} on 
its dual space if considering no analytical difficulties.
These mechanics structures
for Lie groups
were reconsidered 
in the 1960's
(see  \cite{M&R} for the historical information).
Marsden and Weinstein \cite{M&W2}, in 1974,
proposed the 
{\it Marsden-Weinstein reduction method}  
that allows a Hamiltonian system
to be reduced due to
the symmetry determined by an appropriate
Lie group, while
Guillemin and Sternberg \cite{G&S}
introduced
{\it the collective-Hamiltonian method}  that
describes the equation of motion for a Hamiltonian system as
the Lie-Poisson equation of a reduced Lie-Poisson system.

Let 
$G$ be taken to be  a  finite-  or infinite-dimensional Lie group
and ${  g}$ 
the Lie algebra of $G$; i.e.,
the multiplications $ \ \ \cdot \ \  : G \times G \to G:
(\phi _1 , \phi _2 )\to \phi _1 \cdot  \phi _2 $ 
with a unit $e \in G$
satisfy $\phi _1^{-1} \cdot \phi _2 \in G$
and induce
the commutation relation
$[ \ \ , \ \  ] :  {  g}\times  {  g} \to  {  g} : ( v_1 , v_2 )
\to [ v_1 , v_2]  $.
For a function
$F \in C^{\infty }( G , {\bf R}) $,
two types of derivatives  respectively 
define  the left- and the right-invariant vector
field $v^+ $ and $v^- \in {\cal X}(G)$
in the space ${\cal X}(G)$ of all smooth vector fields
on $G$: 
\begin{eqnarray}
v^+  F(\phi) &=& {d \over  d \tau } \vert_{\tau =0}
F( \phi \cdot  e^{\tau v})
\\
v^-  F(\phi) &=& {d \over  d \tau } \vert_{\tau =0}
F( e^{\tau v} \cdot  \phi )  .
\end{eqnarray}
Accordingly, the left- and  the right-invariant element
of the space ${\cal X}(G)$  satisfy 
\begin{equation}
[ v_1^+
, v_2 ^+]= [ v_1 , v_2]^+ 
,\ \ \
[ v_1^-, v_2^- ]=- [ v_1 , v_2]^- 
,\ \ \ 
and \ \ \ \ [ v_1^+, v_2^- ]=0 .
\end{equation} 
In the subsequent formulation,
$+$ and $-$ denote 
left- and right-invariance, respectively.
In addition, $\langle   \ \   ,  \ \  \rangle :
{  g}^* \times {  g} \to {\bf R}:
(\mu  , v )\to \langle \mu   , v \rangle $ denotes 
the  nondegenerate natural pairing (that is weak in general
\cite{A&M}) for the dual space ${  g}^*$ of the Lie algebra ${ 
g}$, defining 
the left- or right-invariant 1-form
$\mu ^{\pm } \in  \Lambda^1 (G) $ corresponding to
$\mu \in {  g}^*$
by introducing the 
natural pairing  $\langle   \ \   ,  \ \  \rangle :
T^*_{\phi }G \times T_{\phi } G \to {\bf R}$
for $\phi \in G$
as
\begin{equation}
  \langle \mu^{\pm } (\phi ), v^{\pm } (\phi ) \rangle 
= \langle  \mu  , v \rangle .
\end{equation}

Let us now consider
how the motion
on a
Poisson manifold $P$
can be represented by the Lie-Poisson 
equation for $G$ (or its central extension \cite{A&M}),
where $P$ is a finite or infinite Poisson manifold
modeled on $C^{\infty }$ Banach spaces
with Poisson bracket $\{  \ \ , \ \  \} :
C^{\infty }(P, {\bf R})\times C^{\infty }(P, {\bf R}) 
\to C^{\infty }(P, {\bf R})$.
Also, $\Psi :G \times P
\to P$ is an action of $G$ on $P$
such that
the mapping $\Psi_{\phi } : P \to P$ is a
Poisson mapping
for each $\phi  \in G$
in which
$\Psi_{\phi }(y)
= \Psi (\phi , y)$ for $ y \in P $.
It is assumed that 
the Hamiltonian mapping $\hat J : {  g} \to C^{\infty }(P, {\bf R})$
is obtained
for this action
s.t. $X_{\hat J (v)}=v_P $
for $ v  \in {  g}$,
where $X_{\hat J (v)}$
and $v _P \in {\cal X}(P)$ denote 
the Hamiltonian vector field for $\hat J (v) \in C^{\infty }(P, {\bf R})$
and
the infinitesimal generator of 
the action on $P$ corresponding to
$v \in {  g} $, respectively.
As such, the momentum (moment) mapping 
$J : P \to {  g}^*$
is defined 
by $\hat J (v) (y) = \langle J(y), v \rangle $. 
 For 
the special case in which 
$(P,\omega )$ is a
symplectic manifold 
with a symplectic 2-form  $\omega \in \Lambda^2 (G)$ 
(i.e., $d \omega =0 $ and $\omega $ is weak nondegenerate),
this momentum mapping 
is equivalent to that defined by 
$ d\hat J ( v ) = v _P \rfloor \omega $.
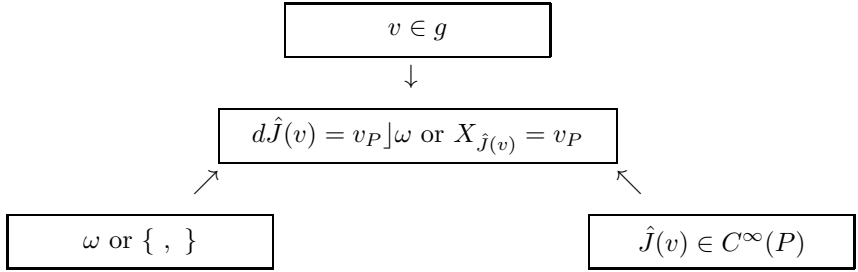
$$ $$
\begin{picture}(400,100)(0,0)
\put (185,80){\framebox(100,20){$v \in {  g} $}}
\put (150,30){$\nearrow $}
\put (310,30){$\nwarrow $}
\put (230,70){$\downarrow $}
\put (300,0){\framebox(100,20){$\hat J (v) \in C^{\infty }(P) $}}
\put (80,0){\framebox(100,20){$\omega $ or
$\{ \ , \ \} $}}
\put (160,40){\framebox(150,20){$ d\hat J ( v ) = v _P \rfloor
\omega $ or $ X_{\hat J
( v )} = v _P $}}
\end{picture}
$$ $$
In twentieth century, 
lots of mathematicians would have based their study
especially on the Poisson structure or the symplectic structure
in
the above diagram,
while the physicists 
would usually have made importance the functions
as the Hamiltonian and the other invariance of motions
as some physical matter.
In  Lie-Poisson mechanics,
the Lie group plays the most important
role as "motion" itself,
while the present theory inherits such an idea.

For the trivial topology
of $G$ (consult \cite{A&M}
in the nontrivial cases),
the Poisson bracket satisfies
\begin{equation} \label{one-cocycle}
\{\hat J (v_1), \hat J(v_2) \} = \pm \hat J([v_1,v_2] )  .
\end{equation}
The {\it Collective Hamiltonian Theorem} \cite{M&R}
concludes
the Poisson bracket for $ A\circ J $ and $ B\circ J
\in C^{\infty }(P, {\bf R})$ 
can be expressed for $\mu = J(y) \in {  g}^*$
 as
\begin{equation} \label{Poisson}
\{ A \circ J , B \circ J \} (y) =   \pm \langle J(y) , 
[{{\partial A }\over {\partial \mu  }}(\mu ),
{{\partial B }\over {\partial
\mu  }}(\mu )] \rangle  ,
\end{equation}
where 
$ 
{{\partial F }\over {\partial \mu  }} :
{  g}^* \to {  g}
$ is the Fr\'echet derivative of 
$ F \in C^{\infty }({  g}^*, {\bf R})$
that  every $\mu \in {  g}^*$ and $ \xi \in {  g}$ satsfies
\begin{equation}
{d \over {d \tau }}\vert_{\tau =0 } F(\mu + \tau \xi )
= \left\langle \xi , {{\partial F }\over {\partial \mu }} (\mu )
\right\rangle .
\end{equation}
Thus,
the collective 
Hamiltonian $H\in  
C^{\infty }( {  g}, {\bf R}) $ such that $ H_P=H\circ J $ 
collects or reduces
 the   Poisson  equation of motion
into the following Lie-Poisson  equation of motion:
\begin{equation} \label{ad*eq}
 {d \over {d t }}\mu_t = 
\pm ad^*_{{{\partial H }\over {\partial  \mu }}(\mu_t )} 
\mu_t ,
\end{equation}
where $ \mu_t = J  (x_t) $ for $x_t \in P$.
We can further obtain the formal solution of Lie-Poisson equation of motion
(\ref{ad*eq}) as
\begin{equation}
 \mu_t   = Ad^*_{\phi_t }\mu_0 ,
\end{equation}
where
generator $\phi_t \in \tilde G$ satisfies
$\{ {{\partial H}\over {\partial \mu }}(\mu_t) \}^{+}
= \phi_t ^{-1}\cdot {{d \phi_t }\over {dt}}  $
or
$\{ {{\partial H}\over {\partial \mu }}(\mu_t) \}^{-}
= {{d \phi_t }\over {dt}}  \cdot \phi_t ^{-1}$
The existence of this solution, however, 
should independently verified
(see \cite{E&M} for example).

In particular,
Arnold \cite{Arnold}
applies such group-theoretic method not only to
the equations of motion of a rigid body
but also to
that of an ideal incompressible fluid,
and constructs them as the motion of a particle
on the three-dimensional special 
orthogonal group $SO(3)$ and as that on the infinite-dimensional Lie group 
$D_v(M)$ of
all $C^{\infty }$ volume-preserving diffeomorphisms
on a compact oriented manifold $M$.
By introducing semidirect products of Lie algebras,
Holm and Kupershmidt \cite{H&K}
and
Marsden {\it et al.} 
\cite{M&R&W} went on to 
complete the method
such that various Hamiltonian systems can be 
treated as Lie-Poisson systems, e.g.,
the motion of a top under gravity and
that of an ideal magnetohydodynamics (MHD) fluid.

For the motion of an isentropic fluid,
the governing Lie group is a semidirect product of 
the Lie group $D
 (M)$
of all $C^{\infty }$-diffeomorphisms
on $M$
with 
$C^{\infty }(M)\times C^{\infty }(M)$, i.e., 
\begin{equation}\label{(3.1)}
G(M)=D(M)\times_{semi.} \left\{ C^{\infty }(M)\times  C^{\infty }(M) 
\right\} .
\end{equation}
For $ \tilde \phi_1=(\phi_1, f_1, g_1)$,
$ \tilde \phi_2=(\phi_2,  f_2, g_2)\in I(M)$, 
the product of two elements of  $I(M)$ is
defined as follows: 
\begin{eqnarray} \nonumber
  \tilde \phi_1 \cdot  \tilde \phi_2 &=&  (\phi_1, f_1,g_1)
\cdot (\phi_2,  f_2,g_1)
\\ \label{(3.2)}
&=&  \left( \phi_1\circ \phi
_2,
\phi_2^* f_1+f_2 ,\phi_2^*
g_1+g_2 
\right) 
\ , 
\end{eqnarray}     
where $\phi^*$ denotes the pullback by $\phi\in
D(M)$ and 
the unit element of $G(M)$ can be denoted as
$(id. ,0, 0) \in G(M)$, where $id. 
\in D (M)$ is the identity mapping from $M$ to itself. 

The
Lie bracket 
for $\tilde v_1 
=( v_1^i\partial
_i, U_1, W_1 ) 
$ and $\tilde v_2 =\left( v_2^i\partial _i ,   U_2, 
W_2 \right)  \in i(M)$ becomes 
\begin{equation}
\left[ \tilde v_1^-, \tilde v_2^-\right] 
=\left( 
 \left[ {v_1^i\partial _i,
v_2^j\partial _j } \right]
,  
v_1^j\partial _jU_2 -
v_2^j\partial _jU_1 ,  
 v_1^j\partial _jW_2 -
v_2^j\partial _jW_1  
\right)  .
\end{equation}
For the volume measure $v$ of $M$,
the element 
of the dual
space $g (M)^*$ 
of the Lie algebra $g (M)$
can
be described  as
\begin{equation}
{\cal J}_t  = \left( dv \ \rho_t \otimes p _t ,
 dv \ \rho_t ,  dv \ \sigma_t \right) ,
\end{equation}
in that $ p _t \in \Lambda^1(M) $,
$ dv \ \rho_t \in \Lambda^3(M)$
 and $ dv \ \sigma_t \in \Lambda^3(M)$
 physically means the 
momentum, the
mass density, and the entropy density.

For the thermodynamic internal energy $ U\left( \rho(x),
\sigma (x)\right) $,
the 
 Hamiltonian 
for the motion of an
isentropic fluid
is introduced as
\begin{equation}
 {\cal H}\left( {\cal J}  \right) =
{1 \over 2}  \int_M dv(x) \ \rho_t (x)  g^{ij}(x) p _{t j} p _{t j}
+
 \int_M dv(x) \ \rho_t (x)  U(\rho_t(x), \sigma_t
(x))     .
\end{equation}
Define
the
operator $\hat F_t =
{{\partial {\cal F} }
\over {\partial {\cal J}}}  \left( {\cal J} _t \right) 
\in g (M)$ 
for every functional $F : g (M)^*
\to {\bf R}$
as
\begin{equation}
\left. {d\over d\epsilon }\right\vert_{\epsilon =0}
{\cal F } \left( {\cal J} _t + \epsilon {\cal K}\right) 
= \left\langle  {\cal K} , \hat F_t  \right\rangle ,
\end{equation}
then,
the Hamiltonian operator $\hat H_t  =
{{\partial {\cal H}_t}\over {\partial {\cal J}}} 
  ({\cal J} _t )\in g(M) $
is calculated for the velocity field $v_t =g^{ij}p_i  \partial_j \in X^1(M)$
as
\begin{equation}
\hat H_t = \left( v^j \partial_j  , 
-{1\over 2}g^{ij }p_{ti} p_{tj} + U\left( \rho_t (x) , \sigma_t (x) \right)
+ \rho_t (x) {{\partial U}\over {\partial \rho }}\left( \rho_t (x) , \sigma_t (x) \right)
,
\rho_t (x) {{\partial U}\over {\partial \sigma }}\left( \rho_t (x) , \sigma_t (x) \right)
\right) .
\end{equation}
The equation of motion
becomes the following Lie-Poisson equation:
\begin{equation}\label{compr.LP}
{{d {\cal J} _t}\over {d t}} 
= ad^*_{\hat H_t  }{\cal J} _t   ,
\end{equation}
which  is  calculated 
as follows:
\begin{enumerate}
\item the conservation laws of
mass and entropy:
\begin{eqnarray}
{{\partial
\bar \rho_t } \over {\partial t }}+  \surd^{-1} 
\partial_j \left( \rho_t v^j _t  \surd \right) &=&0  ,\\
 {{\partial
\bar \sigma_t } \over {\partial t }}+ 
 \surd^{-1} 
\partial_j \left( \sigma_t v^j _t  \surd \right)
 &=&0  ,
\end{eqnarray}
where $\surd = \sqrt{\vert det g^{ij }\vert }$;
\item the conservation law of
momentum:
\begin{equation}
{{\partial  } \over {\partial t}}
\left( \rho  _t  
  p_{tk}    \right)
+
 \surd^{-1} \partial_j \left(
v^j  
\rho  _t    p_{tk}   \surd  
\right)   +  \partial_k P_t =0 ,
\end{equation}
where the pressure $ P_t $ satisfies the 
following condition:
\begin{equation}\label{(3.23)}
 P_t (x) = \rho_t (x) \{
\rho_t (x) { { \partial U } \over {
\partial \rho }}  +\sigma_t (x) { { \partial U } \over
{ \partial \sigma }} \} \left( \rho_t (x) , \sigma_t (x) \right) , 
\end{equation}
which is consistent with the first law of thermodynamics.
\end{enumerate}

Next, we consider $D_{v}(M)  $, being the Lie group of
volume-preserving diffeomorphisms of $M$,
where
every element $\phi \in D_{v}(M) $ satisfies 
$dv \left( \phi (x) \right) =dv (x)$.
Lie group $D_{v}(M)$ is  a subgroup of  $G(M) $,
and inherits its Lie-algebraic structure of.
A right-invariant vector at $T_eD_{v}(M)$
is identified with the corresponding divergence-free
vector field on $M$, i.e.,    
\begin{equation}\label{(2.14)}
u^- (e)=u^i\partial _i
\quad \quad \nabla \cdot {\bf u}=0 \quad for \quad all
\quad x \in M .
\end{equation} 
We can define an operator
$P_{\phi}$ \cite{E&M} that orthogonally projects the elements of
$T_{\phi}G(M)$ onto $T_{\phi}D_v(M)$ for $\phi \in D_v(M)
 \subset G(M)$
such that
\begin{equation}\label{(2.15)}
P_{\phi}[v^- (\phi )]=P[v]^- (\phi )
\end{equation}
and
\begin{equation}\label{(2.16)}
 P[v]^- (e)  =
(v^i - \partial^i \theta )\partial _i, 
\end{equation}
where  $\theta : M \to {\bf R}$ satisfies
$\partial _i(v^i(x) - \partial^i \theta (x))=0$
for every $x \in M$.
This projection
changes   
Lie Poisson equation
(\ref{compr.LP})
into the new Lie-Poisson equation
representing
the Euler equation for the motion
of an incompressible fluid:
\begin{equation}
\quad{{\partial {\bf u}_t} \over {\partial t}}+{\bf
u}_t\cdot \nabla {\bf u}_t+\nabla p=0, 
\end{equation}    
where the pressure $p : M \to {\bf R}$ is 
determined by the condition $\nabla
\cdot {\bf u}_t=0$.

\end{document}